\documentclass[aps,superscriptaddress,showpacs,nofootinbib,11pt]{revtex4-1}
\usepackage{mathrsfs}
\usepackage{bigints}
\usepackage{latexsym,bm}
\usepackage{graphicx}
\usepackage{indentfirst}
\usepackage{slashed} 
\usepackage{amssymb}
\usepackage{amsmath}
\usepackage{bbm}
\usepackage{color}
\usepackage[dvipsnames]{xcolor}
\usepackage{epsfig}
\usepackage[titletoc]{appendix}
\usepackage{rotating}
\usepackage{epstopdf}
\usepackage{extarrows}
\usepackage{tabularx}
\usepackage{soul} 
\usepackage{xcolor}
\usepackage{lipsum}
\usepackage[colorlinks=true,allcolors=BlueViolet,hyperfootnotes=false]{hyperref}

\usepackage[utf8]{inputenc}

\newcommand{\be}{\begin{equation}} \newcommand{\ee}{\end{equation}}
\newcommand{\ba}{\begin{array}{c}} \newcommand{\ea}{\end{array}}
\newcommand{\bea}{\begin{eqnarray}} \newcommand{\eea}{\end{eqnarray}}

\newcommand\bstrut{\rule[-1.2ex]{0pt}{0pt}} 

\begin{document}

\title{\Large New physics effects on  
 $\Lambda_b\to \Lambda^*_c\tau\bar\nu_\tau$  decays}

\author{Meng-Lin Du}
\affiliation{Instituto de F\'{\i}sica Corpuscular (centro mixto CSIC-UV), 
Institutos de Investigaci\'on de Paterna,
C/Catedr\'atico Jos\'e Beltr\'an 2, E-46980 Paterna, Valencia, Spain}

\author{Neus Penalva}
\affiliation{Instituto de F\'{\i}sica Corpuscular (centro mixto CSIC-UV), 
Institutos de Investigaci\'on de Paterna,
C/Catedr\'atico Jos\'e Beltr\'an 2, E-46980 Paterna, Valencia, Spain}

\author{Eliecer Hern\'andez}
\affiliation{Departamento de F\'\i sica Fundamental 
  e IUFFyM,\\ Universidad de Salamanca, Plaza de la Merced s/n, E-37008 Salamanca, Spain}
\author{ Juan Nieves}
\affiliation{Instituto de F\'{\i}sica Corpuscular (centro mixto CSIC-UV), 
Institutos de Investigaci\'on de Paterna,
C/Catedr\'atico Jos\'e Beltr\'an 2, E-46980 Paterna, Valencia, Spain}


\date{\today}
\begin{abstract}
We benefit from a recent lattice determination of the full set of vector, axial and tensor
form factors  for  
the  $\Lambda_b\to \Lambda^*_c(2595)\tau\bar\nu_\tau$ and $\Lambda^*_c(2625)\tau\bar\nu_\tau$
 semileptonic decays to study the  possible 
 role of these two reactions in lepton flavor universality violation studies. Using an effective theory approach, 
 we analyze different observables that can be accessed through the visible kinematics of the charged particles produced
in the tau decay, for which we consider the  $\pi^-\nu_\tau,\rho^-\nu_\tau$
 and $\mu^-\bar\nu_\mu\nu_\tau$ channels.  We compare the results obtained in the Standard Model and  other schemes 
 containing new physics (NP) interactions, with either left-handed or right-handed neutrino operators.
 We find a discriminating power between models similar to the one of the 
 $\Lambda_b\to \Lambda_c$ decay, although
  somewhat hindered in this case by the larger errors of the 
 $\Lambda_b\to\Lambda^*_c$ 
 lattice form factors. Notwithstanding  this,  the analysis of these reactions is already able
 to discriminate  between some of the NP scenarios and
  its potentiality will certainly  improve when more precise form factors are available.
\end{abstract}

%


\maketitle

\section{Introduction}
The experimental observation of the Higgs boson by the ATLAS~\cite{ATLAS:2012yve}
and CMS~\cite{CMS:2012qbp} collaborations announced the completion of the
electroweak sector of the Standard Model (SM).  Despite its enormous success 
in describing many different experimental
data, there are however theoretical indications (see for instance
chapter 10 of Ref.~\cite{langacker:2017}) as well as experimental measurements 
 that hint at the possibility of the SM 
 being just a low
energy effective limit of a more fundamental underlying theory. One of the predictions of the SM is lepton flavour
universality (LFU), which implies that the
couplings to the $W$ and $Z$ gauge bosons is the same for all three 
lepton families. However, this prediction is being challenged by different
semileptonic decays mediated by charged currents (CC) involving the third
lepton and quark generation, i.e. by $b\to c\tau^-\bar\nu_\tau$ transitions.
The strongest evidence in the direction of LFU violation comes from the  ratios
${\cal R}_{D^{(*)}}=\frac{ \Gamma(\bar B\to D^{(*)}\tau^-\bar\nu_\tau)}
{\Gamma(\bar B\to D^{(*)}\mu^-\bar\nu_\mu)}$ measured by the 
BaBar~\cite{BaBar:2012obs, BaBar:2013mob}, Belle~\cite{Belle:2015qfa, 
Belle:2016ure,Belle:2016dyj,Belle:2019rba} and LHCb~\cite{LHCb:2015gmp, 
LHCb:2017smo,LHCb:2017rln} collaborations. Their combined analysis by the HFLAV
collaboration indicates a $3.1\,\sigma$ tension with SM 
predictions~\cite{HFLAV:2019otj}. LHCb~\cite{LHCb:2017vlu} has also measured the ratio
${\cal R}_{J/\psi}=
  \Gamma(\bar B_c\to J/\psi\tau\bar\nu_\tau)/\Gamma(\bar B_c\to
   J/\psi\mu\bar\nu_\mu)$, which deviates from the SM predictions~\cite{Anisimov:1998uk,Ivanov:2006ni,
Hernandez:2006gt,Huang:2007kb,Wang:2008xt,Wen-Fei:2013uea, Watanabe:2017mip, Issadykov:2018myx,Tran:2018kuv,
Hu:2019qcn,Leljak:2019eyw,Azizi:2019aaf,Wang:2018duy} at the $1.8\,\sigma$ level. 
If these differences were finally confirmed they would be a clear indication
for the necessity of new physics (NP) beyond the SM.

 A model-independent way to approach this problem  is to take a phenomenological point of view and
 to carry out an effective field theory analysis, which  includes the most general $b\to c \tau^- \bar\nu_\tau$ dimension-six operators  (for one of the pioneering works on 
this type of approaches, see
Ref.~\cite{Fajfer:2012vx}). These operators are assumed to be generated  by physics beyond the SM. 
Their strengths
  are encoded into  unknown Wilson coefficients { (WCs)} that can be determined by fitting to experimental data. %
 In order to constrain and/or determine the most plausible extension of the
SM, observables beyond the above-mentioned LFU ratios need to be 
considered. Those observables typically include the averaged  tau-polarization  
asymmetry  and the  longitudinal  $D^*$ polarization, which have also been measured
by Belle~\cite{Belle:2016dyj, Belle:2019ewo}, the $\tau$ forward-backward
asymmetry
and the upper bound of the
$\bar B_c\to \tau\bar\nu_\tau$ leptonic decay rate~\cite{Alonso:2016oyd}.
A large number of studies  along these lines have been conducted, not
only for the  
 $\bar B \to D^{(*)}$~\cite{Nierste:2008qe, Tanaka:2012nw, Fajfer:2012vx, 
Duraisamy:2013pia,Duraisamy:2014sna,Becirevic:2016hea, Ligeti:2016npd, Ivanov:2017mrj,
Bernlochner:2017jka, Blanke:2018yud, Bhattacharya:2018kig, Colangelo:2018cnj,Murgui:2019czp, 
Shi:2019gxi, Alok:2019uqc, Mandal:2020htr, Kumbhakar:2020jdz,Iguro:2020cpg, 
Bhattacharya:2020lfm,  Penalva:2021gef,Penalva:2020ftd} and   
$\bar B_c\to J/\psi,\eta_c$~\cite{Dutta:2017xmj,Tran:2018kuv,Leljak:2019eyw,Harrison:2020nrv, Penalva:2020ftd} 
semileptonic decays, but also  for the  $\Lambda_b \to \Lambda_c$ 
transition~\cite{Dutta:2015ueb,Shivashankara:2015cta, Li:2016pdv,Datta:2017aue,Ray:2018hrx,
Blanke:2018yud,Bernlochner:2018bfn,DiSalvo:2018ngq,Blanke:2019qrx,
Boer:2019zmp,Murgui:2019czp,Mu:2019bin,Hu:2020axt, Penalva:2019rgt, 
Penalva:2020xup, Penalva:2021gef}, where  a similar  behavior is to be 
expected. 
A better discriminating power for different models could be achieved if four body
reactions, involving for instance $D^*\to D\pi, D\gamma$
~\cite{Duraisamy:2013pia,Duraisamy:2014sna,Becirevic:2016hea, Ligeti:2016npd, 
Colangelo:2018cnj,
Bhattacharya:2020lfm, Mandal:2020htr} or  
$\Lambda_c\to \Lambda \pi$~\cite{Boer:2019zmp,Hu:2020axt} decays of the final hadron, are analyzed. 

Very recently, the LHCb collaboration has reported 
a measurement of the 
 ${\cal
R}_{\Lambda_c}=\frac{\Gamma(\Lambda_b\to\Lambda_c\tau^-\bar\nu_\tau)}
{\Gamma(\Lambda_b\to\Lambda_c\mu^-\bar\nu_\mu)}$ ratio~\cite{LHCb:2022piu} and
 the experimental value
${\cal R}_{\Lambda_c}=0.242 \pm 0.026 \pm 0.040 \pm 0.059$   turns out to be
in agreement,  within errors,
with the SM prediction ${\cal R}_{\Lambda_c}^{\rm SM}=0.332 \pm 0.007 \pm 
0.007$~\cite{Detmold:2015aaa}. The $\tau^-$ lepton  was reconstructed using 
the  hadronic $\tau^-   \to\pi^-\pi^+\pi^-(\pi^0)\,\nu_\tau$ decay, with the same technique 
 used by the LHCb experiment to obtain 
${\cal R}_{ D^*}=0.291 \pm 0.019 \pm 0.026 \pm 0.013$~\cite{LHCb:2017rln}, also in agreement with the SM prediction. 
A higher value ${\cal R}_{ D^*}=0.336 \pm  0.027 \pm  0.030$, however, 
was obtained by the same experiment when the  $\tau$ lepton was reconstructed
 using its leptonic decay into a muon~\cite{LHCb:2015gmp}.
It is then of great interest to see if the above result for the
$\Lambda_b\to\Lambda_c$ decay is confirmed or not when the muonic reconstruction
channel is used. Such an analysis is under way~\cite{Marco}. As already discussed  in Ref.~\cite{Penalva:2022vxy}, 
 the different deviation  of  the present ${\cal R}_{\Lambda_c}$ and $R_{D^{(*)}}$ ratios with respect to their SM values,  
suppression for ${\cal R}_{\Lambda_c}$ versus enhancement for $R_{D^{(*)}}$,  puts a very stringent test on NP extensions of the SM, since 
scenarios 
leading to different deviations from SM expectations  seem to be required. 
In this respect, in the very recent work of  Ref.~\cite{Bernlochner:2022hyz}, 
it is argued that a more consistent comparison with the SM prediction for 
${\cal R}_{\Lambda_c}$ is achieved if the  recent  $\Gamma(\Lambda_b\to\Lambda_c\tau^-\bar\nu_\tau)$ LHCb
 measurement is normalized against
the SM value for $\Gamma(\Lambda_b\to\Lambda_c\mu^-\bar\nu_\mu)$ instead of the old LEP data used by the 
LHCb collaboration. This analysis gives rise  to a new  ${\cal R}_{\Lambda_c}= |0.04/V_{cb}|^2 
(0.285 \pm 0.073)$ value~\cite{Bernlochner:2022hyz}, also in agreement with the  SM  but with a 
less suppressed central value.

In Refs.~\cite{Penalva:2021wye,Penalva:2022vxy} the 
$\Lambda_b\to\Lambda_c\tau^-\bar\nu_\tau$ and the $\bar B\to
D^{(*)}\tau^-\bar\nu_\tau$ decays were analyzed by employing the
$\tau^-\to\pi^-\nu_\tau,\rho^-\nu_\tau$ and $\tau^-\to\mu^-\bar\nu_\mu\nu_\tau$
reconstruction channels. There, an  special attention is paid to  
different quantities 
that can be measured by looking just at the visible kinematics of the charged particle
produced in the $\tau$ decay~\cite{Penalva:2021wye}.
Given a good-statistics measurement 
of these visible distributions, one has access to the values of the unpolarized differential decay width 
$d\Gamma_{{\rm SL}}(\omega)/d\omega$  and the spin
$\langle P^{\rm CM}_L\rangle(\omega),\langle P^{\rm CM}_T\rangle(\omega)$,
angular $A_{FB}(\omega),A_Q(\omega)$, and angular-spin 
$Z_L(\omega), Z_\perp(\omega),Z_Q(\omega)$ asymmetries.  Here $\omega$ is the product of the
two hadron four-velocities. As shown in
Ref.~\cite{Penalva:2021wye}, in the absence of CP violation, the above 
quantities  provide the maximal information that can be extracted from the analysis 
of the semileptonic $H_b\to H_c\tau^-\bar\nu_\tau$ decay for a polarized 
final $\tau$. The
general expression that links the visible-kinematics differential distributions 
  to the above given asymmetries  was first given
in Ref.~\cite{Asadi:2020fdo} for the $\tau\to\pi^-\nu_\tau,\rho^-\nu_\tau$
hadronic decay modes. Actually, these hadronic channels are more convenient to determine
all the above asymmetries and  the role of the latter in distinguishing among
different extensions of the SM was analyzed in detail in 
Refs.~\cite{Penalva:2021gef,Penalva:2021wye}. Since the full
visible-kinematics differential decay width may suffer from low statistics,  
possible statistically enhanced distributions, which can be obtained by 
integrating in one or more of the related visible-kinematics variables, are analyzed in Ref.~\cite{Penalva:2022vxy} in the search for NP.

In the present work, we will extend this kind  of studies to the $\Lambda_b\to \Lambda^*_c(2595)$ and $ \Lambda_b\to \Lambda^*_c(2625)$ semileptonic decays,  with  the help of the  recent lattice Chromodynamics (LQCD) determination of the full set of vector, axial and tensor
form factors  for  
these two transitions~\cite{Meinel:2021rbm,Meinel:2021mdj}. These two isoscalar odd parity resonances, with $J^P=\frac{1}{2}^-$ and $\frac{3}{2}^-$ respectively, are promising  candidates for the lightest charmed baryon heavy-quark-spin doublet~\cite{Leibovich:1997az, Du:2022fxg}\footnote{Some doubts on this respect have recently been put forward \cite{Nieves:2019nol, Nieves:2019kdh}, and experimental distributions for the semileptonic decay of the ground-state bottom baryon $\Lambda_b$ into both  excited states would definitely contribute to shed light into this issue~\cite{Du:2022fxg}.}.
The LFU analysis of the transitions involving these excited baryons could provide valuable/complementary information on the possible existence 
of NP beyond the SM and on its
preferred extensions. 
The LQCD form factors in Refs.~\cite{Meinel:2021rbm,Meinel:2021mdj} are defined based 
on a helicity decomposition of the amplitudes. After
extrapolation to the  physical point (both the continuum and the physical pion
mass limits), each form factor was parameterized in terms of $\omega$ as
$f(\omega)=F^f+A^f(\omega-1)$, corresponding to the first order Taylor expansion around the zero 
recoil point ($\omega=1$). That was appropriate since lattice data were only available for
 just two kinematics  near zero recoil. Thus, one expects this parameterization  to be reliable only
for small values of $(\omega-1)$ and, in accordance,  
we shall restrict our evaluation of the different
observables to a certain kinematical region near zero recoil.

This work is organized as follows: in  Sec.~\ref{sec:eh} we will introduce the
most general effective Hamiltonian of all possible dimension-six operators for the semileptonic $b\to
c$ transitions. We give general analytical results valid for the production of any lepton in the
final state, although it is generally assumed that the { WCs} are
nonzero only for the third quark and lepton generation. We also provide 
 the general expression for the transition amplitude squared for the production
 of a charged lepton in a given polarization state. In Sec.~\ref{sec:sequential}
 we present the general formula for the visible-kinematics differential decay
 width for the sequential 
 $H_b\to H_c\tau^-(\pi^-\nu_\tau,\rho^-\nu_\tau,\mu^-\bar\nu_\mu\nu_\tau)
 \bar\nu_\tau$ decays and the expressions after integration in one or more of 
 the related
 variables. The results and the discussion are presented in Sec.~\ref{sec:results}. In 
 Appendices~\ref{app:12p12m} and \ref{app:12p32m} we collect the matrix elements (form
 factor decomposition) and the $\widetilde W_\chi$ structure functions needed to construct the hadron tensors  for the $1/2^+\to 1/2^-$ and $1/2^+\to 3/2^-$ transitions, respectively.

\section{$H_b\to H_c\ell^-\bar\nu_\ell$ Effective Hamiltonian and decay amplitude}
\label{sec:eh}
Following Ref.~\cite{Mandal:2020htr}, we use an effective low energy
Hamiltonian that includes 
 all dimension-six semileptonic $b\to c$ operators with both left-handed (L) 
 and right-handed (R)
neutrino fields,
\bea
H_{\rm eff}&=&\frac{4G_F V_{cb}}{\sqrt2}\left[(1+C^V_{LL}){\cal O}^V_{LL}+
C^V_{RL}{\cal O}^V_{RL}+C^S_{LL}{\cal O}^S_{LL}+C^S_{RL}{\cal O}^S_{RL}
+C^T_{LL}{\cal O}^T_{LL}\right.\nonumber \\
&&+\left. C^V_{LR}{\cal O}^V_{LR}+
C^V_{RR}{\cal O}^V_{RR}+C^S_{LR}{\cal O}^S_{LR}+C^S_{RR}{\cal O}^S_{RR}
+C^T_{RR}{\cal O}^T_{RR} \right]+h.c. ,
\label{eq:hnp}
\eea
with\footnote{Note that tensor operators with different lepton and quark chiralities
vanish identically. It directly follows from
\be
\sigma^{\mu\nu}(1+h_\chi \gamma_5)\otimes \sigma_{\mu\nu}(1+h_{\chi'} \gamma_5) = 
(1+h_\chi h_{\chi'})\sigma^{\mu\nu}\otimes \sigma_{\mu\nu}- (h_\chi+h_{\chi'})\frac{i}{2} 
\epsilon^{\mu\nu}_{\ \ \,\alpha\beta}\sigma^{\alpha\beta}\otimes \sigma_{\mu\nu},
\ee
where we use the convention $\epsilon_{0123}=+1$.
} 
\be
{\cal O}^V_{(L,R)L} = (\bar c \gamma^\mu b_{L,R}) 
(\bar \ell \gamma_\mu \nu_{\ell L}), \, {\cal O}^S_{(L,R)L} = 
(\bar c\,  b_{L,R}) (\bar \ell \, \nu_{\ell L}), \, {\cal O}^T_{LL} = 
(\bar c\, \sigma^{\mu\nu} b_{L}) (\bar \ell \sigma_{\mu\nu} \nu_{\ell L}),
\label{eq:hnp2}
\ee
\be
{\cal O}^V_{(L,R)R} = (\bar c \gamma^\mu b_{L,R}) 
(\bar \ell \gamma_\mu \nu_{\ell R}), \, {\cal O}^S_{(L,R)R} = 
(\bar c\,  b_{L,R}) (\bar \ell \, \nu_{\ell R}), \, {\cal O}^T_{RR} = 
(\bar c\, \sigma^{\mu\nu} b_{R}) (\bar \ell \sigma_{\mu\nu} \nu_{\ell R}),
\label{eq:hnp2R}
\ee
and where $\psi_{R,L}= (1 \pm \gamma_5)\psi/2$,  $G_F=1.166\times 10^{-5}$~GeV$^{-2}$  
and $V_{cb}$ is the corresponding Cabibbo-Kobayashi-Maskawa matrix element.

The 10, complex in general,  { WCs} $C^X_{AB}$ ($X= S, V,T$ and 
$A,B=L,R$) parameterize 
the deviations from the SM. 
They could be lepton and flavor dependent although they  are generally assumed to be nonzero 
only for the third 
quark and lepton generation. 
The transition amplitude for a $H_b\to H_c\ell^-\bar\nu_\ell$
decay can be written, in a short-hand notation, as
\be
{\cal M} = \left(J_{H}^\alpha J^{L}_\alpha+ J_{H} J^{L}+ 
J_{H}^{\alpha\beta} J^{L}_{\alpha\beta}\right)_{\nu_{_{\ell L}}}+\left(J_{H}^\alpha J^{L}_\alpha+ J_{H} J^{L}+ 
J_{H}^{\alpha\beta} J^{L}_{\alpha\beta}\right)_{\nu_{_{\ell R}}}, \\
\ee
where the two contributions correspond to the two different neutrino
chiralities. In the $m_{\nu_\ell}=0$ limit there is  no interference 
between these two terms and 
$|{\cal M}|^2$ is given by an incoherent sum of $\nu_{_{\ell L}}$ and 
$\nu_{_{\ell R}}$
contributions.

The  lepton currents for a fully polarized  charged lepton are given by
\bea
J^{L(\alpha\beta)}_{\chi, h S} &=& \frac{1}{\sqrt{2}} 
\bar u_\ell^S(k';h) \Gamma^{(\alpha\beta)} P_5^{h_\chi} v_{\bar\nu_\ell}(k), 
\nonumber\\ \Gamma^{(\alpha\beta)}&=& 1,\gamma^\alpha,\sigma^{\alpha\beta}, \quad
 P_5^{h_\chi} = \frac{1+h_\chi\gamma_5}{2},
  \label{eq:lepton-current}
 \eea
with $u^S_\ell(k'; h)$ the spinor of the final charged lepton corresponding to a state with
$h=\pm 1$ polarization (covariant spin) 
along a certain four-vector $S^\alpha$.\footnote{This $u^S_\ell(k'; h)$ spinor is defined by the
condition
\be
\gamma_5\slashed{S}\,u^S_\ell(k'; h)=h\,u^S_\ell(k'; h).
\ee 
where the four-vector $S^\alpha$ satisfies the constraints $S^{\,2}=-1$ and  $S\cdot k'=0$.
A helicity state corresponds to  $S^\alpha= (|\vec{k}'|, k^{\prime 0}\hat k')
/m_\ell$, with $\hat k'=\vec{k}'/|\vec{k}'|$ and $m_\ell$ the charged lepton
mass.}
$h_\chi=\pm1$ accounts for the two possible neutrino chiralities ($h_\chi=-1$ and $+1$ for $\chi=L$ and $\chi=R$, respectively) considered in the effective
Hamiltonian.
From the lepton currents one can readily obtain the  corresponding lepton  tensors
 needed to evaluate  $|{\cal M}|^2$. 
 They are constructed as
 \bea
 L^{(\alpha\beta)(\rho\lambda)}_{\chi,hS}=J^{L(\alpha\beta)}_{\chi, h
 S}(J^{L(\rho\lambda)}_{\chi, h S})^*=\frac12{\rm
Tr}\,[(\slashed{k'}+m_\ell)\Gamma^{(\alpha\beta)}P_5^{h_\chi}\slashed{k}\gamma^0
\Gamma^{(\rho\lambda)\dagger}\gamma^0P_{S}^h],
 \eea
where we have taken $m_{\nu_\ell}=0$ and $P_{S}^h$ stands for the projector
\bea
P_{S}^h=\frac{1+h\gamma_5\slashed{S}}2.
\eea
The final expressions  for the lepton tensors
have been collected in Appendix~B of Ref.~\cite{Penalva:2021wye}.

The dimensionless  hadron currents are given by 
\be
J_{H rr'\,\chi(=L,R)}^{(\alpha\beta)}(p,p') =  \langle H_c; p',r'| \bar c(0) 
O_{H\chi}^{(\alpha\beta)}b(0) | H_b; p, r\rangle, \label{eq:JH1}
\ee
with the hadron states  normalized as $\langle \vec{p}\,', r'| \vec{p},
r\rangle= (2\pi)^3(E/M)\delta^3(\vec{p}-\vec{p}\,')\delta_{rr'}$ and where 
$r,r'$ represent the spin index.  The different $O_{H\chi}^{(\alpha\beta)}$ operator 
structures are
\be
O_{H\chi}^{(\alpha\beta)}= (C^S_{\chi}+h_\chi C^P_{\chi} \gamma_5),\  (C^V_{\chi}\gamma^\alpha+ h_\chi C^A_{\chi} \gamma^\alpha\gamma_5),
 \  C^T_{\chi}\sigma^{\alpha\beta} (1+h_\chi\gamma_5). \label{eq:Jh}
 \ee
The  { WCs} above are obtained as linear 
combinations of those introduced in the effective Hamiltonian in Eq.~\eqref{eq:hnp} and their expressions  can be found in 
 Appendix~A of Ref.~\cite{Penalva:2021wye}. 
The hadron tensors that enter the evaluation of $|{\cal M}|^2$ are defined as
\be
W_\chi^{(\alpha\beta)(\rho\lambda)}=\overline{\sum_{r,r'}}
\langle H_c; p',r'| \bar c(0) 
O_{H\chi}^{(\alpha\beta)}b(0) | H_b; p, r\rangle 
\langle H_b; p, r|\bar b(0)\gamma^0O_{H\chi}^{(\rho\lambda)\dagger}\gamma^0c(0)| H_c; p',r'\rangle, 
\ee
where we sum (average) over the spin of the final (initial) hadron. As  discussed in detail in
 Ref.~\cite{Penalva:2020xup}, the use of Lorentz, parity and time-reversal  
transformations of the hadron currents and states~\cite{Itzykson:1980rh}  allows 
one to write  general expressions for the hadron tensors valid for any $H_b\to H_c$ transition. They are linear combinations of independent tensor and pseudotensor structures, constructed out of the vectors $p^\mu$, $q^\mu$, the metric tensor $g^{\mu\nu}$ and the Levi-Civita pseudotensor $\epsilon^{\mu\nu\delta\eta}$. The coefficients of the independent structures are  scalar functions of the four-momentum transferred squared $q^2$, denoted by $\widetilde{W}_\chi$ as introduced in Refs.~\cite{Penalva:2021wye}. The different $\widetilde W_\chi$ scalar structure 
 functions (SFs) depend on the  { WCs} $C_\chi^{V,A,S,P,T}$ and on the genuine  hadronic responses, the matrix elements of the involved hadron operators which can be derived from the form factors 
 parameterizing each particular transition. 
 It is shown in Refs.~\cite{Penalva:2021wye,Penalva:2020xup} that there is a total of  16
 independent $\widetilde W_{\chi}$ SFs for each neutrino chirality, with the
  $\widetilde W_{R}$ SFs directly obtained from the  $\widetilde W_{L}$ ones by the replacements
$C^{V,A,S,P,T}_{\chi=L}\to C^{V,A,S,P,T}_{\chi=R}$. 
 The different $W_\chi^{(\alpha\beta)(\rho\lambda)}$ hadron tensors, together with the definition of the 
 $\widetilde W_{\chi}$ SFs are compiled 
 in Appendix~C of Ref.~\cite{Penalva:2021wye}.
 
 As shown here in Appendix~\ref{app:12p12m}, the $\widetilde{W}_\chi$  SFs for the 
 $\Lambda_b\to\Lambda^*_c[J^P=\frac12^-\,]$ transition  can be easily obtained from those in Appendix C 
 of Ref.~\cite{Penalva:2021wye} by replacing $C_\chi^V\longleftrightarrow  C_\chi^A$ and 
 $C_\chi^S\longleftrightarrow  C_\chi^P$. In addition, the genuine hadron $W_{i=1,2,4,5}^{VV,AA}$, 
 $W_{i=3}^{VA}$, $W_{1,2,3,4,5}^T$, $W_{S}$, $W_{P}$, $W_{I1,I2}^{VS,AP}$, $W_{I3}^{ST,PpT}$ and 
 $W_{I4,I5,I6,I7}^{VT,ApT}$ SFs, which are independent of the { WCs}, can be read out from  Eqs.~(E3)-(E5) 
 of Ref.~\cite{Penalva:2020xup}  for the $\Lambda_b\to\Lambda_c$ transition\footnote{Note that  
 the names of the form factors in the parametrizations of  Eqs.~\eqref{eq:ffv}-\eqref{eq:ffpt} 
 are chosen in order to directly use the results of  Eqs.~(E3)-(E5) of Ref.~\cite{Penalva:2020xup}.} . 
 
 On the other hand, the $\widetilde{W}_\chi$ SFs for the $\Lambda_b\to\Lambda_c^* [J^P=\frac32^-\,]$ 
 decay are explicitly calculated in this work and they are given in Eqs.~\eqref{eq:w32-ini}-\eqref{eq:w32-fin} 
 of Appendix~\ref{app:12p32m}.

Going back to the amplitude squared, it was shown in Refs.~\cite{Penalva:2020xup,Penalva:2021gef} 
that for the 
production of a
charged lepton with polarization $h$ along the four vector $S^\alpha$, and for massless
neutrinos, one has that
\begin{eqnarray}
&&\hspace{-1.cm}\frac{2\,\overline\sum\, |{\cal M}|^2 }{M^2}= 
\frac{2\,\overline\sum\, \left(|{\cal M}|_{\nu_{\ell L}}^2 + |{\cal M}|_{\nu_{\ell R}}^2\right) }
{M^2}=
{\cal N}(\omega, p\cdot k) + h\bigg\{ \frac{(p\cdot S)}{M}\,
{\cal N_{H_{\rm 1}}}(\omega, p\cdot k) \nonumber\\
&&\hspace{6cm}+\frac{(q\cdot S)}{M}\,
{\cal N_{H_{\rm 2}}}(\omega, p\cdot k)+\frac{\epsilon^{ S k' qp}}{M^3}\,{\cal N_{H_{\rm 3}}}(\omega, p\cdot k)
 \ \bigg\},\label{eq:pol}
\end{eqnarray}
where we have summed (averaged) over the polarization state of the final (initial) hadron.
As already mentioned, $\omega$ is the product of the two hadron four-velocities and it is related to 
$q^2$ via $q^2=M^2+M^{\prime2}-2MM'\omega$, with $M$ ($M'$) the mass of the
initial (final) hadron. Besides, we have made use of the notation 
$\epsilon^{ S k' qp}=\epsilon^{\alpha\beta\rho\lambda}S_\alpha k'_\beta q_\rho 
p_\lambda$. As for the   
 ${\cal N}$ and  $\cal N_{H_{\rm 123}}$ scalar functions, they are given by
 \bea
 {\cal N}(\omega, k\cdot p)&=&\frac12\Big[{\cal A}(\omega)
+{\cal B}(\omega) \frac{(k\cdot p)}{M^2}+ {\cal C}(\omega) 
\frac{(k\cdot p)^2}{M^4}\Big],\nonumber\\
{\cal N_{H_{\rm 1}}}(\omega, k\cdot p)&=&{\cal A_H}(\omega)
+ {\cal C_H}(\omega)
 \frac{(k\cdot p)}{M^2},\nonumber\\
{\cal N_{H_{\rm 2}}}(\omega, k\cdot p)&=& {\cal B_H}(\omega)
  + {\cal D_H}(\omega) \frac{(k\cdot p)}{M^2}+ {\cal E_H}(\omega) 
  \frac{(k\cdot p)^2}
  {M^4},\nonumber\\ 
 {\cal  N_{H_{\rm 3}}}(\omega, k\cdot p)&=&{\cal F_H}(\omega)+ 
 {\cal G_H}(\omega)\frac{(k\cdot p)}{M^2},\label{eq:pol2}
  \eea
The term ${\cal  N_{H_{\rm 3}}}$ is proportional to
the imaginary part of SFs, which requires the existence of relative phases between some of the complex  { WCs}, thus incorporating
violation of the CP symmetry in the NP effective Hamiltonian.
The expressions of the ${\cal A,\,B,\,C,\, A_H,\,B_H,\,C_H,\,D_H,\,E_H,\,F_H}$ and
${\cal G_H}$ in terms of the $\widetilde W_\chi$ SFs are collected in  
Appendix D of Ref.~\cite{Penalva:2021wye}. As inferred from Eq.~(\ref{eq:pol}), 
${\cal A,\,B,}$ and ${\cal C}$
describe the production of an unpolarized final
charged lepton, while  ${\cal A_H,\,B_H,\,C_H,\,D_H,\,E_H,\,F_H}$ and
${\cal G_H}$ are also needed for the description of decays with a defined
polarization ($h=\pm 1$) of the outgoing charged lepton along the four vector $S^\alpha$.

\section{Sequential $H_b\to
H_c\tau^-(\pi^-\nu_\tau,\rho^-\nu_\tau,\mu^-\bar\nu_\mu\nu_\tau)\bar\nu_\tau$ decays and visible
kinematics}  
\label{sec:sequential}

Due to its short mean life, the $\tau$ produced in a $H_b\to
H_c\tau^-\bar\nu_\tau$ process can not be directly measured  and all the accessible information
on the decay is encoded in the visible kinematics of the $\tau$-decay products. The three  
dominant decay  modes $\tau \to \pi \nu_\tau ,\, \rho \nu_\tau$ and 
$\ell\bar\nu_\ell\nu_\tau$ ($\ell=e,\mu$)  account 
for more than 70\% of the total $\tau$ width. The (visible) differential distributions of the charged particle produced in the tau decay  have been studied extensively for $\bar B\to D^{(*)}$ decays in Refs.~\cite{Kiers:1997zt,Nierste:2008qe,Alonso:2016gym,Alonso:2017ktd,
Asadi:2020fdo}. The general expression for the differential decay width for the $H_b\to H_c\tau^-(d\nu_\tau)\bar\nu_\tau$
decay, with $d=\pi^-,\rho^-,\ell^-\bar\nu_\ell$, reads~\cite{Alonso:2016gym,Alonso:2017ktd,
Asadi:2020fdo,Penalva:2021wye}
\bea
\frac{d^3\Gamma_d}{d\omega  d\xi_d d\cos\theta_d} & = & {\cal B}_{d}
\frac{d\Gamma_{\rm SL}}{d\omega} \Big\{  F^d_0(\omega,\xi_d)+ F^d_1(\omega,\xi_d) 
\cos\theta_d + F^d_2(\omega,\xi_d)P_2(\cos\theta_d)\Big\},
\label{eq:visible-distr}
\eea
 which is given  in terms of $\omega$, $\xi_d=\frac{E_d}{\gamma m_\tau}$ (here $\gamma=\frac{q^2+m_\tau^2}{2m_\tau\sqrt{q^2}}$), which is the 
ratio of the energies of the tau-decay
charged particle and the tau lepton measured in the $\tau^-\bar\nu_\tau$ center of mass frame
(CM), and $\theta_d$, the angle made by the three-momenta of the final hadron and the
tau-decay charged particle measured in the same CM system (see Fig.~1 of Ref.~\cite{Penalva:2022vxy}). The azimuthal angular ($\phi_d$) distribution of the tau decay charged product is sensitive to possible CP odd effects (${\cal  N_{H_{\rm 3}}}$ term in Eq.~\eqref{eq:pol2}). 
However, the measurement of  $\phi_d$ would require the full reconstruction of the tau
three momentum, and this azimuthal angle has been integrated out to obtain the differential decay width of Eq.~\eqref{eq:visible-distr}. That is the reason why the latter visible distribution does not depend on ${\cal  N_{H_{\rm 3}}}$, and thus it does not contain any information on possible CP violation contributions to the effective NP Hamiltonian of Eq.~\eqref{eq:hnp}.

In addition, ${\cal B}_{d}$ in Eq.~\eqref{eq:visible-distr} is the branching ratio for
the $\tau^-\to d^-\nu_\tau$ decay mode and $P_2(\cos\theta_d)$ stands for the Legendre
polynomial of order two. As for $d\Gamma_{\rm SL}/d\omega$, it represents the
differential decay width for the unpolarized semileptonic $H_b\to H_c\tau^-\bar\nu_\tau$ decay.
It reads
\be
\frac{d\Gamma_{\rm SL}}{d\omega}=\frac{G_F^2|V_{cb}|^2M^{\prime3}M^2}{24\pi^3} \label{eq:n0w}
\sqrt{\omega^2-1}\Big(1-\frac{m_\tau^2}{q^2}\Big)^2n_0(\omega),
\ee
where $n_0(\omega)$ contains all the dynamical information, including any
possible NP contribution. It is given by $n_0(\omega) =
3a_0(\omega)+a_2(\omega)$,
 where $a_{0,2}(\omega)$ are linear combinations of 
${\cal A}(\omega),\,{\cal B}(\omega)$ and ${\cal C}(\omega)$, with explicit expressions given in Eq.~(18) of Ref.~\cite{Penalva:2020xup}.
The $F^d_{0,1,2}(\omega,\xi_d)$ functions in Eq.~(\ref{eq:visible-distr}) can be written as \cite{Asadi:2020fdo,Penalva:2022vxy} 
 \begin{eqnarray}
 F^d_0(\omega,\xi_d) &=& C_n^d(\omega,\xi_d)+C_{P_L}^d(\omega,\xi_d)\,\langle P^{\rm CM}_L\rangle(\omega), \nonumber \\
 F^d_1(\omega,\xi_d) &=& C_{A_{FB}}^d(\omega,\xi_d)A_{FB}(\omega)+C_{Z_L}^d(\omega,\xi_d)Z_L(\omega)
 + C_{P_T}^d(\omega,\xi_d)\,\langle P^{\rm CM}_T\rangle(\omega), \nonumber \\ 
 F^d_2(\omega,\xi_d) &=& C_{A_Q}^d(\omega,\xi_d)A_{Q}(\omega)+
 C_{Z_Q}^d(\omega,\xi_d)Z_Q(\omega)+ C_{Z_\perp}^d(\omega,\xi_d)Z_\perp(\omega).
 \label{eq:coeff}
\end{eqnarray}
where the  decay-mode dependent  coefficients $C^d_a(\omega,\xi_d)$ are  purely 
kinematical. Their  analytical expressions 
for the $\pi^-\nu_\tau,\rho^-\nu_\tau$ and $\ell^-\bar\nu_\ell\nu_\tau$ decay 
channels
  can be found in
  Appendix G of Ref.~\cite{Penalva:2021wye}.  The rest of the observables in   Eq.~(\ref{eq:coeff})
 are  the tau-spin ($\langle P^{\rm CM}_{L,T}\rangle(\omega)$), 
 tau-angular  ($A_{FB,Q}(\omega)$) and  tau-angular-spin ($Z_{L,Q,\perp}(\omega)$) 
 asymmetries of the $H_b\to H_c\tau\bar\nu_\tau$ parent
 decay. They can be written~\cite{Penalva:2021wye,Penalva:2020xup} in terms of the
 ${\cal A},\,{\cal B},\,{\cal C},\,{\cal A_H}
 ,\,{\cal B_H},\,{\cal C_H},\,{\cal D_H}$ and ${\cal E_H}$
 functions introduced in Eq.~\eqref{eq:pol2}.
A numerical analysis of the role of each of the observables 
$d\Gamma_{\rm SL}/d\omega$, $\langle P^{\rm CM}_{L,T}\rangle(\omega)$, 
$A_{FB,Q}(\omega)$ and $Z_{L,Q,\perp}(\omega)$ in the context of 
LFU violation was conducted for the  $\Lambda_b\to\Lambda_c\tau^-\bar\nu_\tau$ transition 
in Refs.~\cite{Penalva:2021gef,Penalva:2021wye}. Here we perform an analog analysis 
for the  $\Lambda_b\to\Lambda^*_c(2595)$ and  $\Lambda_b\to\Lambda^*_c(2625)$ semileptonic decays, 
for which the only differences are fully encoded in the form-factor input contained in the $\widetilde{W}_\chi$ SFs. This is because the expressions of ${\cal A},\,{\cal B},\,{\cal C},\,{\cal A_H} ,\,{\cal B_H},\,{\cal C_H},\,{\cal D_H}$ and ${\cal E_H}$ (or equivalently the differential decay width for unpolarized tau,  the tau-spin,  tau-angular and  tau-angular-spin asymmetries) in terms of the latter is independent of the $b\to c$ transition, and they are given by Eqs.~(D1) and (D2) of Ref.~\cite{Penalva:2021wye}. 

Measuring the triple differential decay width in Eq.~(\ref{eq:visible-distr}) could also be
difficult due to  low statistics. An increased statistics is achieved by integrating
in one  or more of the variables $\cos\theta_d$, $\xi_d$ and $\omega$,  at the price that the resulting distributions might not depend on some of the observables in Eq.~(\ref{eq:coeff}). For instance, accumulating in the polar angle leads to  the distribution~\cite{Tanaka:2010se} 
\bea
\frac{d^2\Gamma_d}{d\omega  d\xi_d} & = & 2{\cal B}_{d}
\frac{d\Gamma_{\rm SL}}{d\omega}\Big\{  C_n^d(\omega,\xi_d)+C_{P_L}^d(\omega,\xi_d)\,\langle P^{\rm CM}_L\rangle(\omega)\Big\},\label{eq:wE}
\eea
from where one can only extract, looking at the  dependence on $\xi_d$, $d\Gamma_{\rm SL}/d\omega$ and  the CM $\tau$ longitudinal 
polarization 
[$\langle P^{\rm CM}_L\rangle(\omega)]$. From the latter,  it immediately follows the averaged 
CM tau  
longitudinal polarization asymmetry,
\be
P_\tau = -\frac{1}{\Gamma_{\rm SL}}\int d\omega \frac{d\Gamma_{\rm SL}}{d\omega}
\langle P^{\rm CM}_L\rangle(\omega),
\ee
that has been measured for 
the $\bar B \to D^* \tau \bar\nu_\tau$ 
decay by the Belle collaboration~\cite{Belle:2016dyj}.  

Integrating Eq.~(\ref{eq:visible-distr}) in the $\xi_d$ variable one obtains the double 
differential decay
width~\cite{Penalva:2022vxy}
\be
\frac{d^2\Gamma_d}{d\omega  d\cos\theta_d}  =  {\cal B}_{d}
\frac{d\Gamma_{\rm SL}}{d\omega} \Big[ 
\widetilde F^d_0(\omega)+ \widetilde F^d_1(\omega) 
\cos\theta_d +\widetilde  F^d_2(\omega)P_2(\cos\theta_d)\Big].
\label{eq:visible-distr_theta_d}
\ee
While $\widetilde F^d_0(\omega)=1/2$, losing in this way all information
on $\langle P^{\rm CM}_L\rangle(\omega)$, one has that
\bea
\widetilde F^d_1(\omega)&=&C^d_{A_{FB}}(\omega)\,A_{FB}(\omega)+C^d_{Z_L}(\omega)
\,Z_L(\omega)+C^d_{P_T}(\omega)\,\langle P_T^{\rm CM}\rangle(\omega),
\label{eq:F1} \\
\widetilde F^d_2(\omega)&=&C^d_{A_Q}(\omega)\,A_Q(\omega)+C^d_{Z_Q}(\omega)
\,Z_Q(\omega)+C^d_{Z_\perp}(\omega)\,Z_\perp(\omega),\label{eq:F2}
\eea
which retain all the information on the other six asymmetries, since the kinematical coefficients  $C^d_{i}(\omega)$ are known.

A further integration in $\omega$ additionally enhances the statistics. The obtained angular
distribution~\cite{Penalva:2022vxy}
\bea
\frac{d\Gamma_d}{d\cos\theta_d}={\cal B}_d\Gamma_{\rm SL}\Big[
\frac12+\widehat F_1^d\cos\theta_d+\widehat F_2^d\, 
P_2(\cos\theta_d)\Big], \quad  \widehat F_{1,2}^d=\frac1\Gamma_{\rm SL}\int_1^{\omega_{\rm max}}
\frac{d\Gamma_{\rm SL}}{d\omega}\widetilde F_{1,2}^d(\omega)\,d\omega.\label{eq:Gcos}
\eea
could still be a useful observable in the search for NP beyond the SM.

Finally, from the differential decay width $d^2\Gamma_d/(d\omega  d\xi_d)$ given in 
Eq.~\eqref{eq:wE} one can get~\cite{Penalva:2022vxy}
\bea
\frac{d\Gamma_d}{dE_d} & = & 2{\cal B}_{d} \int_{\omega_{\rm inf}(E_d)}^{\omega_{\rm sup}(E_d)} d\omega
\frac{1}{\gamma m_\tau}\frac{d\Gamma_{\rm SL}}{d\omega}\Big\{ 
C_n^d(\omega,\xi_d)+C_{P_L}^d(\omega,\xi_d)\,\langle P^{\rm CM}_L\rangle(\omega)\Big\},
\label{eq:EdG}
\eea
where the appropriate limits in $\omega$ 
for each of the sequential decays considered are given
in Ref.~\cite{Penalva:2022vxy}. From the latter distribution one can define the
dimensionless observable
\be
\widehat F^d_0(E_d)=\frac{m_\tau}{2{\cal B}_d\Gamma_{\rm SL}}\frac{d\Gamma_d}{dE_d}.
\ee
Although it is normalized for all channels as
\be
\frac{1}{m_\tau}\int_{E_d^{\rm min}}^{E_d^{\rm min}} dE_d \widehat F^d_0(E_d)
= \frac12,
\ee
 its energy dependence is still affected by the CM $\tau$ longitudinal polarization 
$\langle P^{\rm CM}_L\rangle(\omega)$.

Predictions for the $d^2\Gamma/(d\omega\, d\cos\theta_d)$, $d\Gamma/d\cos\theta_d$
and the $\widehat F^d_0(E_d)$ distributions, and their role in distinguishing
among different NP models, were presented and discussed in Ref.~\cite{Penalva:2022vxy} for the 
$\Lambda_b\to\Lambda_c\tau^-(\pi^-\nu_\tau,\rho^-\nu_\tau,\mu^-\bar\nu_\mu\nu_\tau)\bar\nu_\tau$ 
and $\bar B\to D^{(*)}\tau^-(\pi^-\nu_\tau,\rho^-\nu_\tau,\mu^-\bar\nu_\mu\nu_\tau)\bar\nu_\tau$
sequential decays. Here, we will also extend the study to reactions initiated by the  $\Lambda_b\to\Lambda^*_c(2595)$ and  $\Lambda_b\to\Lambda^*_c(2625)$ semileptonic parent decays.

\section{Results  and discussion}
\label{sec:results}

\begin{figure}[t]
\resizebox{14cm}{16cm}{\includegraphics{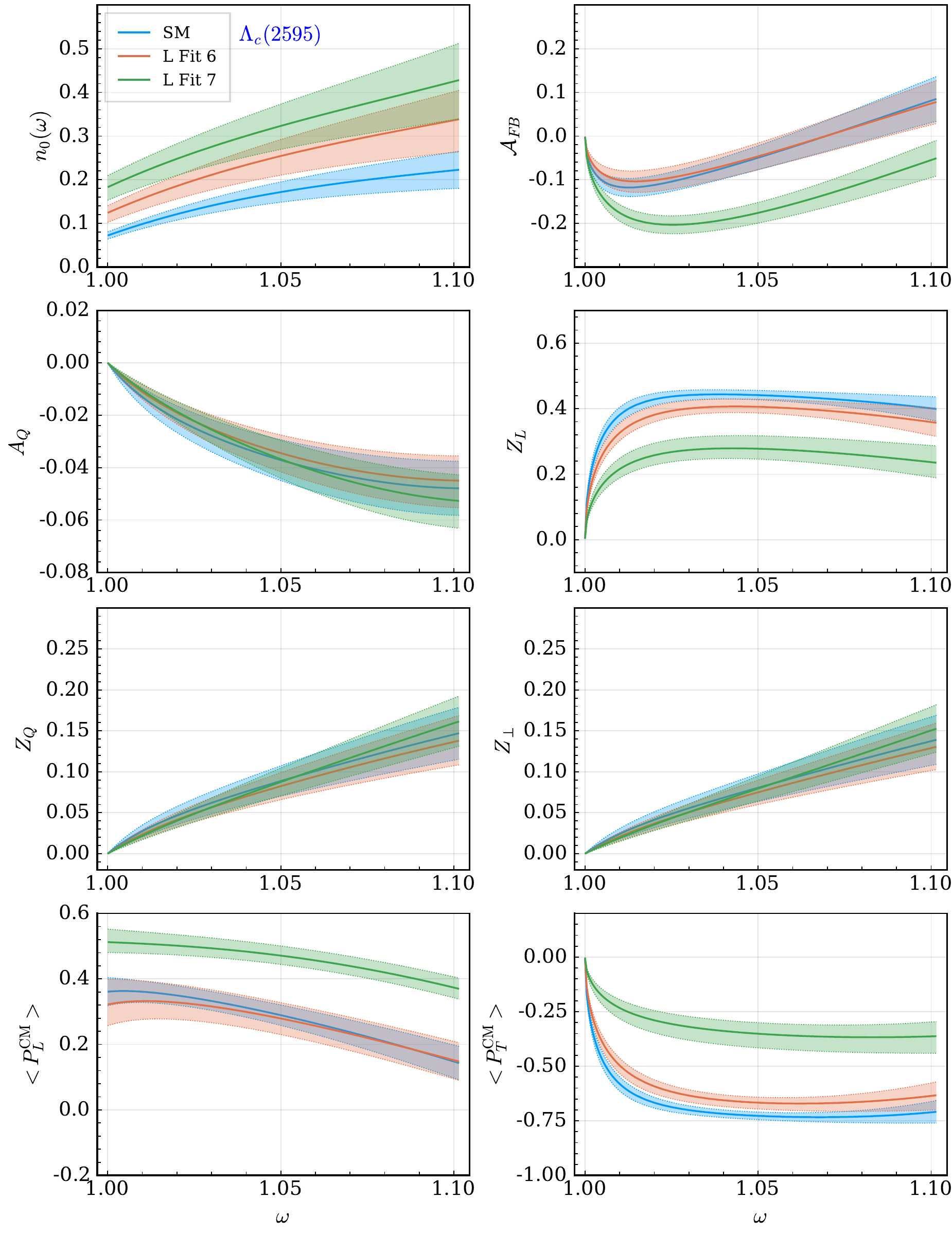}}
\caption{ $d\Gamma_{\rm SL}/d\omega$ differential decay width and
$A_{FB},\,A_Q,\,Z_L,\,Z_Q,\,Z_\perp,\,\langle P^{CM}_L\rangle$
and $\langle P^{CM}_T\rangle$ asymmetries evaluated for the $\Lambda_b\to \Lambda^*_c(2595)\tau^-
 \bar\nu_\tau$ decay  within the SM and the left-handed neutrino NP models corresponding to 
  Fits 6 and 7 of Ref.~\cite{Murgui:2019czp}. The error bands account for 
  uncertainties both in the  { WCs} and form factors, see text for details.}
\label{fig:12lh}
\end{figure}
\begin{figure}
\resizebox{14cm}{16cm}{\includegraphics{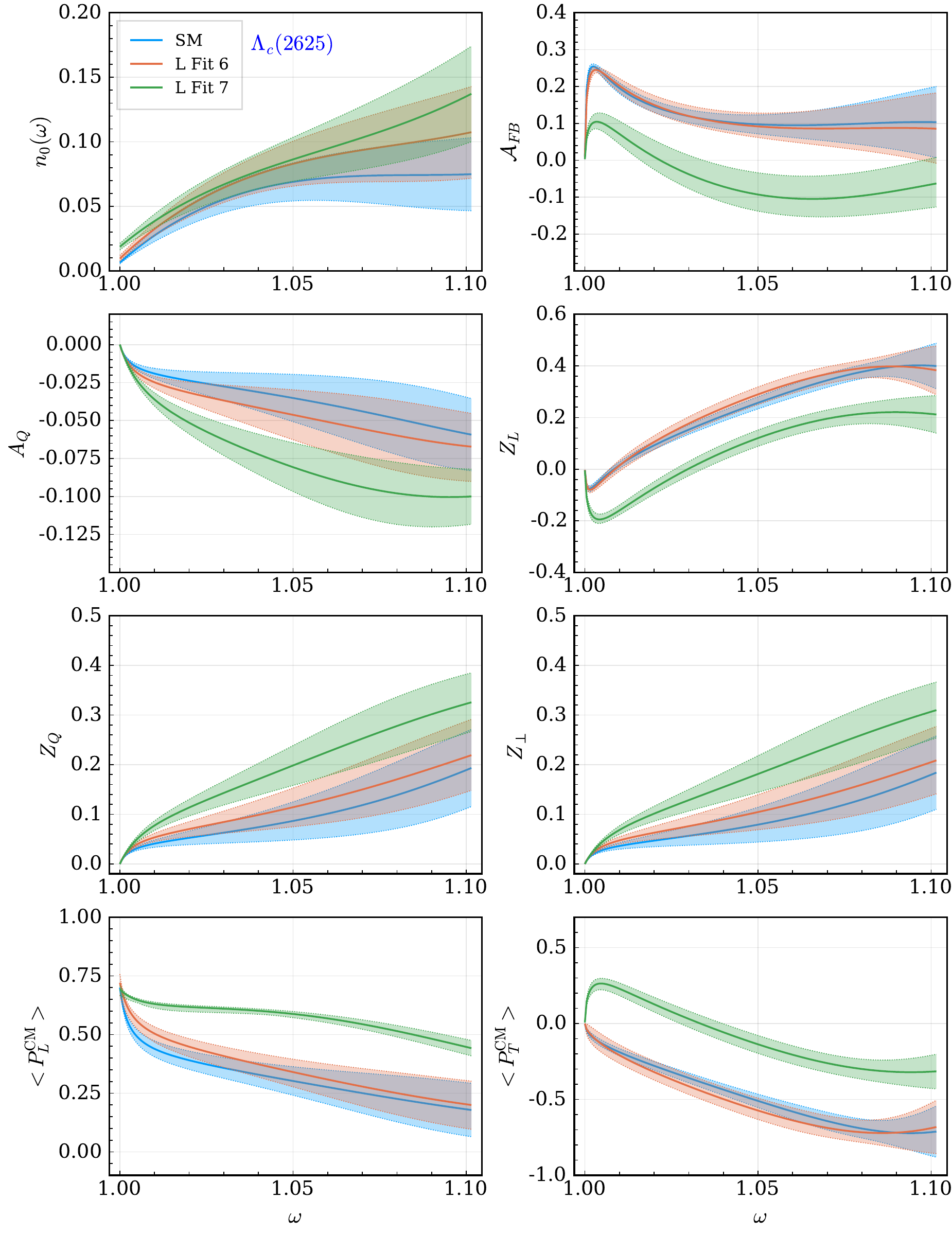}}
\caption{ Same as Fig.~\ref{fig:12lh}, but for the 
$\Lambda_b\to \Lambda^*_c(2625)\tau^-\bar\nu_\tau$ decay.}
\label{fig:32lh}
\end{figure}
In this section we present $\Lambda_b\to \Lambda^*_c(2595),\Lambda^*_c(2625)$   results for   the  observables mentioned
in Sec.~\ref{sec:sequential} above. We will consider  SM and different NP scenarios involving left- and right-handed neutrino fields taken from Refs.~\cite{Murgui:2019czp} and \cite{Mandal:2020htr}. Since the LQCD form factors from Refs.~\cite{Meinel:2021mdj,Meinel:2021rbm} that we use are
 not reliably obtained at high $\omega$ values,
we will restrict ourselves to the $1\le\omega\le1.1$ region. For this latter
reason we
will not show results for $d\Gamma_d/d\cos\theta_d$ or $\widehat F^d_0(E_d)$ since
they involve an integration in the $\omega$ variable over the full available phase space, 
including regions for which the LQCD form factors are not reliable. 

 For each observable, we give central values
plus an error band that we construct by adding in quadrature the form-factor and   {WC} uncertainties.
For the errors related to the { WCs} we shall use
statistical samples of  { WCs} selected such that the $\chi^2$-merit function computed
in Refs.~\cite{Murgui:2019czp} and \cite{Mandal:2020htr}, for left- and right-handed neutrino NP fits, respectively, changes at most by one unit from its value at the fit minimum (for further details  see Sec. III.B of Ref.~\cite{Penalva:2020xup}). For the uncertainty associated to the form factors, we consider two different sources~\cite{Meinel:2021rbm}: statistical and systematic. We 
obtain the statistical error using the appropriate covariance matrix  to Monte-Carlo generate a great number of
form factor samples  from which we evaluate the corresponding quantity and its standard deviation. The systematic
error  is evaluated as explained in Sec.~VI 
of Ref.~\cite{Meinel:2021rbm}. This latter determination makes use of the form
factors obtained  with higher-order fits. Statistical and systematic errors are then  added  in quadrature to get the total error associated to the form factors.

 \begin{table}[tb]
\renewcommand\arraystretch{1.2}
\begin{ruledtabular}
\caption{ {Wilson coefficients entering in the hadron currents of Eqs.~\eqref{eq:JH1}-\eqref{eq:Jh}. We collect numerical values for the NP models fitted in Refs.~\cite{Murgui:2019czp} (L Fits 6 and 7)  and \cite{Mandal:2020htr} (R S7a). Here $C^{V,A}_{\chi=L} = 1+C_{LL}^V\pm C_{RL}^V$, $C^{S,P}_{\chi = L}=C^S_{LL}\pm C^S_{RL}$, $C^T_{\chi=L} = C^T_{LL}$, $C^{V,A}_{\chi=R}=  C_{RR}^V \pm C_{LR}^V$, $C^{S,P}_{\chi=R}=C^S_{RR} \pm C^S_{LR}$ and $C^T_{\chi=R}=C^T_{RR}$, where the coefficients $C^{S,V,T}_{AB}$ ($A,B=L,R$) appear directly in the NP effective Hamiltonian of Eq.~\eqref{eq:hnp}. Note that within the R S7a scalar leptoquark scenario, $C_{\chi=R}^A = C_{\chi=R}^V$ and $C_{\chi=R}^P=C_{\chi=R}^S$ since in that model  $C_{LR}^V=C_{LR}^S=0$~\cite{Mandal:2020htr}.} }\label{tab:wc}
\begin{tabular}{l|ccccc|ccc}
\, & $C^S_{\chi=L}$ & $C^P_{\chi=L}$ & $C^V_{\chi=L}$ & $C^A_{\chi=L}$ & $C_{\chi=L}^T$ & $C_{\chi=R}^S$ & $C^V_{\chi=R}$ & $C_{\chi=R}^T$\bstrut \\ 
\hline
SM & - & - & 1 & 1 & - & - & - & -\\
\hline
L Fit 6 & $-0.16^{+0.10}_{-0.07}$ & $-0.26^{+0.00}_{-0.01}$ & $1.26^{+0.03}_{-0.04}$ & $1.10^{+0.07}_{-0.11}$ & $0.01^{+0.02}_{-0.04}$ & - & - & - \\
\hline
L Fit 7 & $-1.32^{+0.15}_{-0.12}$ & $-0.22^{+0.01}_{-0.01}$ & $1.70^{+0.02}_{-0.02}$ & $1.02^{+0.05}_{-0.07}$ & $-0.01^{+0.02}_{-0.02}$ & - & - & -\\
\hline
R S7a & - & - & 1 & 1 & - & $-0.18^{+0.60}_{-0.32}$ & $0.42^{+0.03}_{-0.20}$ & $0.02^{+0.04}_{-0.08}$
\end{tabular}
\end{ruledtabular}
\end{table}

In Figs.~\ref{fig:12lh} and \ref{fig:32lh} we show, for the
$\Lambda_b\to\Lambda^*_c(2595)\tau^-\bar\nu_\tau$ and
$\Lambda_b\to\Lambda^*_c(2625)\tau^-\bar\nu_\tau$ decays
respectively, the results for 
$n_0(\omega)$ and the
full set of asymmetries introduced in Eq.~\eqref{eq:coeff}. They have been  
obtained within the SM
and the two NP models corresponding to Fits 6 and 7 of
 Ref.~\cite{Murgui:2019czp}. The latter two models include only left-handed (L) neutrino operators {and the corresponding  WCs were obtained from fits to the experimental  evidences of  LFU violations in the $B$-meson sector. In Table~\ref{tab:wc}  we give the values corresponding to the  $C^S_\chi$, $C^P_\chi$, $C_\chi^V$, $C_\chi^A$ and $C_\chi^T$ coefficients introduced in Eq.~\eqref{eq:Jh} since the  $\widetilde W_\chi$ hadronic SFs are written in terms of the latter.}
 Even though these two NP scenarios have been adjusted to reproduce the measured $R_{D^{(*)}}$ ratios, they
 show a different behavior for other quantities. As seen from the figures,  L Fit 6 and 
SM results agree within 
errors for  most of the observables, while the predictions from L Fit 7 are quite different
for the $A_{FB},\,Z_L,\,\langle P^{\rm CM}_L\rangle$ and $\langle P^{\rm CM}_T\rangle$ asymmetries. The latter
 are thus helpful in distinguishing between these two NP models that otherwise 
give very similar results
for the $R_{D^{(*)}}$ ratios.

\begin{figure}
\resizebox{8.15cm}{13cm}{\includegraphics{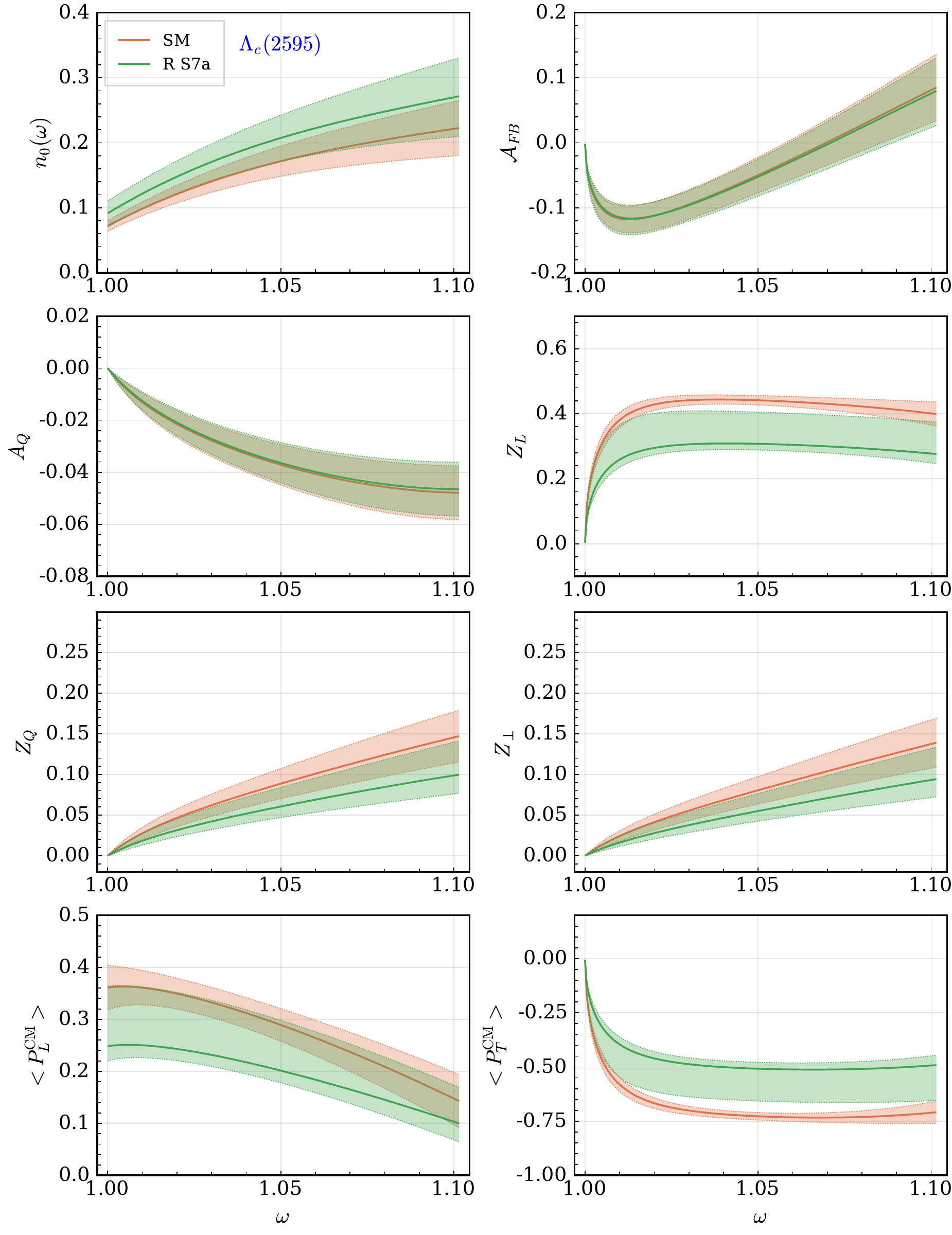}}
\resizebox{8.15cm}{13cm}{\includegraphics{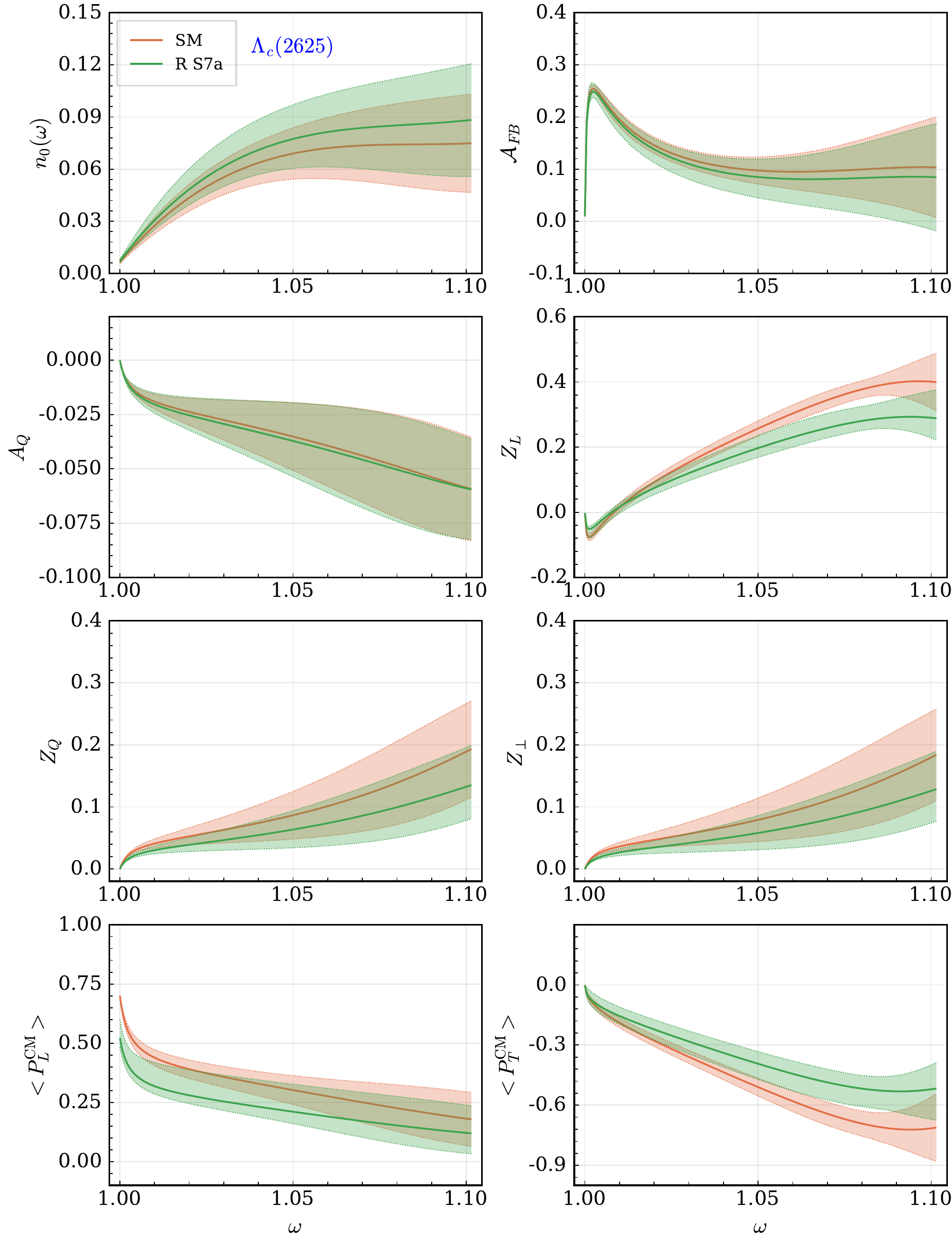}}\\
\caption{ Same as Figs.~\ref{fig:12lh} and \ref{fig:32lh}, but comparing in this
case SM  to the  model corresponding to fit R S7a of 
Ref.~\cite{Mandal:2020htr}, which NP contributions are constructed using right-handed neutrino fields. Two left columns: 
$\Lambda_b\to \Lambda^*_c(2595)\tau^-
 \bar\nu_\tau$ decay. Two right columns: $\Lambda_b\to \Lambda^*_c(2625)\tau^-
 \bar\nu_\tau$ decay. }
\label{fig:1232rh}
\end{figure}

In Fig.~\ref{fig:1232rh} we compare the results obtained
within the SM and fit R S7a  of Ref.~\cite{Mandal:2020htr}. The latter includes
only NP operators constructed with right-handed (R) neutrino fields,  and the corresponding  { WCs} {(see Table~\ref{tab:wc})} have also been adjusted to 
reproduce the measured $R_{D^{(*)}}$ ratios. Among the different R fits conducted in 
Ref.~\cite{Mandal:2020htr}, this is one of the more promising  in terms of the pull
 from the SM hypothesis\footnote{ This NP scenario considers a scalar leptoquark, with non-vanishing $C^V_{RR}$,  $C^T_{RR}$  and $C^S_{RR}$  and $C^S_{RR}\approx -8\, C^T_{RR}$. The fit carried out in Ref.~\cite{Mandal:2020htr} leads to a solution dominated by $C^V_{RR}$, with  $C^T_{RR}$ compatible with zero within one sigma.}. 
However, due to the wide error bands,  we find no significant
difference between the  R S7a model and SM results. The exceptions are the $Z_L$ and 
$\langle P^{\rm CM}_T\rangle$ asymmetries for the $\Lambda_b\to \Lambda^*_c(2595)\tau^-
 \bar\nu_\tau$ decay.   The R S7a  and the L Fit 6 models give
also similar predictions, agreeing within errors. As for the differences with L Fit 7, 
the best observables 
 to distinguish between  the  R S7a and the L Fit 7 models are the $A_{FB}$
and $\langle P^{\rm CM}_L\rangle$ asymmetries for the $\Lambda_b\to \Lambda^*_c(2595)\tau^-
 \bar\nu_\tau$ decay, whereas for $\Lambda_b\to \Lambda^*_c(2625)\tau^-
 \bar\nu_\tau$ one finds  that not only $A_{FB}$
and $\langle P^{\rm CM}_L\rangle$, but also $Z_L$ and 
$\langle P^{\rm CM}_T\rangle$ are adequate observables.

\begin{figure}
\resizebox{8.15cm}{10cm}{\includegraphics{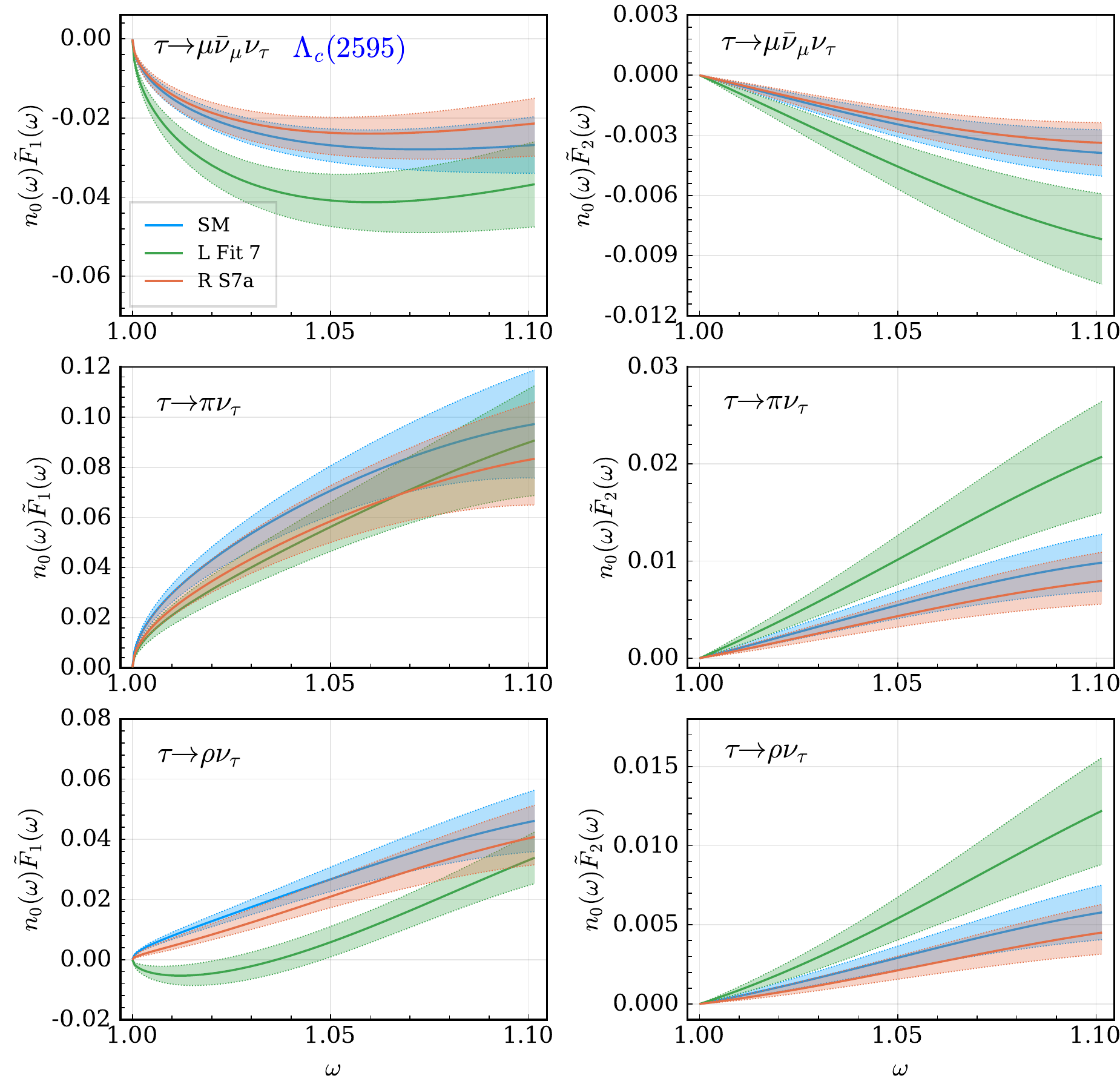}}
\resizebox{8.15cm}{10cm}{\includegraphics{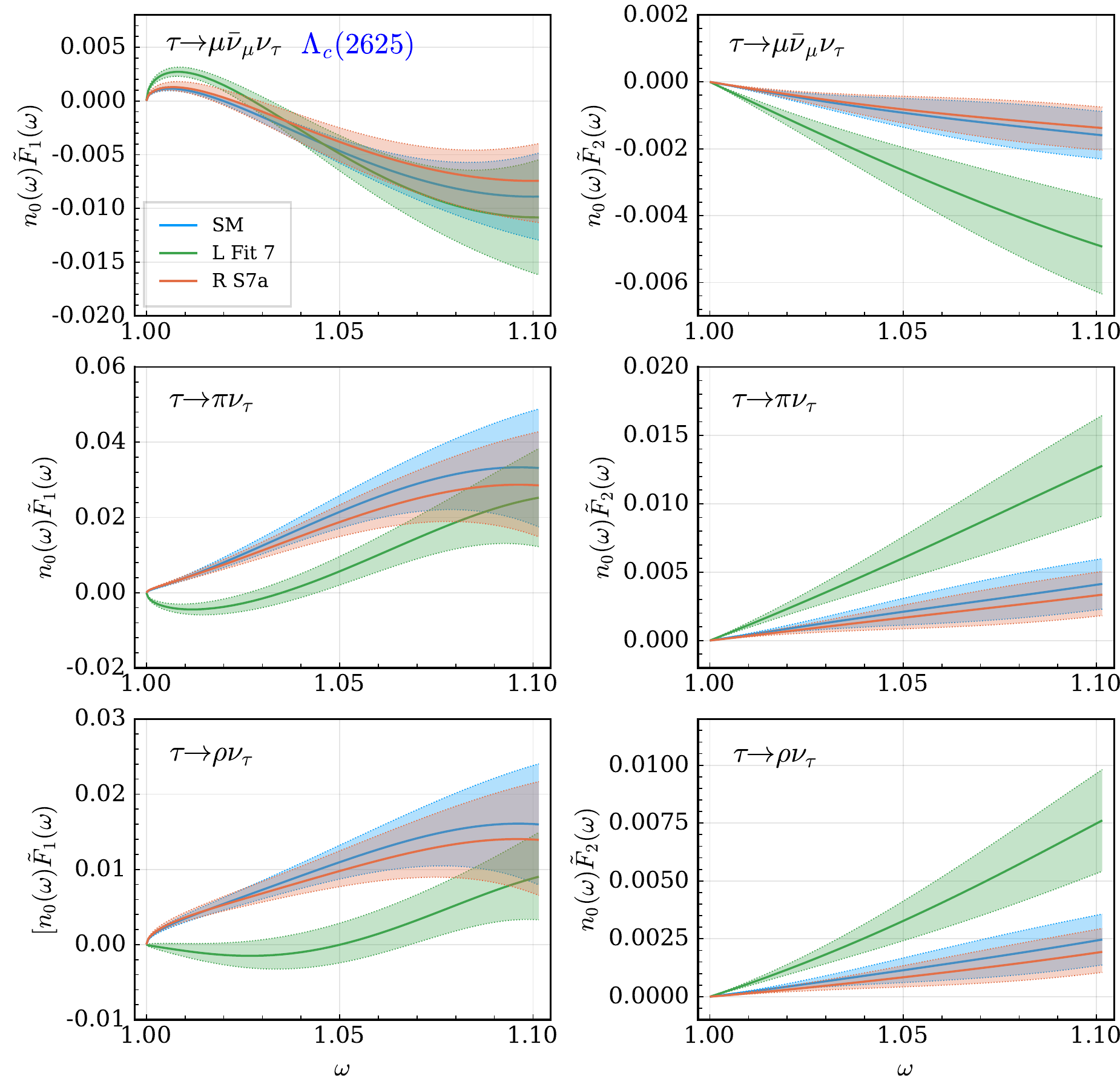}}
\caption{ Two left columns: $n_0(\omega)\widetilde F^{d}_{1,2}(\omega)$ for the three
$\Lambda_b\to
\Lambda^*_c(2595)\tau^-(\mu^-\bar\nu_\mu\nu_\tau,\pi^-\nu_\tau,\rho^-\nu_\tau)
 \bar\nu_\tau$ sequential decays evaluated within the SM, the L Fit 7 model of 
 Ref.~\cite{Murgui:2019czp} and the R S7a model of
Ref.~\cite{Mandal:2020htr}. Two right columns: the same but for the 
$\Lambda_b\to \Lambda^*_c(2625)\tau^-(\mu^-\bar\nu_\mu\nu_\tau,\pi^-\nu_\tau,\rho^-\nu_\tau)
 \bar\nu_\tau$ sequential decays}.
\label{fig:f12_1232}
\end{figure}

We show now the  results for  the
$\widetilde F^{d}_{1,2}(\omega)$  coefficient functions that expand the statistically-enhanced
$d^2\Gamma/(d\omega d\cos\theta_d)$
differential decay width of Eq.~(\ref{eq:visible-distr_theta_d})\footnote{Note 
that $\widetilde F^{d}_{0}(\omega)=1/2$ in all  cases.}. In fact, 
we  show the products
$n_0(\omega)\widetilde F^{d}_{1,2}(\omega)$ since, as mentioned,
$n_0(\omega)$
contains all the dynamical effects included in $d\Gamma_{\rm SL}/d\omega$ which
appears as an overall factor of the $d^2\Gamma/(d\omega d\cos\theta_d)$
distribution. Predictions obtained within the SM, and the L Fit 7  and the R S7a NP models of  
Refs.~\cite{Murgui:2019czp} and \cite{Mandal:2020htr}, respectively, are presented in Fig.~\ref{fig:f12_1232}.
 As it was to be expected from the previous results, SM and R Fit 7a results agree within 
errors in all cases. Similar results (not shown) are found for L Fit 6.  However, for most of the observables plotted in the figure, 
the L Fit 7  predictions are distinguishable from those obtained using  the 
SM or R S7a models, either in the near zero-recoil region or in the upper
part of the  shown $\omega$ interval.

When compared to the $\Lambda_b\to\Lambda_c$ decay considered in 
Refs.~\cite{Penalva:2021wye,Penalva:2022vxy}, we find here a worse discriminating power 
 between  different models due to the large errors in the form factors. Nevertheless, 
with the present values of the latter, these $\Lambda_b\to\Lambda_c^*$ reactions
 are already able to distinguish between the L fit 7 model of Ref.~\cite{Murgui:2019czp} and 
 the L Fit 6  and R S7a models  of Refs.~\cite{Murgui:2019czp} and \cite{Mandal:2020htr}, or between L Fit 7 and the SM.
A more precise determination of the form factors,
with less error and an extended $\omega$ region of validity, would certainly increase the value
of the $\Lambda_b\to\Lambda^*_c(2595),\Lambda^*_c(2625)$ decays in the search for 
NP in LFU violation studies.

Focusing on the SM $n_0(\omega)$ distributions in Figs.~\ref{fig:12lh} and \ref{fig:32lh}, we conclude that  $\Gamma_{\rm SL}$  (or at least the partially integrated width up to $\omega\le 1.1$)  
for the $\Lambda^*_c(2625)$ mode is smaller than for the $\Lambda^*_c(2595)$ final state, contradicting  the expectations 
from heavy-quark spin symmetry~\cite{Du:2022fxg}. Moreover, comparing with the results displayed in the left-upper plot of Fig.~2 of Ref.~\cite{Penalva:2021wye}, both widths are probably around a factor of ten lower than that of the $\Lambda_b$ 
decay into the ground state charmed baryon, $\Lambda_b\to \Lambda_c[J^P=1/2^+]$.  This  reduction does 
not affect  the tau-spin,  tau-angular and  tau-angular-spin asymmetries also shown in these figures, 
since  these observables  should not depend on the overall size of  the semileptonic width 
$\Gamma_{\rm SL}$. Actually, the asymmetries  provide distinct $\omega-$patterns for the $\Lambda_b$ 
decay into each of the charmed final state baryons, which have different spin-parity quantum numbers. 
This makes the comparison of theoretical model predictions, considering jointly all three 
[$\Lambda_c,\Lambda^*_c(2595), \Lambda^*_c(2625)$] modes,   more exhaustive and demanding. 

{ Next, for some of the observables studied so far, we investigate  how the  NP operators
affect the  SM predictions. In Fig.~\ref{fig:wc1}, we pay attention to the $n_0(\omega)$ (which equals to $d\Gamma_{\rm SL}/d\omega$ up to a kinematical factor), and the angular- $A_{FB}(\omega)$ and spin-    $\langle P^{\rm CM}_{L,T}\rangle(\omega)$ asymmetry distributions. These three observables are commonly discussed  in the literature  and, presumably, they are amongst  the easiest ones to be measured since they do not 
involve the second Legendre multipole in  Eq.~(\ref{eq:visible-distr}). We show results  for both  
$\Lambda_b \to \Lambda^*_c(2595)$ and $\Lambda_b \to \Lambda^*_c(2625)$ semileptonic decays (panels in 
 the first two and last two rows, respectively) and for the three beyond the SM scenarios considered in this work. In the left-handed neutrino NP models of Fits 6 and 7  of Ref.~\cite{Murgui:2019czp}, there is a total of five real WCs. In both cases (see Table~\ref{tab:wc})  the tensor coefficient $C^T_{\chi=L}$ is negligible, but even without considering the tensor contributions, we still have ten different contributions  taking interference into account. In the plots collected in  Fig.~\ref{fig:wc1}, we show SM and full results, 
 as well as the predictions obtained when the SM is supplemented only by some of the NP terms\footnote{ Note that for $C^{V,A}_{\chi=L}$,  NP is encoded  in their deviations from one.}. For the sake of simplicity and clarity, we display only the largest contributions and eliminate the $\chi=L$ label. We do the same for the case of the   right-handed neutrino R S7a scenario of Ref.~\cite{Mandal:2020htr}, where always $C^V_{\chi=L}= C^A_{\chi=L}=1$  and for the rest of WCs, the $\chi=R$ subindex is removed.   Thus, for instance the curves denoted in Fig.~\ref{fig:wc1} as $C^S$, both for L or R scenarios,  stand for the results obtained when $C^S$ is fixed to the corresponding value fitted in the NP scheme, and the other WCs are set to the SM values. Likewise, the $C^S+C^V$ lines show the predictions  when $C^S$ and  $C^V$ are fixed to the corresponding values fitted in the NP scheme, while the rest of WCs are set to the SM values.
In Fig.~\ref{fig:wc1}, we see that the main contributions responsible for the differences between the L Fit 7 and SM predictions come from $C^S$ and/or $C^V$. For the L Fit 6, the latter are much smaller (see Table~\ref{tab:wc}), being closer to the SM values. This explains why the predictions from this NP scheme are more difficult to be distinguished from those obtained within the SM, cf. Figs.~\ref{fig:12lh}, \ref{fig:32lh}, and \ref{fig:f12_1232}. For the R S7a scheme, the main NP contributions originate from the right-handed neutrino $C^A$ and $C^V$ terms. 

In Fig.~\ref{fig:wc2}, we present a similar analysis, but in this case for the visible pion-energy accumulated distributions $\widetilde F_1 (\omega)$ and $\widetilde F_2(\omega)$ for the hadron $\tau^- \to \pi^-\nu_\tau$ decay mode.  The last observable, $\widetilde F_2(\omega)$, depends on some of the asymmetries not considered in the previous figure. We do not show  results from the  L Fit 6 of Ref.~\cite{Murgui:2019czp} since its predictions are similar to those obtained within the SM.

On the whole,  we see  that the observed pattern of changes induced by NP  depend on the studied quantity and the information encoded in these two figures might be helpful to disentangle between different extensions of the SM.

\begin{figure}
\includegraphics[width=1.0\textwidth]{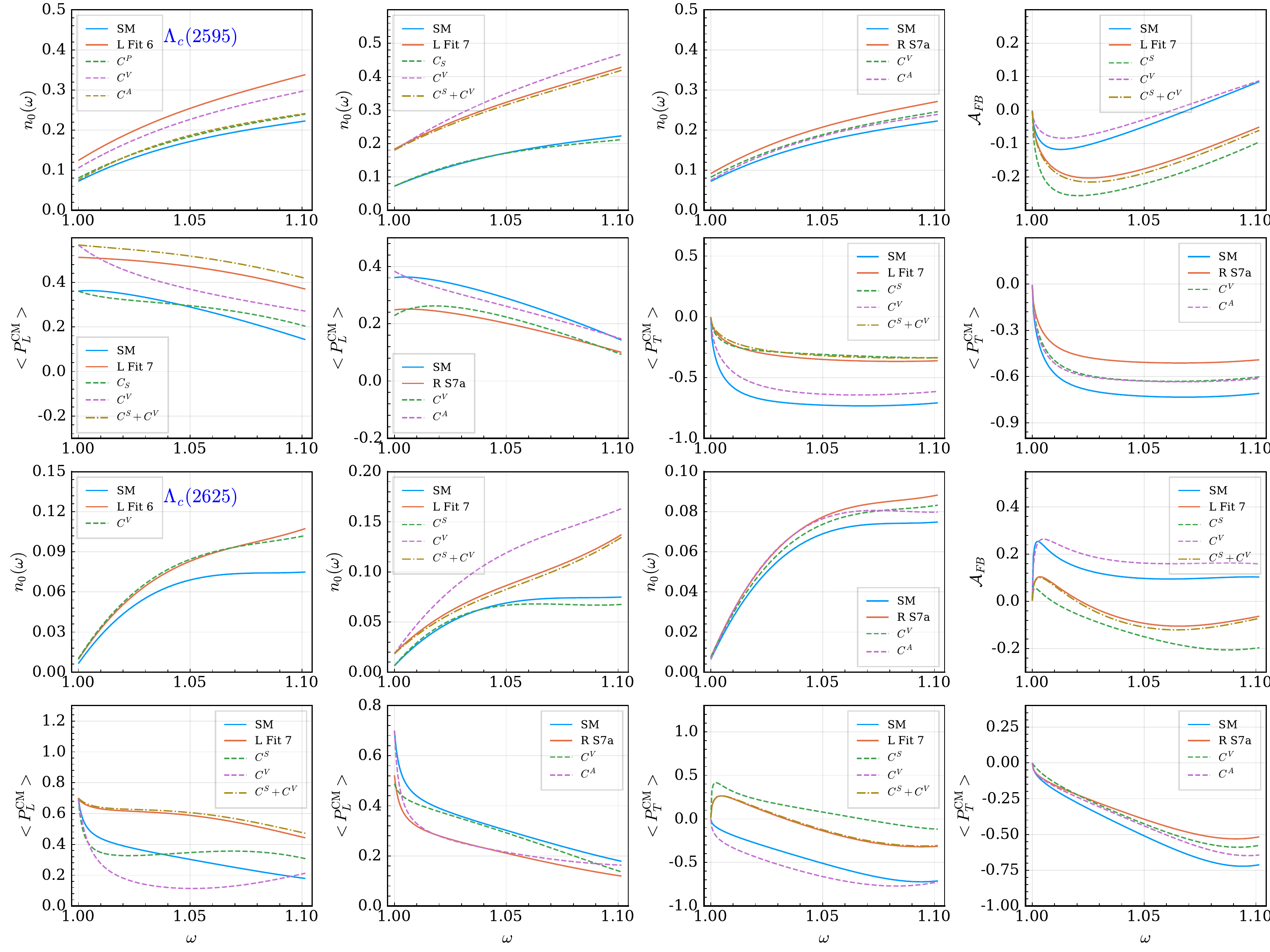}
\caption{ SM, L Fits 6 and 7~\cite{Murgui:2019czp} and R S7a~\cite{Mandal:2020htr} results, as well as the predictions obtained when the SM is supplemented by only some of the NP terms (see text for details), for some selected  observables among those which can be extracted from the visible differential  decay width for the $H_b\to H_c\tau^-(\pi^-\nu_\tau)\bar\nu_\tau$ sequential
decay (see  Eq.~\eqref{eq:visible-distr}).  We show results  for both the  $\Lambda_b \to \Lambda^*_c(2595)$ (two top rows) and 
$\Lambda_b \to \Lambda^*_c(2625)$ (two bottom rows) semileptonic decays, and for  the sake of clarity, we do not display uncertainty bands.}
\label{fig:wc1}
\end{figure}

\begin{figure}
\includegraphics[width=1.0\textwidth]{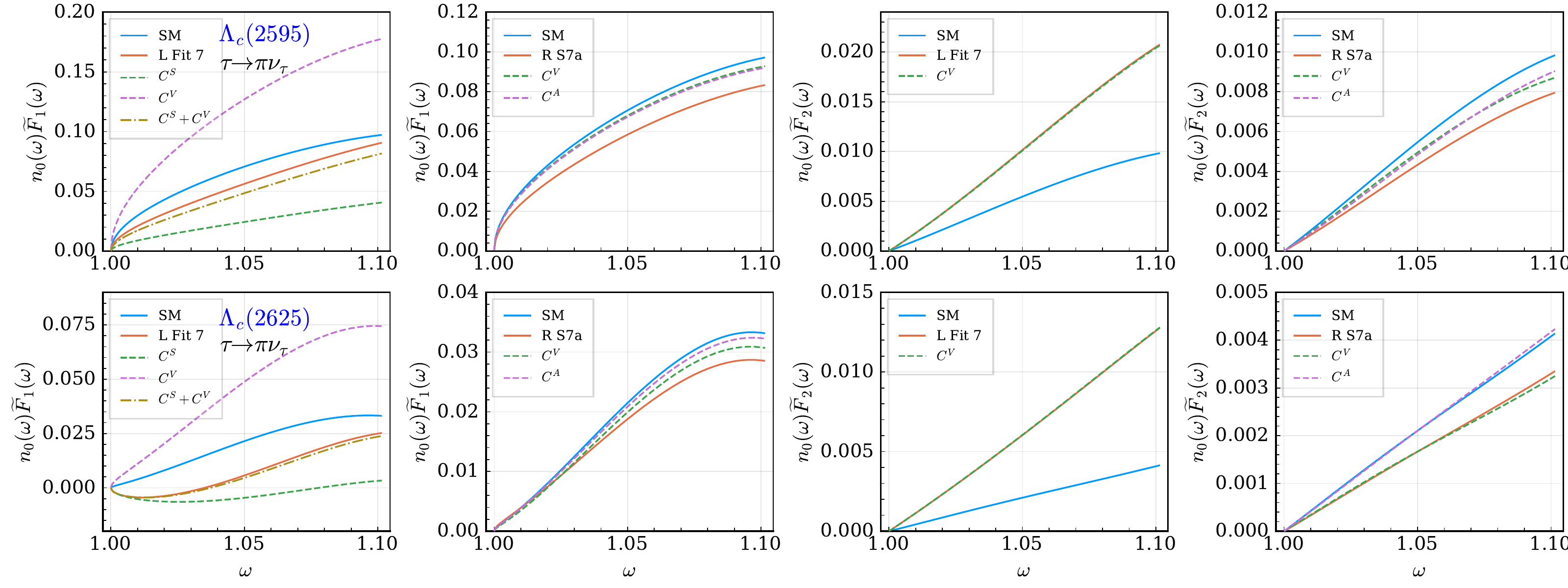}
\caption{ Same as in Fig.~\ref{fig:wc1}, but for the pion-energy accumulated distributions 
$n_0(\omega)\widetilde F_1 (\omega)$ and $n_0(\omega)\widetilde F_2(\omega)$.
 Here, we do not show results from the   L Fit 6 of Ref.~\cite{Murgui:2019czp}.}
\label{fig:wc2}
\end{figure}

}

Finally, we would like to stress that the expressions of  the $\widetilde W_\chi$ SFs in terms of the $1/2^+\to 1/2^-$ and $1/2^+\to 3/2^-$
 form factors derived in this work 
are general, and they do not apply only to the $\Lambda_b\to \Lambda^*_c(2595),\Lambda^*_c(2625)$ transitions studied here. Actually, 
using the appropriate numerical values for the form factors, these $\widetilde W_\chi$ SFs can be used for
any $1/2^+\to 1/2^-$ or $1/2^+\to 3/2^-$ CC semileptonic decay, driven by a $q \to q' \ell^- \bar\nu_\ell$ transition at the quark level. This will allow to systematically analyse NP effects in the charged-lepton unpolarized and polarized differential distributions in all these kind of reactions. 

\section*{Acknowledgements}
 N.P. thanks Physics and Astronomy  at the University of Southampton for hospitality while this work was completed and a Generalitat Valenciana grant CIBEFP/2021/32. This research has been supported  by the Spanish Ministerio de Ciencia e Innovaci\'on (MICINN)
and the European Regional Development Fund (ERDF) under contracts PID2020-112777GB-I00 and PID2019-105439GB-C22, 
the EU STRONG-2020 project under the program H2020-INFRAIA-2018-1, 
grant agreement no. 824093 and by  Generalitat Valenciana under contract PROMETEO/2020/023.

\appendix
\section{Hadronic matrix elements and SFs for   $1/2^+\to 1/2^-$ transitions}
\label{app:12p12m}
\subsection{Form factors}
We parameterize 
the matrix elements of the different $b\to c$ transition
operators for the $\Lambda_b\to\Lambda^*_c(2595)$ decay in such a way that we can make use of the expressions obtained 
in Refs.~\cite{Penalva:2020xup,Penalva:2021wye} for $1/2^+\to 1/2^+$ transitions with a
minimum of changes. To that end,  we use the form factor decompositions\footnote{The form factors defined in this work are related to those in Ref.~\cite{Du:2022fxg} by identifying $G_i=d_{V_i}$, $F_i=d_{A_i}$, $F_P=d_S$, $F_S=d_P$, and $T_i=d_{T_i}$.}
\bea
\langle\Lambda^*_c;\vec p\,',r'|\bar c(0)\gamma^\alpha b(0)
|\Lambda_b;\vec p,r\rangle&=&\bar u_{\Lambda^*_c,r'}(\vec p\,')
\left(G_1\gamma^{\alpha}+G_2\frac{p^{\alpha}}{M}+G_3\frac{p^{\prime\alpha}}{M'}\right)\gamma_5
u_{\Lambda_b,r}(\vec p\,),
\label{eq:ffv}\\
\langle\Lambda^*_c;\vec p\,',r'|\bar c(0)\gamma^\alpha\gamma_5 b(0)
|\Lambda_b;\vec p,r\rangle&=&\bar u_{\Lambda^*_c,r'}(\vec p\,')
\left(F_1\gamma^{\alpha}+F_2\frac{p^{\alpha}}{M}+F_3\frac{p^{\prime\alpha}}{M'}\right)u_{\Lambda_b,r}(\vec p\,),
\label{eq:ffa}
\\
\langle\Lambda^*_c;\vec p\,',r'|\bar c(0) b(0)
|\Lambda_b;\vec p,r\rangle&=&F_P\,\bar u_{\Lambda^*_c,r'}(\vec p\,')\gamma_5
u_{\Lambda_b,r}(\vec p\,),
\label{eq:ffs}\\
\langle\Lambda^*_c;\vec p\,',r'|\bar c(0)\gamma_5 b(0)
|\Lambda_b;\vec p,r\rangle&=&F_S\,\bar u_{\Lambda^*_c,r'}(\vec p\,')
u_{\Lambda_b,r}(\vec p\,),
\label{eq:ffps}
\\
\langle\Lambda^*_c;\vec p\,',r'|\bar c(0)\sigma^{\alpha\beta}\gamma_5 b(0)
|\Lambda_b;\vec p,r\rangle&=&\bar u_{\Lambda^*_c,r'}(\vec p\,')\Big[i \frac{ T_1}{M^2}(p^\alpha p^{\prime\beta}-p^\beta
p^{\prime\alpha})+
i \frac{ T_2}M(\gamma^\alpha p^\beta-\gamma^\beta p^\alpha)\nonumber\\
&&\hspace{1.75cm}+i \frac{ T_3}M
(\gamma^\alpha p^{\prime\beta}-\gamma^\beta p^{\prime\alpha})
+  T_4 \sigma^{\alpha\beta}\Big]u_{\Lambda_b,r}(\vec p\,)
\label{eq:fft} \\
\langle\Lambda^*_c;\vec p\,',r'|\bar c(0)\sigma^{\alpha\beta} b(0)
|\Lambda_b;\vec p,r\rangle&=&\bar u_{\Lambda^*_c,r'}(\vec p\,')
\epsilon^{\alpha\beta}_{\ \ \ \rho\lambda}\Big[
\frac{ T_1}{M^2}p^\rho p^{\prime\lambda}+
\frac{ T_2}M\gamma^\rho p^\lambda\nonumber\\
&&\hspace{2.5cm}+ \frac{ T_3}M
\gamma^\rho p^{\prime\lambda}
+ \frac12 T_4 \gamma^{\rho}\gamma^{\lambda}\Big]u_{\Lambda_b,r}(\vec p\,)
\label{eq:ffpt}
\eea
where  $p$ and $p'$ ($M$ and $M'$) are the four-momenta (masses) of the 
$\Lambda_b$ and $\Lambda^*_c$ baryons, respectively,  
$u_{\Lambda_b,\Lambda^*_c}$ are Dirac spinors, and we have made use of 
 $\sigma^{\alpha\beta}\gamma_5=-\frac i2\epsilon^{\alpha\beta}_{\ \ \
\rho\lambda}\sigma^{\rho\lambda}$. The form factors are Lorentz scalar functions of $q^2$ or equivalently of $\omega$,  the product of the four-velocities of the initial and final hadrons.  

The form factors used in this work are related to the helicity ones evaluated in the LQCD 
simulation of Refs.~\cite{Meinel:2021rbm,Meinel:2021mdj} by 
\bea
&&G_1=-f^{(\frac12^-)}_\perp\nonumber\\
&&G_2=M\Big(f^{(\frac12^-)}_0\frac{M+M'}{q^2}
+f^{(\frac12^-)}_+\frac{M-M'}{s_-}(1-\frac{M^2-M^{\prime2}}{q^2})
+f^{(\frac12^-)}_\perp\frac{2M'}{s_-}\Big)\nonumber\\
&&G_3=M'\Big(-f^{(\frac12^-)}_0\frac{M+M'}{q^2}
+f^{(\frac12^-)}_+\frac{M-M'}{s_-}(1+\frac{M^2-M^{\prime2}}{q^2})
-f^{(\frac12^-)}_\perp\frac{2M}{s_-}\Big)\\
&&F_1=-g^{(\frac12^-)}_\perp\nonumber\\
&&F_2=M\Big(-g^{(\frac12^-)}_0\frac{M-M'}{q^2}
-g^{(\frac12^-)}_+\frac{M+M'}{s_+}(1-\frac{M^2-M^{\prime2}}{q^2})
+g^{(\frac12^-)}_\perp\frac{2M'}{s_+}\Big)\nonumber\\
&&F_3=M'\Big(g^{(\frac12^-)}_0\frac{M-M'}{q^2}
-g^{(\frac12^-)}_+\frac{M+M'}{s_+}(1+\frac{M^2-M^{\prime2}}{q^2})
+g^{(\frac12^-)}_\perp\frac{2M}{s_+}\Big)\\
&&T_1=\frac{2M^2}{s_+}[h^{(\frac12^-)}_+-\tilde
h^{(\frac12^-)}_+-\frac{s_+(M-M')^2}{q^2s_-}(h^{(\frac12^-)}_\perp-h^{(\frac12^-)}_+)+\frac{(M+M')^2}{q^2}
(\tilde h^{(\frac12^-)}_\perp-h^{(\frac12^-)}_+)]\nonumber\\
&&T_2=-\frac{2Mq\cdot p'}{q^2s_-}(h^{(\frac12^-)}_\perp-h^{(\frac12^-)}_+)(M-M')+\frac{M}{q^2}
(\tilde h^{(\frac12^-)}_\perp-h^{(\frac12^-)}_+)(M+M')\nonumber\\
&&T_3=\frac{2Mq\cdot p}{q^2s_-}(h^{(\frac12^-)}_\perp-h^{(\frac12^-)}_+)(M-M')-\frac{M}{q^2}
(\tilde h^{(\frac12^-)}_\perp-h^{(\frac12^-)}_+)(M+M')\nonumber\\
&&T_4=h^{(\frac12^-)}_+
\eea
where $s_\pm=(M\pm M')^2-q^2=2p\cdot p'\pm2MM'=2MM'(\omega\pm1)$. Finally, thanks to  the equations 
of motion of the heavy quarks, one can relate  $F_P$ and $F_S$ to the vector and  axial form factors  as
\bea
F_P&=&\frac1{m_b-m_c}[-(M+M')G_1+(M-M'\omega)G_2+(M\omega-M')G_3],\nonumber\\
F_S&=&-\frac1{m_b+m_c}[(M-M')F_1+(M-M'\omega)F_2+(M\omega-M')F_3],
\eea
with $m_b$ and $m_c$ the masses of the $b$ and $c$ quarks respectively.

\subsection{Hadron tensors and $\widetilde W_\chi$ SFs}

In Eqs.~(\ref{eq:ffv})-(\ref{eq:ffpt}), we have interchanged the
form factor decomposition of the $\bar c(0){ O}^{(\alpha\beta)}b(0)$ and 
$\bar c(0){ O}^{(\alpha\beta)}\gamma_5b(0)$ matrix elements, with ${O}^{(\alpha\beta)}=
I,\gamma^\alpha,\sigma^{\alpha\beta}$, with respect to the ones used
in Refs.~\cite{Penalva:2020xup,Penalva:2021wye} for the $1/2^+\to 1/2^+$ case due to the opposite parity here of the final charmed baryon.  
In this way, when comparing the 
vector ($J_{HVrr'}^\alpha$), axial ($J_{HArr'}^\alpha$), scalar 
($J_{HSrr'}$), pseudoscalar ($J_{HPrr'}$), tensor 
($J_{HTrr'}^{\alpha\beta}$) and pseudotensor ($J_{HpTrr'}^{\alpha\beta}$)
hadronic matrix elements here with those for the
$1/2^+\to 1/2^+$ transition, and apart from the obvious differences in 
the actual values of the form
factors, we only   have to implement the following  changes 
\bea
J^\alpha_{Hrr'\chi}=C^V_\chi J_{HVrr'}^\alpha+h_\chi C^A_\chi J_{HArr'}^\alpha&\to &C^V_\chi
J_{HArr'}^\alpha+h_\chi C^A_\chi J_{HVrr'}^\alpha=h_\chi [C^A_\chi 
J_{HVrr'}^\alpha+h_\chi  C^V_\chi J_{HArr'}^\alpha],\nonumber\\
J_{Hrr'\chi}=\, C^S_\chi J_{HSrr'}+h_\chi C^P_\chi J_{HPrr'}&\to& C^S_\chi J_{HPrr'}
+h_\chi C^P_\chi J_{HSrr'}\,=h_\chi [C^P_\chi J_{HSrr'}
+h_\chi  C^S_\chi J_{HPrr'}],\nonumber\\
J^{\alpha\beta}_{Hrr'\chi}=\ C^T_\chi(J_{HTrr'}^{\alpha\beta}+h_\chi J_{HpTrr'}^{\alpha\beta})&\to& 
C^T_\chi(J_{HpTrr'}^{\alpha\beta}+h_\chi J_{HTrr'}^{\alpha\beta})
\ =h_\chi[ C^T_\chi (J_{HTrr'}^{\alpha\beta}+h_\chi J_{HpTrr'}^{\alpha\beta})].\nonumber\\
\eea
Since there is no left-right interference for massless neutrinos and all 
hadronic tensors are quadratic in the {WCs}, the global factor $h_\chi$ 
 is irrelevant and, to get the $\widetilde W_\chi$ SFs for the $1/2^+\to 1/2^-$ decay, it suffices to do the changes
\bea
C^V_\chi\longleftrightarrow C^A_\chi\ \ ,\ \ 
C^S_\chi\longleftrightarrow C^P_\chi
\eea
in our original expressions of Appendix C of Ref.~\cite{Penalva:2021wye}. In addition, the genuine hadron $W_{i=1,2,4,5}^{VV,AA}$, $W_{i=3}^{VA}$, $W_{1,2,3,4,5}^T$, $W_{S}$, $W_{P}$, $W_{I1,I2}^{VS,AP}$, $W_{I3}^{ST,PpT}$ and $W_{I4,I5,I6,I7}^{VT,ApT}$ SFs, which are independent of the {WCs}, 
can be read out from  Eqs.~(E3)-(E5) of Ref.~\cite{Penalva:2020xup}  obtained for the 
$\Lambda_b\to\Lambda_c$ ($1/2^+\to 1/2^+$) transition.


\section{Hadronic matrix elements for   $1/2^+\to 3/2^-$ transition}
\label{app:12p32m}

\subsection{Form factors}

For the $\Lambda_b\to\Lambda^*_c(2625)$ decay, we use the
following form factor decompositions\footnote{The form factors below are related to those in Ref.~\cite{Du:2022fxg} as $F_i^{V,A,T}=l_{V,A,T_i}$ and $F_{S,P}^{(3/2)}=l_{S,P}$.}
\bea
\langle\Lambda^*_c;\vec p\,',r'|\bar c(0)\gamma^\alpha b(0)
|\Lambda_b;\vec p,r\rangle&=&\bar u^\mu_{\Lambda^*_c,r'}(\vec p\,')\Big[
 \frac{F_1^V}{M}p_\mu\gamma^\alpha+ \frac{F^V_2}{M^2} p_\mu p^{\alpha}+
\frac{F_3^V}{MM'} p_\mu p^{\prime\alpha}\nonumber \\
&&\hspace{1.75cm}+ F_4^V g_\mu^{\ \alpha}\Big]
u_{\Lambda_b,r}(\vec p\,),
\label{eq:ffv32}\eea
\bea
\langle\Lambda^*_c;\vec p\,',r'|\bar c(0)\gamma^\alpha b(0)
|\Lambda_b;\vec p,r\rangle&=&\bar u^\mu_{\Lambda^*_c,r'}(\vec p\,')
\Big [ \frac{F_1^A}{M}p_\mu\gamma^\alpha+ \frac{F_2^A}{M^2} p_\mu p^{\alpha}+
\frac{F_3^A}{MM'} p_\mu p^{\prime\alpha}\nonumber\\
&&\hspace{1.75cm}+ F_4^A g_\mu^{\ \alpha}\Big ]\gamma_5u_{\Lambda_b,r}(\vec p\,) ,
\label{eq:ffa32} 
\eea
\bea
\langle\Lambda^*_c;\vec p\,',r'|\bar c(0) b(0)
|\Lambda_b;\vec p,r\rangle=\bar u^\mu_{\Lambda^*_c,r'}(\vec p\,')
 p_\mu \frac{F_S^{(3/2)}}M
u_{\Lambda_b,r}(\vec p\,) ,
\label{eq:ffs32} \eea
\bea
\langle\Lambda^*_c;\vec p\,',r'|\bar c(0)\gamma_5 b(0)
|\Lambda_b;\vec p,r\rangle=\bar u^\mu_{\Lambda^*_c,r'}(\vec p\,')
 p_\mu \frac{F_P^{(3/2)}}M
\gamma_5u_{\Lambda_b,r}(\vec p\,) ,
\label{eq:ffp32} 
\eea
 \bea
&&\hspace{-1.cm}\langle\Lambda^*_c;\vec p\,',r'|\bar c(0)\sigma^{\alpha\beta} b(0)
|\Lambda_b;\vec p,r\rangle=\bar u^\mu_{\Lambda^*_c,r'}(\vec p\,')
\Big[i\frac{F^T_1}{M^3} p_{\mu}(p^\alpha p^{\prime\beta}-
 p^\beta p^{\prime\alpha})+
i\frac{F^T_2}{M^2} p_{\mu}(\gamma^\alpha p^{\beta}-
 \gamma^\beta p^{\alpha})\nonumber\\
&&\hspace{6.25cm}+ i\frac{F^T_3}{M^2} p_{\mu}(\gamma^\alpha p^{\prime\beta}-
 \gamma^\beta p^{\prime\alpha})+\frac{F^T_4}M p_{\mu}\sigma^{\alpha\beta}\nonumber\\
&&\hspace{6.25cm}+iF^T_5(g_\mu^{\ \alpha}\gamma^\beta
-g_\mu^{\ \beta}\gamma^\alpha)+i\frac{F^T_6}M(g_\mu^{\ \alpha}p^\beta
-g_\mu^{\ \beta}p^\alpha)\nonumber\\
&&\hspace{6.25cm}+i\frac{F^T_7}M(g_\mu^{\ \alpha}p^{\prime\beta}
-g_\mu^{\ \beta}p^{\prime\alpha})\Big]
u_{\Lambda_b,r}(\vec p\,)
\label{eq:fft32} 
\\
&&\hspace{-1.cm}\langle\Lambda^*_c;\vec p\,',r'|\bar c(0)\sigma^{\alpha\beta}\gamma_5 b(0)
|\Lambda_b;\vec p,r\rangle=\bar u^\mu_{\Lambda^*_c,r'}(\vec p\,')
\Big[\frac{F^T_1}{M^3} p_{\mu}\epsilon^{\alpha\beta}_{\ \ \rho\lambda}p^\rho 
p^{\prime\lambda}+
\frac{F^T_2}{M^2} p_{\mu}\epsilon^{\alpha\beta}_{\ \ \rho\lambda}\gamma^\rho 
p^{\lambda}\nonumber\\
&&\hspace{6.75cm}+ 
\frac{F^T_3}{M^2} p_{\mu}
\epsilon^{\alpha\beta}_{\ \ \rho\lambda} \gamma^\rho p^{\prime\lambda}-
i\frac{F^T_4}{M} p_{\mu}\frac12\epsilon^{\alpha\beta}_{\ \
\rho\lambda}\sigma^{\rho\lambda}\nonumber\\
&&\hspace{6.75cm}+ F^T_5\epsilon^{\alpha\beta}_{\ \ \ \mu \lambda}\gamma^\lambda
+\frac{F^T_6}M\epsilon^{\alpha\beta}_{\ \ \ \mu \lambda}\,p^\lambda
\nonumber\\
&&\hspace{6.75cm}+\frac{F^T_7}M\epsilon^{\alpha\beta}_{\ \ \ \mu \lambda}\,p^{\prime\lambda}
\Big]
u_{\Lambda_b,r}(\vec p\,).
\label{eq:ffpt32}
 \eea
Here, $u^\mu(\vec p\,')$ is the Rarita-Schwinger spinor satisfying
$\slashed{p}'u^\mu(p')
=M'u^\mu(p')$ and the orthogonality conditions $\gamma_\mu u^\mu(p')=
p'_{\mu} u^\mu(p')=0$. Using these relations, together with
$\slashed{p}u(\vec p\,)=Mu(\vec p\,)$ and the identity
\bea
 \epsilon^{\alpha\mu\nu\lambda}=-\gamma_5(-i\sigma^{\alpha\mu}
 \sigma^{\nu\lambda}-ig^{\alpha \lambda}
 g^{\mu\nu}+ig^{\alpha\nu}
 g^{\mu\lambda}+g^{\alpha\nu}\sigma^{\mu \lambda}+
 g^{\mu \lambda}\sigma^{\alpha\nu}-g^{\alpha \lambda}\sigma^{\mu\nu}-
 g^{\mu\nu}\sigma^{\alpha \lambda})
\eea
one can  rewrite Eq. (\ref{eq:ffpt32}) as
 \bea
&&\hspace{-1cm}\langle\Lambda^*_c;\vec p\,',r'|\bar c(0)\sigma^{\alpha\beta}\gamma_5 b(0)
|\Lambda_b;\vec p,r\rangle\nonumber\\&&=\bar u^\mu_{\Lambda^*_c,r'}(\vec p\,')
\gamma_5\Big[- ip_{\mu}(p^\alpha p^{\prime\beta}-
 p^\beta p^{\prime\alpha})\frac{F^T_1}{M^3}+
 ip_{\mu}(\gamma^\alpha p^{\beta}-
 \gamma^\beta p^{\alpha})\frac{M'F^T_1-MF^T_2}{M^3}\nonumber\\
&&\hspace{2.5cm}+ 
 ip_{\mu}(\gamma^\alpha p^{\prime\beta}-
 \gamma^\beta p^{\prime\alpha})\frac{F^T_1+F^T_3}{M^2}\nonumber\\
&&\hspace{2.5cm}+ p_{\mu}\sigma^{\alpha\beta}[\frac{F^T_6}M-\frac{F^T_1}{M^3}(p\cdot
p'+MM')+\frac{F^T_2}M-\frac{M'}{M^2}F^T_3+\frac{F^T_4}M]\nonumber\\
&&\hspace{2.5cm}+i(g_\mu^{\ \alpha}\gamma^\beta
-g_\mu^{\ \beta}\gamma^\alpha)(F^T_5+F^T_6+\frac{M'}MF^T_7)-i(g_\mu^{\ \alpha}p^\beta
-g_\mu^{\ \beta}p^\alpha)\frac{F^T_6}M\nonumber\\
&&\hspace{2.5cm}+i(g_\mu^{\ \alpha}p^{\prime\beta}
-g_\mu^{\ \beta}p^{\prime\alpha})\frac{F^T_7}M\Big]
u_{\Lambda_b,r}(\vec p\,)
 \eea
which will be the form used in the rest of the appendix.

The vector and axial form factors used in this work are related to the helicity ones computed in  Refs.~\cite{Meinel:2021rbm,Meinel:2021mdj}
via
\bea
&&F_1^V=(f_\perp^{(\frac32^-)}+f_{\perp'}^{(\frac32^-)})\frac{MM'}{s_-},\nonumber\\
&&F_2^V=M^2\Big[
f_0^{(\frac32^-)}\frac{M'}{s_+}\frac{(M-M')}{q^2}+
f_+^{(\frac32^-)}\frac{M'}{s_-}\frac{(M+M')[q^2-
(M^2-M^{\prime2})]}{q^2s_+}\nonumber\\
&&\hspace{2cm}-
(f_\perp^{(\frac32^-)}-f_{\perp'}^{(\frac32^-)})
\frac{2M^{\prime2}}{s_-s_+}\Big]\nonumber\\
&&F_3^V=M^{\prime2}\Big[
-f_0^{(\frac32^-)}\frac{M}{s_+}\frac{(M-M')}{q^2}+
f_+^{(\frac32^-)}\frac{M}{s_-}\frac{(M+M')[q^2+
(M^2-M^{\prime2})]}{q^2s_+}\nonumber\\
&&\hspace{2cm}-
[f_\perp^{(\frac32^-)}-f_{\perp'}^{(\frac32^-)}(1-\frac{s_+}{MM'})]
\frac{2M^{2}}{s_-s_+}\Big],
\nonumber\\
&&F_4^V=f_{\perp'}^{(\frac32^-)}.
\eea
\bea
&&F_1^A=(g_\perp^{(\frac32^-)}+g_{\perp'}^{(\frac32^-)})\frac{MM'}{s_+},\nonumber\\
&&F_2^A=M^2\Big[
-g_0^{(\frac32^-)}\frac{M'}{s_-}\frac{(M+M')}{q^2}-
g_+^{(\frac32^-)}\frac{M'}{s_+}\frac{(M-M')[q^2-
(M^2-M^{\prime2})]}{q^2s_-}\nonumber\\
&&\hspace{2cm}-
(g_\perp^{(\frac32^-)}-g_{\perp'}^{(\frac32^-)})
\frac{2M^{\prime2}}{s_-s_+}\Big]\nonumber\\
&&F_3^A=M^{\prime2}\Big[
g_0^{(\frac32^-)}\frac{M}{s_-}\frac{(M+M')}{q^2}-
g_+^{(\frac32^-)}\frac{M}{s_+}\frac{(M-M')[q^2+
(M^2-M^{\prime2})]}{q^2s_-}\nonumber\\
&&\hspace{2cm}+
[g_\perp^{(\frac32^-)}-g_{\perp'}^{(\frac32^-)}(1+\frac{s_-}{MM'})]
\frac{2M^{2}}{s_-s_+}\Big],\nonumber\\
&&F_4^A=g_{\perp'}^{(\frac32^-)}.
\eea
Moreover, using the equations of motion, one can relate $F_S^{(3/2)}$ and $F_P^{(3/2)}$ 
to the vector and axial form factors through
 \bea
 F_S^{(3/2)}&=&\frac{1}{m_b-m_c}[(M-M')F_1^V+(M-M'\omega)F_2^V
 +(M\omega-M')F_3^V+MF_4^V],\nonumber\\
 F_P^{(3/2)}&=&-\frac{1}{m_b+m_c}[-(M+M')F_1^A+(M-M'\omega)F_2^A
 +(M\omega-M')F_3^A+MF_4^A].
 \eea

In the case of the matrix elements of the tensor operators, although there are seven 
different structures of 
the tensor (or pseudotensor) form,
one of them can be removed without any loss of generality. As shown in 
Ref.~\cite{Papucci:2021pmj}, there is a combination of these
structures that does not enter the physical amplitude. The argument goes as
follows. Let us consider the contraction of the matrix element $J^{\alpha\beta}_{HTrr'} 
(p,q)$ of the tensor operator $\bar c(0)\sigma^{\alpha\beta}b(0)$
with a general
 tensor $ F_{\alpha\beta}$. One would then have
\bea
J^{\alpha\beta}_{HTrr'} (p,q) F_{\alpha\beta}&=&
J^{\alpha\beta}_{HTrr'} (p,q)\, 
g^{\ \alpha'}_{\alpha}g^{\ \beta'}_{\beta} F_{\alpha'\beta'}
=
J^{\alpha\beta}_{HTrr'} (p,q)\, g_{rr}\epsilon^{\alpha'*}_r 
\epsilon_{r\alpha}\,
g_{ss}\epsilon^{\beta'*}_s \epsilon_{s\beta}F_{\alpha'\beta'},
\eea
 with $\epsilon_{r=0,\pm1}$ the usual polarization vectors of a vector particle with
four-momentum $q$ and invariant mass $\sqrt{q^2}$, 
$\epsilon_{r=t}=\frac q{\sqrt{q^2}}$
and $-g_{tt}=g_{00}=g_{\pm1\pm1}=-1$. Since $J^{\alpha\beta}_{HTrr'} (p,q)$ is
antisymmetric in the $\alpha,\,\beta$ indexes, one has
\bea
J^{\alpha\beta}_{HTrr'}
(p,q)\epsilon_{r\alpha}\epsilon_{s\beta}=
J^{\alpha\beta}_{HTrr'}
(p,q)\frac12(\epsilon_{r\alpha}\epsilon_{s\beta}-\epsilon_{r\beta}
\epsilon_{s\alpha})
\eea
and then only six different products that correspond to the values
$
(r,s)=\{(t,0),\,(t,-1),\,(t,+1)$, $(0,-1),\,(0,+1),\,(-1,+1)\}
$
could appear. One can find $\lambda_{1-7}(q)$ scalar functions such that the 
linear combination
\bea
&&\hspace{-.5cm}{\Lambda}^{\alpha\beta}_{HTrr'} (p,q,\vec \lambda)
=\bar u_{r'\mu}(\vec p\,')\Big[
\frac{\lambda_1}{M^3} p^{\mu}(p^\alpha p^{\prime\beta}-
 p^\beta p^{\prime\alpha})+
\frac{\lambda_2}{M^2} p^{\mu}(\gamma^\alpha p^{\beta}-
 \gamma^\beta p^{\alpha})+ \frac{\lambda_3}{M^2} p^{\mu}(\gamma^\alpha p^{\prime\beta}-
 \gamma^\beta p^{\prime\alpha})\nonumber\\
&&\hspace{3.75cm}-i\frac{\lambda_4}M
 p^{\mu}\sigma^{\alpha\beta}+\lambda_5(g^{\mu\alpha}
\gamma^\beta
-g^{\mu\beta}\gamma^\alpha)+\frac{\lambda_6}M(g^{\mu\alpha}p^\beta
-g^{\mu\beta}p^\alpha)\nonumber\\
&&\hspace{3.75cm}+\frac{\lambda_7}M(g^{\mu\alpha}p^{\prime\beta}
-g^{\mu\beta}p^{\prime\alpha})\Big]u_r(\vec p\,)
\eea
is orthogonal to the six $\frac12(\epsilon_{r\alpha}\epsilon_{s\beta}
-\epsilon_{r\beta}\epsilon_{s\alpha})$ antisymmetric tensors\footnote{ Using that
\bea
\epsilon^{\mu\nu\alpha\beta}\epsilon_{0\alpha}\epsilon_{t\beta}&=&
i(\epsilon^\mu_{+1}\epsilon^\nu_{-1}-\epsilon^\nu_{+1}\epsilon^\mu_{-1}),\\
\epsilon^{\mu\nu\alpha\beta}\epsilon_{\pm1\alpha}\epsilon_{t\beta}&=&
\pm i(\epsilon^\mu_{\pm1}\epsilon^\nu_{0}-\epsilon^\nu_{\pm1}\epsilon^\mu_{0}),
\eea
it is enough to ask for the orthogonality of both
${\Lambda}^{\alpha\beta}_{HTrr'}$ and ${\Lambda}^{\alpha\beta}_{HpTrr'}=-\frac i 2
\epsilon^{\alpha\beta}_{\ \ \ \rho\lambda}{\Lambda}^{\rho\lambda}_{HTrr'}$
to the combinations $\epsilon_{0}\epsilon_t$ and $\epsilon_{\pm1}\epsilon_t$.}. A choice of such functions is given in Ref.~\cite{Papucci:2021pmj} as
\bea
\vec \lambda=\Lambda\left(0,0,\frac{M}{M'},1,(\omega+1),-1,-\frac{M}{M'}\right)
\eea
where  $\Lambda$ is an arbitrary scalar function of $q^2$. Thus, no physical observable changes 
if one modifies
\bea
F^T_3&\to& F^{T'}_3=F^T_3+\Lambda\frac{M}{M'},\nonumber\\
F^T_4&\to& F^{T'}_4=F^T_3+\Lambda,\nonumber\\
F^T_5&\to& F^{T'}_5=F^T_5+\Lambda (\omega+1),\nonumber\\
F^T_6&\to& F^{T'}_6=F^T_6-\Lambda,\nonumber\\
F^T_7&\to& F^{T'}_7=F^T_7-\Lambda\frac{M}{M'},
\eea
and $\Lambda$ can be chosen so as to cancel one of the above form factors. For simplicity we omit the
prime in what follows and take $F^T_7=0$. Then one has the following relations between the tensor form factors
here and the ones defined and evaluated in  the LQCD simulation of Refs.~\cite{Meinel:2021rbm,Meinel:2021mdj} 
\bea
F^T_1&=&-\frac{2M^3M^{\prime}}{s_+s_-} (h^{(\frac32^-)}_{+}-\tilde h^{(\frac32^-)}_{+})-
\frac{2M^3M^{\prime}(M-M')^2}{s_+s_-q^2}\tilde h^{(\frac32^-)}_{\perp}
+\frac{2M^3(M-M')(M^2-MM'-q^2)}{s_+s_-q^2}\tilde h^{(\frac32^-)}_{\perp'}\nonumber\\
&&+\frac{2M^3M'(M+M')^2}{s_+s_-q^2} h^{(\frac32^-)}_{\perp}
+\frac{2M^3(M+M')(M^2+MM'-q^2)}{s_+s_-q^2} h^{(\frac32^-)}_{\perp'},\nonumber\\
F^T_2&=&\frac{2M^2M^{\prime2}}{s_+s_-}\tilde h^{(\frac32^-)}_{+}-
\frac{M^2M'(M-M')(M^2-M^{\prime2}-q^2)}{s_+s_-q^2}(\tilde h^{(\frac32^-)}_{\perp}-
\tilde h^{(\frac32^-)}_{\perp'})\nonumber\\
&&
+\frac{M^2M'(M+M')}{s_-q^2}
 (h^{(\frac32^-)}_{\perp}+ h^{(\frac32^-)}_{\perp'}),\nonumber\\
F^T_3&=&-\frac{2M^3M'}{s_+s_-}\tilde h^{(\frac32^-)}_{+}+
\frac{M^2M'(M-M')(M^2-M^{\prime2}+q^2)}{s_+s_-q^2}\tilde h^{(\frac32^-)}_{\perp}\nonumber\\
&&
-\frac{M(M-M')(M^2+M^{\prime2}-MM'-q^2)(M^2-M^{\prime2}+q^2)}{s_+s_-q^2}
\tilde h^{(\frac32^-)}_{\perp'}
-\frac{M^2M'(M+M')}{s_-q^2}
 h^{(\frac32^-)}_{\perp}\nonumber\\
&&
-\frac{M(M^3+M^{\prime3}-q^2(M+M'))}{s_-q^2}
 h^{(\frac32^-)}_{\perp'},
\nonumber \\
F^T_4&=&\frac{MM'}{s_+}\tilde h^{(\frac32^-)}_{+}-\frac{M'(M+M')}{q^2}h^{(\frac32^-)}_{\perp'}-
\frac{M'(M-M')(M^2-M^{\prime2}+q^2)}{q^2s_+}\tilde h^{(\frac32^-)}_{\perp'},
\nonumber\\
 F^T_5&=&-\frac1{2M q^2}\Big[h^{(\frac32^-)}_{\perp'} (M+M') s_+
+\tilde h^{(\frac32^-)}_{\perp'}(M-M')(M^2-M'^2+q^2) \Big],\nonumber\\
F^T_6&=&\frac1{q^2}\Big[h^{(\frac32^-)}_{\perp'}(M+M')^2
+\tilde h^{(\frac32^-)}_{\perp'}(M-M')^2\Big].
\eea

 
\subsection{Hadron tensors and $\widetilde W_\chi$ SFs}
\label{app:wchi32}
As already mentioned, in  Refs.~\cite{Penalva:2021wye,Penalva:2020xup} we derived general 
expressions for the hadronic tensors that are  
  valid for any CC transition, the differences being encoded in the actual
 values of the $\widetilde W_\chi$ SFs. In this case, writing 
 \bea
J^{\mu\,(\alpha\beta)}_{H\chi rr'}= \bar u_{r'\mu}(\vec p\,')
\Gamma^{\mu\,(\alpha\beta)}_{H\chi}u_{r}(\vec p\,'),
 \eea
we have that  the hadronic tensors  are given by the traces
\bea
 W_\chi^{(\alpha\beta)(\rho\lambda))}=\frac12{\rm Tr}\Big[\Big(\sum_{r'}u_{r'\nu}(\vec p\,')
 \bar u_{r'\mu}(\vec p\,')\Big)\Gamma^{\mu\,(\alpha\beta)}_{H\chi} \Big(\sum_r
 u_r(\vec p\,)\bar u_r(\vec p\,)\Big)\gamma^0
 \Gamma^{\nu\,(\rho\lambda)\dagger}_{H\chi}\gamma^0\Big],
 \eea
 with
 \bea
 \sum_r
 u_r(\vec p\,)\bar u_r(\vec p\,)&=&(\slashed{p}+M),\\
 \sum_{r'}u_{r'\nu}(\vec p\,')
\bar u_{r'\mu}(\vec p\,')&=&-(\slashed{p}'+M')\Big[g_{\nu\mu}
-\frac13\gamma_\nu\gamma_\mu-\frac23\frac{p'_\nu p'_\mu}{M^{\prime2}}
+\frac13\frac{p'_\nu\gamma_\mu-p'_\mu\gamma_\nu}{M'}\Big]
 \eea
and
  \bea
\Gamma^{\mu\,(\alpha\beta)}_{H\chi}= C^V_\chi\Gamma^{\mu\,\alpha}_{HV}+h_\chi C^A_\chi\Gamma^{\mu\alpha}_{HA},\ 
 C^S_\chi\Gamma^\mu_{HS}+h_\chi C^P_\chi\Gamma^\mu_{HP},
 \ C^T_\chi(\Gamma^{\mu\,\alpha\beta}_{HT}+h_\chi
 \Gamma^{\mu\,\alpha\beta}_{HpT}),
 \eea
 where the  $\Gamma$'s can be easily read out from Eqs.~(\ref{eq:ffv32})-(\ref{eq:ffpt32}).

From a direct comparison of the results for those traces with the general form of 
the different $W^{(\alpha\beta)(\rho\lambda)}_\chi$ tensors in 
Refs.~\cite{Penalva:2021wye,Penalva:2020xup} we extract the  corresponding 
$1/2^+\to 3/2^-$  $\widetilde W_\chi$ SFs. They have been obtained with the use of  
the FeynCalc package~\cite{MERTIG1991345,Shtabovenko:2016sxi,Shtabovenko:2020gxv} on Mathematica~\cite{Mathematica} and they 
are given in terms of the { WCs} and form factors by the following expressions\footnote{Here we have
kept the explicit dependence on $F^T_7$.}. 

\bea
 \widetilde{W}_{1\chi}&=&\frac1{3}\Big\{\hspace{0.25cm} |C^V_\chi|^2(\omega+1)\Big[
 (F^V_4)^2+(F^V_1)^2(\omega-1)^2-F^V_1F^V_4(\omega-1)\Big]
 \nonumber\\
 &&\hspace{0.5cm}+|C^A_\chi|^2(\omega-1)
 \Big[(G^V_4)^2+(G^V_1)^2(\omega+1)^2-G^V_1G^V_4(\omega+1)\Big]\Big\},\label{eq:w32-ini} \\
  \widetilde{W}_{2\chi}&=&\frac1{3M^{\prime2}}\Big\{\ |C^V_\chi|^2
  \Big[2(F^V_1)^2M M'(\omega^2-1)+F^V_1[F^V_4(M^2(1+2\omega)-2MM'-M^{\prime2})
 \nonumber\\
 &&\hspace{2.5cm}+2(M+M')(\omega^2-1)(F^V_3 M+F^V_2 M')]
 +(\omega+1)[(F^V_4)^2M^2\nonumber\\
 &&\hspace{2.5cm}-2(F^V_3M+F^V_2M')F^V_4(M'-M\omega)
+(\omega^2-1)
(F^V_3M+F^V_2M^{\prime})^2] \Big]\nonumber\\
 &&\hspace{1.25cm}+|C^A_\chi|^2
\Big[2(F^A_1)^2M M'(\omega^2-1)+F^A_1[F^A_4(M^2(1-2\omega)+2MM'-M^{\prime2})
\nonumber\\
 &&\hspace{2.5cm} -2(M-M')(\omega^2-1)(F^A_3 M+F^A_2 M')]
 +(\omega-1)[(F^A_4)^2M^2\nonumber\\
&&\hspace{2.5cm}-2(F^A_3 M+F^A_2M')F^A_4(M'-M\omega)
+(\omega^2-1)
(F^A_3M+F^A_2M^{\prime})^2] \Big]
\Big\},\nonumber\\ \\
  \widetilde{W}_{3\chi}&=&-\frac{2{\rm Re}(C^V_\chi C^{A*}_\chi)M}{3M'}\Big\{
  F^V_4[F^A_1(\omega+1)+F^A_4]+(\omega-1)F^V_1[F^A_4-2F^A_1(\omega+1)]\Big\},\\
  \widetilde{W}_{4\chi}&=& \frac{M^2}{3M^{\prime2}}\Big\{\ |C^V_\chi|^2\Big[
   F^V_1[F^V_4(1+2\omega)+2F^V_3(\omega^2-1)]\nonumber\\
  &&\hspace{2.5cm} +(\omega+1)
   [(F^V_4)^2+2F^V_3F^V_4\omega+(F^V_3)^2(\omega^2-1)]\Big]\nonumber\\
&&\hspace{1.1cm}+|C^A_\chi|^2 \Big[
   F^A_1[F^A_4(1-2\omega)-2F^A_3(\omega^2-1)]\nonumber\\
  &&\hspace{2.5cm}+(\omega-1)
   [(F^A_4)^2+2F^A_3F^A_4\omega+(F^A_3)^2(\omega^2-1)]\Big] 
  \Big\},\\
    \widetilde{W}_{5\chi}&=& \frac{2M}{3M'}\Big\{\ |C^V_\chi|^2\Big[
  -(\omega^2-1){ [}(F^V_1+F^V_2)(F^V_1+F^V_3)+F^V_2F^V_3\omega]\nonumber\\
   &&\hspace{2.5cm}+F^V_4
  [F^V_1+(F^V_3-F^V_2\omega)(\omega+1)]\nonumber\\
   &&\hspace{2.5cm}-\frac{M}{M'}\{F^V_1[2F^V_3(\omega^2-1)+F^V_4(1+2\omega)]\nonumber\\
   &&\hspace{2.5cm}
   +(\omega+1)[(F^V_4)^2+2F^V_3F^V_4\omega+(F^V_3)^2(\omega^2-1)]\}\Big]\nonumber\\
&&\hspace{1.05cm}+|C^A_\chi|^2\Big[
  -(\omega^2-1)[(F^A_1-F^A_2)(F^A_1+F^A_3)+F^A_2F^A_3\omega]\nonumber\\
   &&\hspace{2.5cm}-F^A_4
  [F^A_1 {-} (F^A_3-F^A_2\omega)(\omega-1)]\nonumber\\
   &&\hspace{2.5cm}+\frac{M}{M'}\{F^A_1[2F^A_3(\omega^2-1)+F^A_4(2\omega-1)]\nonumber\\
   &&\hspace{2.5cm}
   -(\omega-1)[(F^A_4)^2+2F^A_3F^A_4\omega+(F^A_3)^2(\omega^2-1)]\}\Big] 
  \Big\},\\
  \widetilde{W}_{SP\chi}&=&\frac{\omega^2-1}3\Big[|C^S_\chi|^2\left(F^{(3/2)}_S\right)^2(\omega+1)+
  |C^P_\chi|^2\left(F^{(3/2)}_P\right)^2(\omega-1)\Big],
  \eea
  \bea
   \widetilde{W}_{I1\chi}&=& \frac{2}{3M'}\Big\{
  C^V_\chi C^{S*}_\chi
  F_S^{(3/2)}(\omega+1)\Big[F^V_1(M+M')(\omega-1)+(F^V_2M'+F^V_3M)(\omega^2-1)\nonumber\\
  &&\hspace{1.cm}+
  F^V_4(M\omega-M')\Big]+C^A_\chi C^{P*}_\chi
  F_P^{(3/2)}(\omega-1)\Big[F^A_1(M'-M)(\omega+1)\nonumber\\
  &&\hspace{1.cm}+(F^A_2M'+F^A_3M)(\omega^2-1)+
  F^A_4(M\omega-M')\Big]\Big\},\\
   \widetilde{W}_{I2\chi}&=& \frac{2M}{3M'}\Big\{
  -C^V_\chi C^{S*}_\chi
  F_S^{(3/2)}(\omega+1)\Big[F^V_1(\omega-1)+F^V_3(\omega^2-1)+
  F^V_4\omega\Big]\nonumber\\
  &&\hspace{1.15cm}+C^A_\chi C^{P*}_\chi
  F_P^{(3/2)}(\omega-1)\Big[F^A_1(\omega+1)-F^A_3(\omega^2-1)-
  F^A_4\omega
  \Big]\Big\},\\
   \widetilde{W}_{I3\chi}&=&-\frac{2}{3M'}\Big\{C^S_\chi C^{T*}_\chi F_S^{(3/2)}(\omega+1)
   \Big[M[F^T_5+\omega F^T_6+(F^T_2+F^T_4)(\omega-1)]\nonumber\\
   &&\hspace{1.25cm}+M'[F^T_7-(\omega-1)(F^T_1(\omega+1)+F^T_3)]\Big]\nonumber\\
   &&\hspace{1.25cm}+
  C^P_\chi C^{T*}_\chi F_P^{(3/2)}(\omega-1)M[-F^T_5+F^T_4(\omega+1)]\Big\},\\ 
   \widetilde{W}_{I4\chi}&=&\frac1{3M^{\prime2}}
 \Big\{C^V_\chi C^{T*}_\chi \Big[-F^V_1\{F^T_5 M(M'+M(2\omega+1))+
F^T_6 M(M'\omega+M(2\omega+1))\nonumber\\
&&\hspace{2.75cm}+
M'[(2MF^T_2-2MF^T_3-2(M+M')F^T_1)(\omega^2-1)\nonumber\\
&&\hspace{2.75cm}+F^T_7(M(2+\omega)+M')]\}-2(F^V_2M'+F^V_3M)(\omega+1)\nonumber\\
&&\hspace{3.75cm}\times\{-F^T_1M'(\omega^2-1)+
[M(F^T_2+F^T_4)-M'F^T_3](\omega-1)\nonumber\\
&&\hspace{4.5cm}+M(F^T_5+F^T_6\omega)+M'F^T_7\}\nonumber\\
&&\hspace{2.75cm}-F^V_4\{2F^T_1M'(\omega+1)[M^{\prime}-M\omega]+
F^T_2M[M(1+2\omega)-M'(2+\omega)]\nonumber\\
&&\hspace{2.75cm}+F^T_3M'(M'-M\omega)+F^T_4M[M(1+2\omega)-M']
\nonumber\\
&&\hspace{2.75cm}+M^2[F^T_5+2F^T_6(\omega+1)]
\}\Big]\nonumber\\
&&\hspace{1.cm}+C^A_\chi C^{T*}_\chi M
 \Big[F^A_1\{M'(\omega+1)[-2(F^T_2+F^T_3)(\omega-1)+F^T_6+F^T_7]\nonumber\\
&&\hspace{3cm}+
 F^T_5(M'+M(2\omega-1))\}\nonumber\\
&&\hspace{3cm}-2(F^A_2M'+F^A_3M)(\omega-1)[F^T_4(\omega+1)-F^T_5]\nonumber\\
&&\hspace{3cm}+F^A_4\{MF^T_5+M'[F^T_6+F^T_7+(\omega-1)(F^T_2+F^T_3)]\nonumber\\
&&\hspace{3cm}+
F_4(M(1-2\omega)+M')\}\Big]\Big\},\\
   \widetilde{W}_{I5\chi}&=&\frac M{3M^{\prime2}}
 \Big\{C^V_\chi C^{T*}_\chi  \Big[
F^V_1\{(F^T_5+F^T_6)M(1+2\omega)+M'[F^T_7(\omega+2)\nonumber\\
&&\hspace{2.5cm}-2(F^T_1+F^T_3)
(\omega^2-1)]\}+2F^V_3(\omega+1)\{M'[-F^T_1(\omega^2-1)\nonumber\\
&&\hspace{2.5cm}-F^T_3(\omega-1)+
F^T_7]+M[F^T_5+(F^T_2+F^T_4)(\omega-1)+F^T_6\omega]\}\nonumber\\
&&\hspace{2.5cm}+F^V_4\{-M'[2F^T_1\omega(\omega+1)+F^T_3\omega]
+M[F^T_5+2F^T_6(1+\omega)\nonumber\\
&&\hspace{2.5cm}+(F^T_2+F^T_4)(1+2\omega)]
\}
\Big]\nonumber\\
&&\hspace{1.25cm}+C^A_\chi C^{T*}_\chi  \Big[-F_1^A \{F^T_5M(2 \omega-1)+
 M'(\omega+1)[-2F^T_3(\omega-1)+F^T_7]\}\nonumber\\
	&&\hspace{3.05cm}+2F_3^A M(\omega-1)[F^T_4(\omega+1)-F^T_5]\nonumber\\
&&\hspace{3.05cm}-F_4^A\{F^T_5M+M'[F^T_3(\omega-1)+F^T_7]+
F^T_4M(1-2\omega)\}\Big]\Big\},
\eea
%
%
\bea
   \widetilde{W}_{I6\chi}&=&\frac1{3MM^{\prime}}
 \Big\{ C^V_\chi C^{T*}_\chi M
\Big[F^V_1(\omega-1)\{F^T_5[M'-M(1+2\omega)]
+(\omega+1)[2F^T_4(M-M')\nonumber\\
&&\hspace{3.5cm}-M'[-2(F^T_2+F^T_3)(\omega-1)+F^T_6+F^T_7]]\}\nonumber\\
&&\hspace{3.5cm}+F^V_4\{(\omega+1)[F^T_4(M'-M)+M'[(F^T_2+F^T_3)(1-\omega)
\nonumber\\
&&\hspace{3.5cm}+2(F^T_6+F^T_7)]]+F^T_5[M'+M(2+\omega)]\}
\Big]\nonumber\\
&&\hspace{1.5cm}-C^A_\chi C^{T*}_\chi
\Big[F^A_1(\omega+1)\{2F^T_2M(\omega-1)(M-M'\omega)+2F^T_3M'(\omega-1)
(M\omega-M')\nonumber\\
 &&\hspace{3.5cm}+2F^T_4M(M+M')(\omega-1)+F^T_5M[M'+M(1-2\omega)]\nonumber\\
 &&\hspace{3.5cm}-F^T_6M(M-M'\omega)+F^T_7M'(M'-M\omega)\}\nonumber\\
&&\hspace{3.5cm}+F^A_4\{-F^T_2M(\omega-1)(M-M'\omega)+F^T_3M^{\prime}(\omega-1)
(M'-M\omega)
\nonumber\\
 &&\hspace{3.5cm}-F^T_4M(M+M')(\omega-1)+F^T_5M[M'+M(\omega-2)]\nonumber\\
 &&\hspace{3.5cm}-F^T_6M(M-M'\omega)+F^T_7M'(M'-M\omega)\}
	\Big]
\Big\},\\
   \widetilde{W}_{I7\chi}&=&\frac1{3M^{\prime}}
 \Big\{ - C^V_\chi C^{T*}_\chi \Big[F^V_1(\omega-1)
\{(\omega+1)[2F^T_4M-M'[F^T_7-2F^T_3(\omega-1)]]-F^T_5M(1+2\omega)\}\nonumber\\
 &&\hspace{3cm}+F^V_4\{(\omega+1)
 [-F^T_4M+M'[2F^T_7-F^T_3(\omega-1)]]+
F^T_5M(2+\omega) \}
\Big]\nonumber\\
&&\hspace{1.25cm}+C^A_\chi C^{T*}_\chi
 \Big[F^A_1(\omega+1)\{2[(F^T_2+F^T_4)M+F^T_3M'\omega](\omega-1)+F^T_5M(1-2\omega)\nonumber\\
 &&\hspace{3cm}-F^T_6M-F^T_7M'\omega\}
 -F^A_4\{[(F^T_2+F^T_4)M+F^T_3M'\omega](\omega-1)\nonumber\\
 &&\hspace{3cm}
 +F^T_5M(2-\omega)+F^T_6M+F^T_7M'\omega\}
	\Big]
\Big\},\\
%
%
%
%
\widetilde
W_{1\chi}^T&=&\frac{|C^T_\chi|^2}{3M^2}\Big\{M^2\Big[2\omega
(F^{T}_5)^2+2(\omega+1)F^T_5(F^T_6\omega+F^T_2(\omega-1))
+(\omega+1)[\omega^2((F^{T}_4)^2\nonumber\\
&&\hspace{1.25cm}+
(F^T_6+F^T_2+F^T_4)^2)-2\omega(F^T_2+F^T_4)(F^T_6+F^T_2+F^T_4)+F^T_2(F^T_2+2F^T_4)]\Big]\nonumber\\
&&\hspace{1.25cm}+2MM'(\omega+1)[F^T_5+\omega F^T_6-(F^T_2+F^T_4)(1-\omega)]\nonumber\\
&&\hspace{1.75cm}\times
[F^T_7-(\omega-1)(F^T_1(1+\omega)+F^T_3)]\nonumber\\
&&\hspace{1.25cm}+
M^{\prime2}(\omega+1)[F^T_7-(\omega-1)(F^T_1(1+\omega)+F^T_3)]^2\Big\},\\
\widetilde
W_{3\chi}^T&=&\frac{|C^T_\chi|^2}{3M^{\prime2}}\Big\{M^2\Big[-2\omega
(F^T_5)^2+F^T_5(F^T_6+F^T_2(1+2\omega)+4\omega F^T_4)
+F^T_6[F^T_6(1+\omega)\nonumber\\
&&\hspace{1cm}+(F^T_2+F^T_4)(1+2\omega)]\Big]
+MM'\Big[-2\omega^2[F^T_6F^T_1+F^T_2(F^T_1+F^T_3)+F^T_4(F^T_1+2F^T_3)]
\nonumber\\
&&\hspace{1cm}-\omega F^T_6(2F^T_1+F^T_3)+2F^T_2(F^T_1+F^T_3)+2F^T_4(F^T_1+2F^T_3)\nonumber\\
&&\hspace{1cm}+F^T_5
[-2F^T_7(2+\omega)-2F^T_1(1+\omega)-F^T_3(3+2\omega-4\omega^2)]\nonumber\\
&&\hspace{1cm}+F^T_7[F^T_2(\omega+2)+F^T_4(2\omega+3)]\Big]-
M^{\prime2}\Big[2(\omega+1)(F^T_7)^2+F^T_7[2(\omega+1)F^T_1\nonumber\\
&&\hspace{1cm}+F^T_3(3-2\omega^2)]-
(\omega^2-1)[(\omega+1)(F^T_1)^2+2F^T_1F^T_3-2(\omega-1)(F^T_3)^2]
\Big]\Big\},
\eea
\bea
\widetilde
W_{2\chi}^T&=&-\frac{|C^T_\chi|^2}{3M^2M^{\prime2}}\Big\{M^4\Big[2\omega
(F^T_5)^2-F^T_5(F^T_6+F^T_2(1+2\omega)+4\omega F^T_4)
-F^T_6[F^T_6(1+\omega)\nonumber\\
&&\hspace{2cm}+(F^T_2+F^T_4)(1+2\omega)]\Big]+M^3M'\Big[2(F^T_5)^2-4(F^T_4)^2
+F^T_5[4F^T_6(\omega+1)\nonumber\\
&&\hspace{2cm}+2F^T_7(2+\omega)+2F^T_1(1+\omega)
+2F^T_2(1+2\omega)+F^T_3(3+2\omega-4\omega^2)+4F^T_4]\nonumber\\
&&\hspace{2cm}+2\omega^2[(F^T_6)^2+F^T_6(F^T_1+2(F^T_2+F^T_4))+2(F^T_4)^2+F^T_2(F^T_1+F^T_3)\nonumber\\
&&\hspace{2cm}+
F^T_4(F^T_1+2(F^T_2+F^T_3))]-2F^T_2(F^T_1+F^T_3)
-2F^T_4(F^T_6+F^T_1+2(F^T_2+F^T_3))\nonumber\\
&&\hspace{2cm}+\omega F^T_6(2(F^T_6+F^T_1+F^T_2)+F^T_3)
-F^T_7[(\omega+2)F^T_2+F^T_4(2\omega+3)]
\Big]\nonumber\\
&&\hspace{2cm}+M^2M^{\prime2}\Big[-\omega^3[(F^T_1)^2+4F^T_1(F^T_6+F^T_2+F^T_4)-2((F^T_2)^2
+(F^T_3)^2)]\nonumber\\
&&\hspace{2cm}-\omega^2[(F^T_1)^2+2F^T_1(2F^T_5+F^T_3)+2((F^T_2)^2
+(F^T_3)^2)+2F^T_3(F^T_7+2F^T_2)\nonumber\\
&&\hspace{2cm}+2F^T_6(2F^T_1+F^T_2+2F^T_3)+4F^T_4(F^T_2+F^T_3)]\nonumber\\
&&\hspace{2cm}+\omega[(F^T_6)^2+2F^T_6(2F^T_7-F^T_2)+2((F^T_7)^2-(F^T_2)^2-(F^T_3)^2)+(F^T_1)^2\nonumber\\
&&\hspace{2cm}+2F^T_1(-2F^T_5+F^T_7+2F^T_2)+4F^T_7F^T_2+4F^T_4(F^T_7+F^T_1)]\nonumber\\
&&\hspace{2cm}+(F^T_6)^2+(F^T_1)^2+2((F^T_7)^2+(F^T_2)^2+(F^T_3)^2)
\nonumber\\
&&\hspace{2cm}
+F^T_5(F^T_6+2F^T_7-3F^T_2-2F^T_3)+F^T_6(4F^T_7+F^T_2+2F^T_3+F^T_4)\nonumber\\
&&\hspace{2cm}+F^T_7(2F^T_1+2F^T_2+3F^T_3+2F^T_4)+2F^T_3(F^T_1+2F^T_2)+4F^T_4(F^T_2+F^T_3)\Big]\nonumber\\
&&\hspace{2cm}+MM^{\prime3}\Big[2F^T_2F^T_3\omega^2+2(F^T_1)^2\omega(\omega-1)
(\omega+1)^2+\omega(-F^T_7F^T_2+F^T_6F^T_3\nonumber\\
&&\hspace{2cm}-2F^T_7F^T_3)-2F^T_2F^T_3-F^T_7(2F^T_2+F^T_4)+F^T_5(F^T_3+2F^T_1(\omega+1))\nonumber\\
&&\hspace{2cm}+2(\omega+1)F^T_1[(\omega-1)(F^T_2+F^T_4)+\omega(F^T_6-2(F^T_7-F^T_3(\omega-1)))]\Big]\nonumber\\
&&\hspace{2cm}+M^{\prime4}\Big[F^T_7(F^T_3+2F^T_1(\omega+1))-(\omega^2-1)F^T_1
(F^T_1(\omega+1)+2F^T_3)\Big]\Big\},\\
\widetilde
W_{4\chi}^T&=&\frac{|C^T_\chi|^2}{3MM^{\prime2}}\Big\{M^3\Big[2\omega
(F^T_5)^2-F^T_5(F^T_6+F^T_2(1+2\omega)+4\omega F^T_4)
-F^T_6[F^T_6(1+\omega)\nonumber\\
&&\hspace{1.75cm}+(F^T_2+F^T_4)(1+2\omega)]\Big]
+M^2M'\Big[(F^T_5)^2+F^T_5[-4F^T_3\omega^2\nonumber\\
&&\hspace{1.75cm}+2\omega(F^T_6+F^T_7+F^T_1+F^T_2+F^T_3)+
2F^T_6+4F^T_7+2F^T_1\nonumber\\
&&\hspace{1.75cm}+F^T_2+3F^T_3+2F^T_4]
+\omega^2[(F^T_6)^2+2F^T_6(F^T_1+F^T_2+F^T_4)]
\nonumber\\
&&\hspace{1.75cm}+2(\omega^2-1)[(F^T_4)^2+F^T_4(F^T_1+F^T_2+2F^T_3)
+F^T_2(F^T_1+F^T_3)]\nonumber\\
&&\hspace{1.75cm}+\omega F^T_6(2F^T_1+F^T_2+F^T_3)+F^T_6(\omega F^T_6-F^T_4)\nonumber\\
&&\hspace{1.75cm}-F^T_7[(\omega+2)F^T_2+(2\omega+3)F^T_4]\Big]
+MM^{\prime2}\Big[-\omega^3[F^T_1(F^T_1+2(F^T_6+F^T_2+F^T_4))
\nonumber\\
&&\hspace{1.75cm}-2(F^T_3)^2]-\omega^2[2F^T_5F^T_1
+2F^T_6(F^T_1+F^T_3)+2F^T_7F^T_3+F^T_1(F^T_1+2F^T_3)\nonumber\\
&&\hspace{1.75cm}+
2F^T_3(F^T_2+F^T_3+F^T_4)]+\omega[2F^T_7(F^T_6+F^T_7+F^T_1+F^T_2+F^T_4)-2(F^T_3)^2
\nonumber\\
&&\hspace{1.75cm}+
F^T_1(-2F^T_5+F^T_1+2(F^T_2+F^T_4))]
+F^T_7(2F^T_7+F^T_5+2F^T_6+2F^T_1+F^T_2\nonumber\\
&&\hspace{1.75cm}+3F^T_3+F^T_4)+F^T_1(F^T_1+2F^T_3)+F^T_3(2F^T_3-F^T_5+F^T_6+2F^T_2+2F^T_4)\Big]\nonumber\\
&&\hspace{1.75cm}+M^{\prime3}\omega\Big[
(\omega^2-1)F^T_1
(F^T_1(\omega+1)+2F^T_3)-F^T_7(F^T_3+2F^T_1(\omega+1))\Big]\Big\}.
\label{eq:w32-fin}
\eea

As shown in Refs.~\cite{Penalva:2020xup,Penalva:2021wye}, one has the
general  constraint
\bea
2M^2\widetilde W^T_{1\chi}+p^2\widetilde W^T_{2\chi}+q^2\widetilde W^T_{3\chi}+
2p\cdot q
\widetilde W^T_{4\chi}=0,
\eea
that allows to eliminate  $\widetilde W^T_{1\chi}$ in terms of the
 other three SFs. In fact, as shown in Refs.~\cite{Penalva:2020xup,Penalva:2021wye}, 
 the term in $\widetilde W^T_{1\chi}$ of the hadron tensor does not 
contribute when contracted with the corresponding lepton tensor.

\bibliography{B2Dbib}

\begin{thebibliography}{88}%
\makeatletter
\providecommand \@ifxundefined [1]{%
 \@ifx{#1\undefined}
}%
\providecommand \@ifnum [1]{%
 \ifnum #1\expandafter \@firstoftwo
 \else \expandafter \@secondoftwo
 \fi
}%
\providecommand \@ifx [1]{%
 \ifx #1\expandafter \@firstoftwo
 \else \expandafter \@secondoftwo
 \fi
}%
\providecommand \natexlab [1]{#1}%
\providecommand \enquote  [1]{``#1''}%
\providecommand \bibnamefont  [1]{#1}%
\providecommand \bibfnamefont [1]{#1}%
\providecommand \citenamefont [1]{#1}%
\providecommand \href@noop [0]{\@secondoftwo}%
\providecommand \href [0]{\begingroup \@sanitize@url \@href}%
\providecommand \@href[1]{\@@startlink{#1}\@@href}%
\providecommand \@@href[1]{\endgroup#1\@@endlink}%
\providecommand \@sanitize@url [0]{\catcode `\\12\catcode `\$12\catcode
  `\&12\catcode `\#12\catcode `\^12\catcode `\_12\catcode `\%12\relax}%
\providecommand \@@startlink[1]{}%
\providecommand \@@endlink[0]{}%
\providecommand \url  [0]{\begingroup\@sanitize@url \@url }%
\providecommand \@url [1]{\endgroup\@href {#1}{\urlprefix }}%
\providecommand \urlprefix  [0]{URL }%
\providecommand \Eprint [0]{\href }%
\providecommand \doibase [0]{http://dx.doi.org/}%
\providecommand \selectlanguage [0]{\@gobble}%
\providecommand \bibinfo  [0]{\@secondoftwo}%
\providecommand \bibfield  [0]{\@secondoftwo}%
\providecommand \translation [1]{[#1]}%
\providecommand \BibitemOpen [0]{}%
\providecommand \bibitemStop [0]{}%
\providecommand \bibitemNoStop [0]{.\EOS\space}%
\providecommand \EOS [0]{\spacefactor3000\relax}%
\providecommand \BibitemShut  [1]{\csname bibitem#1\endcsname}%
\let\auto@bib@innerbib\@empty
\bibitem [{\citenamefont {Aad}\ \emph {et~al.}(2012)\citenamefont {Aad} \emph
  {et~al.}}]{ATLAS:2012yve}%
  \BibitemOpen
  \bibfield  {author} {\bibinfo {author} {\bibfnamefont {G.}~\bibnamefont
  {Aad}} \emph {et~al.} (\bibinfo {collaboration} {ATLAS}),\ }\href {\doibase
  10.1016/j.physletb.2012.08.020} {\bibfield  {journal} {\bibinfo  {journal}
  {Phys. Lett. B}\ }\textbf {\bibinfo {volume} {716}},\ \bibinfo {pages} {1}
  (\bibinfo {year} {2012})},\ \Eprint {http://arxiv.org/abs/1207.7214}
  {arXiv:1207.7214 [hep-ex]} \BibitemShut {NoStop}%
\bibitem [{\citenamefont {Chatrchyan}\ \emph {et~al.}(2012)\citenamefont
  {Chatrchyan} \emph {et~al.}}]{CMS:2012qbp}%
  \BibitemOpen
  \bibfield  {author} {\bibinfo {author} {\bibfnamefont {S.}~\bibnamefont
  {Chatrchyan}} \emph {et~al.} (\bibinfo {collaboration} {CMS}),\ }\href
  {\doibase 10.1016/j.physletb.2012.08.021} {\bibfield  {journal} {\bibinfo
  {journal} {Phys. Lett. B}\ }\textbf {\bibinfo {volume} {716}},\ \bibinfo
  {pages} {30} (\bibinfo {year} {2012})},\ \Eprint
  {http://arxiv.org/abs/1207.7235} {arXiv:1207.7235 [hep-ex]} \BibitemShut
  {NoStop}%
\bibitem [{\citenamefont {Langacker}(2017)}]{langacker:2017}%
  \BibitemOpen
  \bibfield  {author} {\bibinfo {author} {\bibfnamefont {P.}~\bibnamefont
  {Langacker}},\ }\href@noop {} {\emph {\bibinfo {title} {{The Standard Model
  and Beyond}}}}\ (\bibinfo  {publisher} {CRC Press, Taylor \& Francis},\
  \bibinfo {address} {Boca Raton, FL},\ \bibinfo {year} {2017})\BibitemShut
  {NoStop}%
\bibitem [{\citenamefont {Lees}\ \emph {et~al.}(2012)\citenamefont {Lees} \emph
  {et~al.}}]{BaBar:2012obs}%
  \BibitemOpen
  \bibfield  {author} {\bibinfo {author} {\bibfnamefont {J.~P.}\ \bibnamefont
  {Lees}} \emph {et~al.} (\bibinfo {collaboration} {BaBar}),\ }\href {\doibase
  10.1103/PhysRevLett.109.101802} {\bibfield  {journal} {\bibinfo  {journal}
  {Phys. Rev. Lett.}\ }\textbf {\bibinfo {volume} {109}},\ \bibinfo {pages}
  {101802} (\bibinfo {year} {2012})},\ \Eprint {http://arxiv.org/abs/1205.5442}
  {arXiv:1205.5442 [hep-ex]} \BibitemShut {NoStop}%
\bibitem [{\citenamefont {Lees}\ \emph {et~al.}(2013)\citenamefont {Lees} \emph
  {et~al.}}]{BaBar:2013mob}%
  \BibitemOpen
  \bibfield  {author} {\bibinfo {author} {\bibfnamefont {J.~P.}\ \bibnamefont
  {Lees}} \emph {et~al.} (\bibinfo {collaboration} {BaBar}),\ }\href {\doibase
  10.1103/PhysRevD.88.072012} {\bibfield  {journal} {\bibinfo  {journal} {Phys.
  Rev. D}\ }\textbf {\bibinfo {volume} {88}},\ \bibinfo {pages} {072012}
  (\bibinfo {year} {2013})},\ \Eprint {http://arxiv.org/abs/1303.0571}
  {arXiv:1303.0571 [hep-ex]} \BibitemShut {NoStop}%
\bibitem [{\citenamefont {Huschle}\ \emph {et~al.}(2015)\citenamefont {Huschle}
  \emph {et~al.}}]{Belle:2015qfa}%
  \BibitemOpen
  \bibfield  {author} {\bibinfo {author} {\bibfnamefont {M.}~\bibnamefont
  {Huschle}} \emph {et~al.} (\bibinfo {collaboration} {Belle}),\ }\href
  {\doibase 10.1103/PhysRevD.92.072014} {\bibfield  {journal} {\bibinfo
  {journal} {Phys. Rev. D}\ }\textbf {\bibinfo {volume} {92}},\ \bibinfo
  {pages} {072014} (\bibinfo {year} {2015})},\ \Eprint
  {http://arxiv.org/abs/1507.03233} {arXiv:1507.03233 [hep-ex]} \BibitemShut
  {NoStop}%
\bibitem [{\citenamefont {Sato}\ \emph {et~al.}(2016)\citenamefont {Sato} \emph
  {et~al.}}]{Belle:2016ure}%
  \BibitemOpen
  \bibfield  {author} {\bibinfo {author} {\bibfnamefont {Y.}~\bibnamefont
  {Sato}} \emph {et~al.} (\bibinfo {collaboration} {Belle}),\ }\href {\doibase
  10.1103/PhysRevD.94.072007} {\bibfield  {journal} {\bibinfo  {journal} {Phys.
  Rev. D}\ }\textbf {\bibinfo {volume} {94}},\ \bibinfo {pages} {072007}
  (\bibinfo {year} {2016})},\ \Eprint {http://arxiv.org/abs/1607.07923}
  {arXiv:1607.07923 [hep-ex]} \BibitemShut {NoStop}%
\bibitem [{\citenamefont {Hirose}\ \emph {et~al.}(2017)\citenamefont {Hirose}
  \emph {et~al.}}]{Belle:2016dyj}%
  \BibitemOpen
  \bibfield  {author} {\bibinfo {author} {\bibfnamefont {S.}~\bibnamefont
  {Hirose}} \emph {et~al.} (\bibinfo {collaboration} {Belle}),\ }\href
  {\doibase 10.1103/PhysRevLett.118.211801} {\bibfield  {journal} {\bibinfo
  {journal} {Phys. Rev. Lett.}\ }\textbf {\bibinfo {volume} {118}},\ \bibinfo
  {pages} {211801} (\bibinfo {year} {2017})},\ \Eprint
  {http://arxiv.org/abs/1612.00529} {arXiv:1612.00529 [hep-ex]} \BibitemShut
  {NoStop}%
\bibitem [{\citenamefont {Caria}\ \emph {et~al.}(2020)\citenamefont {Caria}
  \emph {et~al.}}]{Belle:2019rba}%
  \BibitemOpen
  \bibfield  {author} {\bibinfo {author} {\bibfnamefont {G.}~\bibnamefont
  {Caria}} \emph {et~al.} (\bibinfo {collaboration} {Belle}),\ }\href {\doibase
  10.1103/PhysRevLett.124.161803} {\bibfield  {journal} {\bibinfo  {journal}
  {Phys. Rev. Lett.}\ }\textbf {\bibinfo {volume} {124}},\ \bibinfo {pages}
  {161803} (\bibinfo {year} {2020})},\ \Eprint
  {http://arxiv.org/abs/1910.05864} {arXiv:1910.05864 [hep-ex]} \BibitemShut
  {NoStop}%
\bibitem [{\citenamefont {Aaij}\ \emph {et~al.}(2015)\citenamefont {Aaij} \emph
  {et~al.}}]{LHCb:2015gmp}%
  \BibitemOpen
  \bibfield  {author} {\bibinfo {author} {\bibfnamefont {R.}~\bibnamefont
  {Aaij}} \emph {et~al.} (\bibinfo {collaboration} {LHCb}),\ }\href {\doibase
  10.1103/PhysRevLett.115.111803} {\bibfield  {journal} {\bibinfo  {journal}
  {Phys. Rev. Lett.}\ }\textbf {\bibinfo {volume} {115}},\ \bibinfo {pages}
  {111803} (\bibinfo {year} {2015})},\ \bibinfo {note} {[Erratum:
  Phys.Rev.Lett. 115, 159901 (2015)]},\ \Eprint
  {http://arxiv.org/abs/1506.08614} {arXiv:1506.08614 [hep-ex]} \BibitemShut
  {NoStop}%
\bibitem [{\citenamefont {Aaij}\ \emph
  {et~al.}(2018{\natexlab{a}})\citenamefont {Aaij} \emph
  {et~al.}}]{LHCb:2017smo}%
  \BibitemOpen
  \bibfield  {author} {\bibinfo {author} {\bibfnamefont {R.}~\bibnamefont
  {Aaij}} \emph {et~al.} (\bibinfo {collaboration} {LHCb}),\ }\href {\doibase
  10.1103/PhysRevLett.120.171802} {\bibfield  {journal} {\bibinfo  {journal}
  {Phys. Rev. Lett.}\ }\textbf {\bibinfo {volume} {120}},\ \bibinfo {pages}
  {171802} (\bibinfo {year} {2018}{\natexlab{a}})},\ \Eprint
  {http://arxiv.org/abs/1708.08856} {arXiv:1708.08856 [hep-ex]} \BibitemShut
  {NoStop}%
\bibitem [{\citenamefont {Aaij}\ \emph
  {et~al.}(2018{\natexlab{b}})\citenamefont {Aaij} \emph
  {et~al.}}]{LHCb:2017rln}%
  \BibitemOpen
  \bibfield  {author} {\bibinfo {author} {\bibfnamefont {R.}~\bibnamefont
  {Aaij}} \emph {et~al.} (\bibinfo {collaboration} {LHCb}),\ }\href {\doibase
  10.1103/PhysRevD.97.072013} {\bibfield  {journal} {\bibinfo  {journal} {Phys.
  Rev. D}\ }\textbf {\bibinfo {volume} {97}},\ \bibinfo {pages} {072013}
  (\bibinfo {year} {2018}{\natexlab{b}})},\ \Eprint
  {http://arxiv.org/abs/1711.02505} {arXiv:1711.02505 [hep-ex]} \BibitemShut
  {NoStop}%
\bibitem [{\citenamefont {Amhis}\ \emph {et~al.}(2021)\citenamefont {Amhis}
  \emph {et~al.}}]{HFLAV:2019otj}%
  \BibitemOpen
  \bibfield  {author} {\bibinfo {author} {\bibfnamefont {Y.~S.}\ \bibnamefont
  {Amhis}} \emph {et~al.} (\bibinfo {collaboration} {HFLAV}),\ }\href {\doibase
  10.1140/epjc/s10052-020-8156-7} {\bibfield  {journal} {\bibinfo  {journal}
  {Eur. Phys. J. C}\ }\textbf {\bibinfo {volume} {81}},\ \bibinfo {pages} {226}
  (\bibinfo {year} {2021})},\ \Eprint {http://arxiv.org/abs/1909.12524}
  {arXiv:1909.12524 [hep-ex]} \BibitemShut {NoStop}%
\bibitem [{\citenamefont {Aaij}\ \emph
  {et~al.}(2018{\natexlab{c}})\citenamefont {Aaij} \emph
  {et~al.}}]{LHCb:2017vlu}%
  \BibitemOpen
  \bibfield  {author} {\bibinfo {author} {\bibfnamefont {R.}~\bibnamefont
  {Aaij}} \emph {et~al.} (\bibinfo {collaboration} {LHCb}),\ }\href {\doibase
  10.1103/PhysRevLett.120.121801} {\bibfield  {journal} {\bibinfo  {journal}
  {Phys. Rev. Lett.}\ }\textbf {\bibinfo {volume} {120}},\ \bibinfo {pages}
  {121801} (\bibinfo {year} {2018}{\natexlab{c}})},\ \Eprint
  {http://arxiv.org/abs/1711.05623} {arXiv:1711.05623 [hep-ex]} \BibitemShut
  {NoStop}%
\bibitem [{\citenamefont {Anisimov}\ \emph {et~al.}(1999)\citenamefont
  {Anisimov}, \citenamefont {Narodetsky}, \citenamefont {Semay},\ and\
  \citenamefont {Silvestre-Brac}}]{Anisimov:1998uk}%
  \BibitemOpen
  \bibfield  {author} {\bibinfo {author} {\bibfnamefont {A.~Y.}\ \bibnamefont
  {Anisimov}}, \bibinfo {author} {\bibfnamefont {I.~M.}\ \bibnamefont
  {Narodetsky}}, \bibinfo {author} {\bibfnamefont {C.}~\bibnamefont {Semay}}, \
  and\ \bibinfo {author} {\bibfnamefont {B.}~\bibnamefont {Silvestre-Brac}},\
  }\href {\doibase 10.1016/S0370-2693(99)00273-7} {\bibfield  {journal}
  {\bibinfo  {journal} {Phys. Lett.}\ }\textbf {\bibinfo {volume} {B452}},\
  \bibinfo {pages} {129} (\bibinfo {year} {1999})},\ \Eprint
  {http://arxiv.org/abs/hep-ph/9812514} {arXiv:hep-ph/9812514 [hep-ph]}
  \BibitemShut {NoStop}%
\bibitem [{\citenamefont {Ivanov}\ \emph {et~al.}(2006)\citenamefont {Ivanov},
  \citenamefont {Korner},\ and\ \citenamefont {Santorelli}}]{Ivanov:2006ni}%
  \BibitemOpen
  \bibfield  {author} {\bibinfo {author} {\bibfnamefont {M.~A.}\ \bibnamefont
  {Ivanov}}, \bibinfo {author} {\bibfnamefont {J.~G.}\ \bibnamefont {Korner}},
  \ and\ \bibinfo {author} {\bibfnamefont {P.}~\bibnamefont {Santorelli}},\
  }\href {\doibase 10.1103/PhysRevD.73.054024} {\bibfield  {journal} {\bibinfo
  {journal} {Phys. Rev.}\ }\textbf {\bibinfo {volume} {D73}},\ \bibinfo {pages}
  {054024} (\bibinfo {year} {2006})},\ \Eprint
  {http://arxiv.org/abs/hep-ph/0602050} {arXiv:hep-ph/0602050 [hep-ph]}
  \BibitemShut {NoStop}%
\bibitem [{\citenamefont {Hern\'andez}\ \emph {et~al.}(2006)\citenamefont
  {Hern\'andez}, \citenamefont {Nieves},\ and\ \citenamefont
  {Verde-Velasco}}]{Hernandez:2006gt}%
  \BibitemOpen
  \bibfield  {author} {\bibinfo {author} {\bibfnamefont {E.}~\bibnamefont
  {Hern\'andez}}, \bibinfo {author} {\bibfnamefont {J.}~\bibnamefont {Nieves}},
  \ and\ \bibinfo {author} {\bibfnamefont {J.}~\bibnamefont {Verde-Velasco}},\
  }\href {\doibase 10.1103/PhysRevD.74.074008} {\bibfield  {journal} {\bibinfo
  {journal} {Phys. Rev. D}\ }\textbf {\bibinfo {volume} {74}},\ \bibinfo
  {pages} {074008} (\bibinfo {year} {2006})},\ \Eprint
  {http://arxiv.org/abs/hep-ph/0607150} {arXiv:hep-ph/0607150} \BibitemShut
  {NoStop}%
\bibitem [{\citenamefont {Huang}\ and\ \citenamefont
  {Zuo}(2007)}]{Huang:2007kb}%
  \BibitemOpen
  \bibfield  {author} {\bibinfo {author} {\bibfnamefont {T.}~\bibnamefont
  {Huang}}\ and\ \bibinfo {author} {\bibfnamefont {F.}~\bibnamefont {Zuo}},\
  }\href {\doibase 10.1140/epjc/s10052-007-0333-4} {\bibfield  {journal}
  {\bibinfo  {journal} {Eur. Phys. J.}\ }\textbf {\bibinfo {volume} {C51}},\
  \bibinfo {pages} {833} (\bibinfo {year} {2007})},\ \Eprint
  {http://arxiv.org/abs/hep-ph/0702147} {arXiv:hep-ph/0702147 [HEP-PH]}
  \BibitemShut {NoStop}%
\bibitem [{\citenamefont {Wang}\ \emph {et~al.}(2009)\citenamefont {Wang},
  \citenamefont {Shen},\ and\ \citenamefont {Lu}}]{Wang:2008xt}%
  \BibitemOpen
  \bibfield  {author} {\bibinfo {author} {\bibfnamefont {W.}~\bibnamefont
  {Wang}}, \bibinfo {author} {\bibfnamefont {Y.-L.}\ \bibnamefont {Shen}}, \
  and\ \bibinfo {author} {\bibfnamefont {C.-D.}\ \bibnamefont {Lu}},\ }\href
  {\doibase 10.1103/PhysRevD.79.054012} {\bibfield  {journal} {\bibinfo
  {journal} {Phys. Rev.}\ }\textbf {\bibinfo {volume} {D79}},\ \bibinfo {pages}
  {054012} (\bibinfo {year} {2009})},\ \Eprint {http://arxiv.org/abs/0811.3748}
  {arXiv:0811.3748 [hep-ph]} \BibitemShut {NoStop}%
\bibitem [{\citenamefont {Wang}\ \emph {et~al.}(2013)\citenamefont {Wang},
  \citenamefont {Fan},\ and\ \citenamefont {Xiao}}]{Wen-Fei:2013uea}%
  \BibitemOpen
  \bibfield  {author} {\bibinfo {author} {\bibfnamefont {W.-F.}\ \bibnamefont
  {Wang}}, \bibinfo {author} {\bibfnamefont {Y.-Y.}\ \bibnamefont {Fan}}, \
  and\ \bibinfo {author} {\bibfnamefont {Z.-J.}\ \bibnamefont {Xiao}},\ }\href
  {\doibase 10.1088/1674-1137/37/9/093102} {\bibfield  {journal} {\bibinfo
  {journal} {Chin. Phys.}\ }\textbf {\bibinfo {volume} {C37}},\ \bibinfo
  {pages} {093102} (\bibinfo {year} {2013})},\ \Eprint
  {http://arxiv.org/abs/1212.5903} {arXiv:1212.5903 [hep-ph]} \BibitemShut
  {NoStop}%
\bibitem [{\citenamefont {Watanabe}(2018)}]{Watanabe:2017mip}%
  \BibitemOpen
  \bibfield  {author} {\bibinfo {author} {\bibfnamefont {R.}~\bibnamefont
  {Watanabe}},\ }\href {\doibase 10.1016/j.physletb.2017.11.016} {\bibfield
  {journal} {\bibinfo  {journal} {Phys. Lett. B}\ }\textbf {\bibinfo {volume}
  {776}},\ \bibinfo {pages} {5} (\bibinfo {year} {2018})},\ \Eprint
  {http://arxiv.org/abs/1709.08644} {arXiv:1709.08644 [hep-ph]} \BibitemShut
  {NoStop}%
\bibitem [{\citenamefont {Issadykov}\ and\ \citenamefont
  {Ivanov}(2018)}]{Issadykov:2018myx}%
  \BibitemOpen
  \bibfield  {author} {\bibinfo {author} {\bibfnamefont {A.}~\bibnamefont
  {Issadykov}}\ and\ \bibinfo {author} {\bibfnamefont {M.~A.}\ \bibnamefont
  {Ivanov}},\ }\href {\doibase 10.1016/j.physletb.2018.06.056} {\bibfield
  {journal} {\bibinfo  {journal} {Phys. Lett.}\ }\textbf {\bibinfo {volume}
  {B783}},\ \bibinfo {pages} {178} (\bibinfo {year} {2018})},\ \Eprint
  {http://arxiv.org/abs/1804.00472} {arXiv:1804.00472 [hep-ph]} \BibitemShut
  {NoStop}%
\bibitem [{\citenamefont {Tran}\ \emph {et~al.}(2018)\citenamefont {Tran},
  \citenamefont {Ivanov}, \citenamefont {K{\"o}rner},\ and\ \citenamefont
  {Santorelli}}]{Tran:2018kuv}%
  \BibitemOpen
  \bibfield  {author} {\bibinfo {author} {\bibfnamefont {C.-T.}\ \bibnamefont
  {Tran}}, \bibinfo {author} {\bibfnamefont {M.~A.}\ \bibnamefont {Ivanov}},
  \bibinfo {author} {\bibfnamefont {J.~G.}\ \bibnamefont {K{\"o}rner}}, \ and\
  \bibinfo {author} {\bibfnamefont {P.}~\bibnamefont {Santorelli}},\ }\href
  {\doibase 10.1103/PhysRevD.97.054014} {\bibfield  {journal} {\bibinfo
  {journal} {Phys. Rev.}\ }\textbf {\bibinfo {volume} {D97}},\ \bibinfo {pages}
  {054014} (\bibinfo {year} {2018})},\ \Eprint
  {http://arxiv.org/abs/1801.06927} {arXiv:1801.06927 [hep-ph]} \BibitemShut
  {NoStop}%
\bibitem [{\citenamefont {Hu}\ \emph {et~al.}(2020)\citenamefont {Hu},
  \citenamefont {Jin},\ and\ \citenamefont {Xiao}}]{Hu:2019qcn}%
  \BibitemOpen
  \bibfield  {author} {\bibinfo {author} {\bibfnamefont {X.-Q.}\ \bibnamefont
  {Hu}}, \bibinfo {author} {\bibfnamefont {S.-P.}\ \bibnamefont {Jin}}, \ and\
  \bibinfo {author} {\bibfnamefont {Z.-J.}\ \bibnamefont {Xiao}},\ }\href
  {\doibase 10.1088/1674-1137/44/2/023104} {\bibfield  {journal} {\bibinfo
  {journal} {Chin. Phys.}\ }\textbf {\bibinfo {volume} {C44}},\ \bibinfo
  {pages} {023104} (\bibinfo {year} {2020})},\ \Eprint
  {http://arxiv.org/abs/1904.07530} {arXiv:1904.07530 [hep-ph]} \BibitemShut
  {NoStop}%
\bibitem [{\citenamefont {Leljak}\ \emph {et~al.}(2019)\citenamefont {Leljak},
  \citenamefont {Melic},\ and\ \citenamefont {Patra}}]{Leljak:2019eyw}%
  \BibitemOpen
  \bibfield  {author} {\bibinfo {author} {\bibfnamefont {D.}~\bibnamefont
  {Leljak}}, \bibinfo {author} {\bibfnamefont {B.}~\bibnamefont {Melic}}, \
  and\ \bibinfo {author} {\bibfnamefont {M.}~\bibnamefont {Patra}},\ }\href
  {\doibase 10.1007/JHEP05(2019)094} {\bibfield  {journal} {\bibinfo  {journal}
  {JHEP}\ }\textbf {\bibinfo {volume} {05}},\ \bibinfo {pages} {094} (\bibinfo
  {year} {2019})},\ \Eprint {http://arxiv.org/abs/1901.08368} {arXiv:1901.08368
  [hep-ph]} \BibitemShut {NoStop}%
\bibitem [{\citenamefont {Azizi}\ \emph {et~al.}(2019)\citenamefont {Azizi},
  \citenamefont {Sarac},\ and\ \citenamefont {Sundu}}]{Azizi:2019aaf}%
  \BibitemOpen
  \bibfield  {author} {\bibinfo {author} {\bibfnamefont {K.}~\bibnamefont
  {Azizi}}, \bibinfo {author} {\bibfnamefont {Y.}~\bibnamefont {Sarac}}, \ and\
  \bibinfo {author} {\bibfnamefont {H.}~\bibnamefont {Sundu}},\ }\href
  {\doibase 10.1103/PhysRevD.99.113004} {\bibfield  {journal} {\bibinfo
  {journal} {Phys. Rev.}\ }\textbf {\bibinfo {volume} {D99}},\ \bibinfo {pages}
  {113004} (\bibinfo {year} {2019})},\ \Eprint
  {http://arxiv.org/abs/1904.08267} {arXiv:1904.08267 [hep-ph]} \BibitemShut
  {NoStop}%
\bibitem [{\citenamefont {Wang}\ and\ \citenamefont
  {Zhu}(2019)}]{Wang:2018duy}%
  \BibitemOpen
  \bibfield  {author} {\bibinfo {author} {\bibfnamefont {W.}~\bibnamefont
  {Wang}}\ and\ \bibinfo {author} {\bibfnamefont {R.}~\bibnamefont {Zhu}},\
  }\href {\doibase 10.1142/S0217751X19501951} {\bibfield  {journal} {\bibinfo
  {journal} {Int. J. Mod. Phys. A}\ }\textbf {\bibinfo {volume} {34}},\
  \bibinfo {pages} {1950195} (\bibinfo {year} {2019})},\ \Eprint
  {http://arxiv.org/abs/1808.10830} {arXiv:1808.10830 [hep-ph]} \BibitemShut
  {NoStop}%
\bibitem [{\citenamefont {Fajfer}\ \emph {et~al.}(2012)\citenamefont {Fajfer},
  \citenamefont {Kamenik},\ and\ \citenamefont {Nisandzic}}]{Fajfer:2012vx}%
  \BibitemOpen
  \bibfield  {author} {\bibinfo {author} {\bibfnamefont {S.}~\bibnamefont
  {Fajfer}}, \bibinfo {author} {\bibfnamefont {J.~F.}\ \bibnamefont {Kamenik}},
  \ and\ \bibinfo {author} {\bibfnamefont {I.}~\bibnamefont {Nisandzic}},\
  }\href {\doibase 10.1103/PhysRevD.85.094025} {\bibfield  {journal} {\bibinfo
  {journal} {Phys. Rev.}\ }\textbf {\bibinfo {volume} {D85}},\ \bibinfo {pages}
  {094025} (\bibinfo {year} {2012})},\ \Eprint {http://arxiv.org/abs/1203.2654}
  {arXiv:1203.2654 [hep-ph]} \BibitemShut {NoStop}%
\bibitem [{\citenamefont {Abdesselam}\ \emph {et~al.}(2019)\citenamefont
  {Abdesselam} \emph {et~al.}}]{Belle:2019ewo}%
  \BibitemOpen
  \bibfield  {author} {\bibinfo {author} {\bibfnamefont {A.}~\bibnamefont
  {Abdesselam}} \emph {et~al.} (\bibinfo {collaboration} {Belle}),\ }in\
  \href@noop {} {\emph {\bibinfo {booktitle} {{10th International Workshop on
  the CKM Unitarity Triangle}}}}\ (\bibinfo {year} {2019})\ \Eprint
  {http://arxiv.org/abs/1903.03102} {arXiv:1903.03102 [hep-ex]} \BibitemShut
  {NoStop}%
\bibitem [{\citenamefont {Alonso}\ \emph
  {et~al.}(2017{\natexlab{a}})\citenamefont {Alonso}, \citenamefont
  {Grinstein},\ and\ \citenamefont {Martin~Camalich}}]{Alonso:2016oyd}%
  \BibitemOpen
  \bibfield  {author} {\bibinfo {author} {\bibfnamefont {R.}~\bibnamefont
  {Alonso}}, \bibinfo {author} {\bibfnamefont {B.}~\bibnamefont {Grinstein}}, \
  and\ \bibinfo {author} {\bibfnamefont {J.}~\bibnamefont {Martin~Camalich}},\
  }\href {\doibase 10.1103/PhysRevLett.118.081802} {\bibfield  {journal}
  {\bibinfo  {journal} {Phys. Rev. Lett.}\ }\textbf {\bibinfo {volume} {118}},\
  \bibinfo {pages} {081802} (\bibinfo {year} {2017}{\natexlab{a}})},\ \Eprint
  {http://arxiv.org/abs/1611.06676} {arXiv:1611.06676 [hep-ph]} \BibitemShut
  {NoStop}%
\bibitem [{\citenamefont {Nierste}\ \emph {et~al.}(2008)\citenamefont
  {Nierste}, \citenamefont {Trine},\ and\ \citenamefont
  {Westhoff}}]{Nierste:2008qe}%
  \BibitemOpen
  \bibfield  {author} {\bibinfo {author} {\bibfnamefont {U.}~\bibnamefont
  {Nierste}}, \bibinfo {author} {\bibfnamefont {S.}~\bibnamefont {Trine}}, \
  and\ \bibinfo {author} {\bibfnamefont {S.}~\bibnamefont {Westhoff}},\ }\href
  {\doibase 10.1103/PhysRevD.78.015006} {\bibfield  {journal} {\bibinfo
  {journal} {Phys. Rev. D}\ }\textbf {\bibinfo {volume} {78}},\ \bibinfo
  {pages} {015006} (\bibinfo {year} {2008})},\ \Eprint
  {http://arxiv.org/abs/0801.4938} {arXiv:0801.4938 [hep-ph]} \BibitemShut
  {NoStop}%
\bibitem [{\citenamefont {Tanaka}\ and\ \citenamefont
  {Watanabe}(2013)}]{Tanaka:2012nw}%
  \BibitemOpen
  \bibfield  {author} {\bibinfo {author} {\bibfnamefont {M.}~\bibnamefont
  {Tanaka}}\ and\ \bibinfo {author} {\bibfnamefont {R.}~\bibnamefont
  {Watanabe}},\ }\href {\doibase 10.1103/PhysRevD.87.034028} {\bibfield
  {journal} {\bibinfo  {journal} {Phys. Rev. D}\ }\textbf {\bibinfo {volume}
  {87}},\ \bibinfo {pages} {034028} (\bibinfo {year} {2013})},\ \Eprint
  {http://arxiv.org/abs/1212.1878} {arXiv:1212.1878 [hep-ph]} \BibitemShut
  {NoStop}%
\bibitem [{\citenamefont {Duraisamy}\ and\ \citenamefont
  {Datta}(2013)}]{Duraisamy:2013pia}%
  \BibitemOpen
  \bibfield  {author} {\bibinfo {author} {\bibfnamefont {M.}~\bibnamefont
  {Duraisamy}}\ and\ \bibinfo {author} {\bibfnamefont {A.}~\bibnamefont
  {Datta}},\ }\href {\doibase 10.1007/JHEP09(2013)059} {\bibfield  {journal}
  {\bibinfo  {journal} {JHEP}\ }\textbf {\bibinfo {volume} {09}},\ \bibinfo
  {pages} {059} (\bibinfo {year} {2013})},\ \Eprint
  {http://arxiv.org/abs/1302.7031} {arXiv:1302.7031 [hep-ph]} \BibitemShut
  {NoStop}%
\bibitem [{\citenamefont {Duraisamy}\ \emph {et~al.}(2014)\citenamefont
  {Duraisamy}, \citenamefont {Sharma},\ and\ \citenamefont
  {Datta}}]{Duraisamy:2014sna}%
  \BibitemOpen
  \bibfield  {author} {\bibinfo {author} {\bibfnamefont {M.}~\bibnamefont
  {Duraisamy}}, \bibinfo {author} {\bibfnamefont {P.}~\bibnamefont {Sharma}}, \
  and\ \bibinfo {author} {\bibfnamefont {A.}~\bibnamefont {Datta}},\ }\href
  {\doibase 10.1103/PhysRevD.90.074013} {\bibfield  {journal} {\bibinfo
  {journal} {Phys. Rev. D}\ }\textbf {\bibinfo {volume} {90}},\ \bibinfo
  {pages} {074013} (\bibinfo {year} {2014})},\ \Eprint
  {http://arxiv.org/abs/1405.3719} {arXiv:1405.3719 [hep-ph]} \BibitemShut
  {NoStop}%
\bibitem [{\citenamefont {Becirevic}\ \emph {et~al.}(2019)\citenamefont
  {Becirevic}, \citenamefont {Fajfer}, \citenamefont {Nisandzic},\ and\
  \citenamefont {Tayduganov}}]{Becirevic:2016hea}%
  \BibitemOpen
  \bibfield  {author} {\bibinfo {author} {\bibfnamefont {D.}~\bibnamefont
  {Becirevic}}, \bibinfo {author} {\bibfnamefont {S.}~\bibnamefont {Fajfer}},
  \bibinfo {author} {\bibfnamefont {I.}~\bibnamefont {Nisandzic}}, \ and\
  \bibinfo {author} {\bibfnamefont {A.}~\bibnamefont {Tayduganov}},\ }\href
  {\doibase 10.1016/j.nuclphysb.2019.114707} {\bibfield  {journal} {\bibinfo
  {journal} {Nucl. Phys. B}\ }\textbf {\bibinfo {volume} {946}},\ \bibinfo
  {pages} {114707} (\bibinfo {year} {2019})},\ \Eprint
  {http://arxiv.org/abs/1602.03030} {arXiv:1602.03030 [hep-ph]} \BibitemShut
  {NoStop}%
\bibitem [{\citenamefont {Ligeti}\ \emph {et~al.}(2017)\citenamefont {Ligeti},
  \citenamefont {Papucci},\ and\ \citenamefont {Robinson}}]{Ligeti:2016npd}%
  \BibitemOpen
  \bibfield  {author} {\bibinfo {author} {\bibfnamefont {Z.}~\bibnamefont
  {Ligeti}}, \bibinfo {author} {\bibfnamefont {M.}~\bibnamefont {Papucci}}, \
  and\ \bibinfo {author} {\bibfnamefont {D.~J.}\ \bibnamefont {Robinson}},\
  }\href {\doibase 10.1007/JHEP01(2017)083} {\bibfield  {journal} {\bibinfo
  {journal} {JHEP}\ }\textbf {\bibinfo {volume} {01}},\ \bibinfo {pages} {083}
  (\bibinfo {year} {2017})},\ \Eprint {http://arxiv.org/abs/1610.02045}
  {arXiv:1610.02045 [hep-ph]} \BibitemShut {NoStop}%
\bibitem [{\citenamefont {Ivanov}\ \emph {et~al.}(2017)\citenamefont {Ivanov},
  \citenamefont {K\"orner},\ and\ \citenamefont {Tran}}]{Ivanov:2017mrj}%
  \BibitemOpen
  \bibfield  {author} {\bibinfo {author} {\bibfnamefont {M.~A.}\ \bibnamefont
  {Ivanov}}, \bibinfo {author} {\bibfnamefont {J.~G.}\ \bibnamefont
  {K\"orner}}, \ and\ \bibinfo {author} {\bibfnamefont {C.-T.}\ \bibnamefont
  {Tran}},\ }\href {\doibase 10.1103/PhysRevD.95.036021} {\bibfield  {journal}
  {\bibinfo  {journal} {Phys. Rev. D}\ }\textbf {\bibinfo {volume} {95}},\
  \bibinfo {pages} {036021} (\bibinfo {year} {2017})},\ \Eprint
  {http://arxiv.org/abs/1701.02937} {arXiv:1701.02937 [hep-ph]} \BibitemShut
  {NoStop}%
\bibitem [{\citenamefont {Bernlochner}\ \emph {et~al.}(2017)\citenamefont
  {Bernlochner}, \citenamefont {Ligeti}, \citenamefont {Papucci},\ and\
  \citenamefont {Robinson}}]{Bernlochner:2017jka}%
  \BibitemOpen
  \bibfield  {author} {\bibinfo {author} {\bibfnamefont {F.~U.}\ \bibnamefont
  {Bernlochner}}, \bibinfo {author} {\bibfnamefont {Z.}~\bibnamefont {Ligeti}},
  \bibinfo {author} {\bibfnamefont {M.}~\bibnamefont {Papucci}}, \ and\
  \bibinfo {author} {\bibfnamefont {D.~J.}\ \bibnamefont {Robinson}},\ }\href
  {\doibase 10.1103/PhysRevD.95.115008, 10.1103/PhysRevD.97.059902} {\bibfield
  {journal} {\bibinfo  {journal} {Phys. Rev.}\ }\textbf {\bibinfo {volume}
  {D95}},\ \bibinfo {pages} {115008} (\bibinfo {year} {2017})},\ \bibinfo
  {note} {[erratum: Phys. Rev.D97,no.5,059902(2018)]},\ \Eprint
  {http://arxiv.org/abs/1703.05330} {arXiv:1703.05330 [hep-ph]} \BibitemShut
  {NoStop}%
\bibitem [{\citenamefont {Blanke}\ \emph
  {et~al.}(2019{\natexlab{a}})\citenamefont {Blanke}, \citenamefont
  {Crivellin}, \citenamefont {de~Boer}, \citenamefont {Kitahara}, \citenamefont
  {Moscati}, \citenamefont {Nierste},\ and\ \citenamefont
  {Ni\v{s}and\v{z}i\'c}}]{Blanke:2018yud}%
  \BibitemOpen
  \bibfield  {author} {\bibinfo {author} {\bibfnamefont {M.}~\bibnamefont
  {Blanke}}, \bibinfo {author} {\bibfnamefont {A.}~\bibnamefont {Crivellin}},
  \bibinfo {author} {\bibfnamefont {S.}~\bibnamefont {de~Boer}}, \bibinfo
  {author} {\bibfnamefont {T.}~\bibnamefont {Kitahara}}, \bibinfo {author}
  {\bibfnamefont {M.}~\bibnamefont {Moscati}}, \bibinfo {author} {\bibfnamefont
  {U.}~\bibnamefont {Nierste}}, \ and\ \bibinfo {author} {\bibfnamefont
  {I.}~\bibnamefont {Ni\v{s}and\v{z}i\'c}},\ }\href {\doibase
  10.1103/PhysRevD.99.075006} {\bibfield  {journal} {\bibinfo  {journal} {Phys.
  Rev.}\ }\textbf {\bibinfo {volume} {D99}},\ \bibinfo {pages} {075006}
  (\bibinfo {year} {2019}{\natexlab{a}})},\ \Eprint
  {http://arxiv.org/abs/1811.09603} {arXiv:1811.09603 [hep-ph]} \BibitemShut
  {NoStop}%
\bibitem [{\citenamefont {Bhattacharya}\ \emph {et~al.}(2019)\citenamefont
  {Bhattacharya}, \citenamefont {Nandi},\ and\ \citenamefont
  {Kumar~Patra}}]{Bhattacharya:2018kig}%
  \BibitemOpen
  \bibfield  {author} {\bibinfo {author} {\bibfnamefont {S.}~\bibnamefont
  {Bhattacharya}}, \bibinfo {author} {\bibfnamefont {S.}~\bibnamefont {Nandi}},
  \ and\ \bibinfo {author} {\bibfnamefont {S.}~\bibnamefont {Kumar~Patra}},\
  }\href {\doibase 10.1140/epjc/s10052-019-6767-7} {\bibfield  {journal}
  {\bibinfo  {journal} {Eur. Phys. J. C}\ }\textbf {\bibinfo {volume} {79}},\
  \bibinfo {pages} {268} (\bibinfo {year} {2019})},\ \Eprint
  {http://arxiv.org/abs/1805.08222} {arXiv:1805.08222 [hep-ph]} \BibitemShut
  {NoStop}%
\bibitem [{\citenamefont {Colangelo}\ and\ \citenamefont
  {De~Fazio}(2018)}]{Colangelo:2018cnj}%
  \BibitemOpen
  \bibfield  {author} {\bibinfo {author} {\bibfnamefont {P.}~\bibnamefont
  {Colangelo}}\ and\ \bibinfo {author} {\bibfnamefont {F.}~\bibnamefont
  {De~Fazio}},\ }\href {\doibase 10.1007/JHEP06(2018)082} {\bibfield  {journal}
  {\bibinfo  {journal} {JHEP}\ }\textbf {\bibinfo {volume} {06}},\ \bibinfo
  {pages} {082} (\bibinfo {year} {2018})},\ \Eprint
  {http://arxiv.org/abs/1801.10468} {arXiv:1801.10468 [hep-ph]} \BibitemShut
  {NoStop}%
\bibitem [{\citenamefont {Murgui}\ \emph {et~al.}(2019)\citenamefont {Murgui},
  \citenamefont {PeÃ±uelas}, \citenamefont {Jung},\ and\ \citenamefont
  {Pich}}]{Murgui:2019czp}%
  \BibitemOpen
  \bibfield  {author} {\bibinfo {author} {\bibfnamefont {C.}~\bibnamefont
  {Murgui}}, \bibinfo {author} {\bibfnamefont {A.}~\bibnamefont {PeÃ±uelas}},
  \bibinfo {author} {\bibfnamefont {M.}~\bibnamefont {Jung}}, \ and\ \bibinfo
  {author} {\bibfnamefont {A.}~\bibnamefont {Pich}},\ }\href {\doibase
  10.1007/JHEP09(2019)103} {\bibfield  {journal} {\bibinfo  {journal} {JHEP}\
  }\textbf {\bibinfo {volume} {09}},\ \bibinfo {pages} {103} (\bibinfo {year}
  {2019})},\ \Eprint {http://arxiv.org/abs/1904.09311} {arXiv:1904.09311
  [hep-ph]} \BibitemShut {NoStop}%
\bibitem [{\citenamefont {Shi}\ \emph {et~al.}(2019)\citenamefont {Shi},
  \citenamefont {Geng}, \citenamefont {Grinstein}, \citenamefont {J{\"a}ger},\
  and\ \citenamefont {Martin~Camalich}}]{Shi:2019gxi}%
  \BibitemOpen
  \bibfield  {author} {\bibinfo {author} {\bibfnamefont {R.-X.}\ \bibnamefont
  {Shi}}, \bibinfo {author} {\bibfnamefont {L.-S.}\ \bibnamefont {Geng}},
  \bibinfo {author} {\bibfnamefont {B.}~\bibnamefont {Grinstein}}, \bibinfo
  {author} {\bibfnamefont {S.}~\bibnamefont {J{\"a}ger}}, \ and\ \bibinfo
  {author} {\bibfnamefont {J.}~\bibnamefont {Martin~Camalich}},\ }\href
  {\doibase 10.1007/JHEP12(2019)065} {\bibfield  {journal} {\bibinfo  {journal}
  {JHEP}\ }\textbf {\bibinfo {volume} {12}},\ \bibinfo {pages} {065} (\bibinfo
  {year} {2019})},\ \Eprint {http://arxiv.org/abs/1905.08498} {arXiv:1905.08498
  [hep-ph]} \BibitemShut {NoStop}%
\bibitem [{\citenamefont {Alok}\ \emph {et~al.}(2020)\citenamefont {Alok},
  \citenamefont {Kumar}, \citenamefont {Kumbhakar},\ and\ \citenamefont
  {Uma~Sankar}}]{Alok:2019uqc}%
  \BibitemOpen
  \bibfield  {author} {\bibinfo {author} {\bibfnamefont {A.~K.}\ \bibnamefont
  {Alok}}, \bibinfo {author} {\bibfnamefont {D.}~\bibnamefont {Kumar}},
  \bibinfo {author} {\bibfnamefont {S.}~\bibnamefont {Kumbhakar}}, \ and\
  \bibinfo {author} {\bibfnamefont {S.}~\bibnamefont {Uma~Sankar}},\ }\href
  {\doibase 10.1016/j.nuclphysb.2020.114957} {\bibfield  {journal} {\bibinfo
  {journal} {Nucl. Phys. B}\ }\textbf {\bibinfo {volume} {953}},\ \bibinfo
  {pages} {114957} (\bibinfo {year} {2020})},\ \Eprint
  {http://arxiv.org/abs/1903.10486} {arXiv:1903.10486 [hep-ph]} \BibitemShut
  {NoStop}%
\bibitem [{\citenamefont {Mandal}\ \emph {et~al.}(2020)\citenamefont {Mandal},
  \citenamefont {Murgui}, \citenamefont {Pe\~nuelas},\ and\ \citenamefont
  {Pich}}]{Mandal:2020htr}%
  \BibitemOpen
  \bibfield  {author} {\bibinfo {author} {\bibfnamefont {R.}~\bibnamefont
  {Mandal}}, \bibinfo {author} {\bibfnamefont {C.}~\bibnamefont {Murgui}},
  \bibinfo {author} {\bibfnamefont {A.}~\bibnamefont {Pe\~nuelas}}, \ and\
  \bibinfo {author} {\bibfnamefont {A.}~\bibnamefont {Pich}},\ }\href {\doibase
  10.1007/JHEP08(2020)022} {\bibfield  {journal} {\bibinfo  {journal} {JHEP}\
  }\textbf {\bibinfo {volume} {08}},\ \bibinfo {pages} {022} (\bibinfo {year}
  {2020})},\ \Eprint {http://arxiv.org/abs/2004.06726} {arXiv:2004.06726
  [hep-ph]} \BibitemShut {NoStop}%
\bibitem [{\citenamefont {Kumbhakar}(2021)}]{Kumbhakar:2020jdz}%
  \BibitemOpen
  \bibfield  {author} {\bibinfo {author} {\bibfnamefont {S.}~\bibnamefont
  {Kumbhakar}},\ }\href {\doibase 10.1016/j.nuclphysb.2020.115297} {\bibfield
  {journal} {\bibinfo  {journal} {Nucl. Phys. B}\ }\textbf {\bibinfo {volume}
  {963}},\ \bibinfo {pages} {115297} (\bibinfo {year} {2021})},\ \Eprint
  {http://arxiv.org/abs/2007.08132} {arXiv:2007.08132 [hep-ph]} \BibitemShut
  {NoStop}%
\bibitem [{\citenamefont {Iguro}\ and\ \citenamefont
  {Watanabe}(2020)}]{Iguro:2020cpg}%
  \BibitemOpen
  \bibfield  {author} {\bibinfo {author} {\bibfnamefont {S.}~\bibnamefont
  {Iguro}}\ and\ \bibinfo {author} {\bibfnamefont {R.}~\bibnamefont
  {Watanabe}},\ }\href {\doibase 10.1007/JHEP08(2020)006} {\bibfield  {journal}
  {\bibinfo  {journal} {JHEP}\ }\textbf {\bibinfo {volume} {08}},\ \bibinfo
  {pages} {006} (\bibinfo {year} {2020})},\ \Eprint
  {http://arxiv.org/abs/2004.10208} {arXiv:2004.10208 [hep-ph]} \BibitemShut
  {NoStop}%
\bibitem [{\citenamefont {Bhattacharya}\ \emph {et~al.}(2020)\citenamefont
  {Bhattacharya}, \citenamefont {Datta}, \citenamefont {Kamali},\ and\
  \citenamefont {London}}]{Bhattacharya:2020lfm}%
  \BibitemOpen
  \bibfield  {author} {\bibinfo {author} {\bibfnamefont {B.}~\bibnamefont
  {Bhattacharya}}, \bibinfo {author} {\bibfnamefont {A.}~\bibnamefont {Datta}},
  \bibinfo {author} {\bibfnamefont {S.}~\bibnamefont {Kamali}}, \ and\ \bibinfo
  {author} {\bibfnamefont {D.}~\bibnamefont {London}},\ }\href {\doibase
  10.1007/JHEP07(2020)194} {\bibfield  {journal} {\bibinfo  {journal} {JHEP}\
  }\textbf {\bibinfo {volume} {07}},\ \bibinfo {pages} {194} (\bibinfo {year}
  {2020})},\ \Eprint {http://arxiv.org/abs/2005.03032} {arXiv:2005.03032
  [hep-ph]} \BibitemShut {NoStop}%
\bibitem [{\citenamefont {Penalva}\ \emph
  {et~al.}(2021{\natexlab{a}})\citenamefont {Penalva}, \citenamefont
  {Hern\'andez},\ and\ \citenamefont {Nieves}}]{Penalva:2021gef}%
  \BibitemOpen
  \bibfield  {author} {\bibinfo {author} {\bibfnamefont {N.}~\bibnamefont
  {Penalva}}, \bibinfo {author} {\bibfnamefont {E.}~\bibnamefont
  {Hern\'andez}}, \ and\ \bibinfo {author} {\bibfnamefont {J.}~\bibnamefont
  {Nieves}},\ }\href {\doibase 10.1007/JHEP06(2021)118} {\bibfield  {journal}
  {\bibinfo  {journal} {JHEP}\ }\textbf {\bibinfo {volume} {06}},\ \bibinfo
  {pages} {118} (\bibinfo {year} {2021}{\natexlab{a}})},\ \Eprint
  {http://arxiv.org/abs/2103.01857} {arXiv:2103.01857 [hep-ph]} \BibitemShut
  {NoStop}%
\bibitem [{\citenamefont {Penalva}\ \emph
  {et~al.}(2020{\natexlab{a}})\citenamefont {Penalva}, \citenamefont
  {Hern\'andez},\ and\ \citenamefont {Nieves}}]{Penalva:2020ftd}%
  \BibitemOpen
  \bibfield  {author} {\bibinfo {author} {\bibfnamefont {N.}~\bibnamefont
  {Penalva}}, \bibinfo {author} {\bibfnamefont {E.}~\bibnamefont
  {Hern\'andez}}, \ and\ \bibinfo {author} {\bibfnamefont {J.}~\bibnamefont
  {Nieves}},\ }\href {\doibase 10.1103/PhysRevD.102.096016} {\bibfield
  {journal} {\bibinfo  {journal} {Phys. Rev. D}\ }\textbf {\bibinfo {volume}
  {102}},\ \bibinfo {pages} {096016} (\bibinfo {year} {2020}{\natexlab{a}})},\
  \Eprint {http://arxiv.org/abs/2007.12590} {arXiv:2007.12590 [hep-ph]}
  \BibitemShut {NoStop}%
\bibitem [{\citenamefont {Dutta}\ and\ \citenamefont
  {Bhol}(2017)}]{Dutta:2017xmj}%
  \BibitemOpen
  \bibfield  {author} {\bibinfo {author} {\bibfnamefont {R.}~\bibnamefont
  {Dutta}}\ and\ \bibinfo {author} {\bibfnamefont {A.}~\bibnamefont {Bhol}},\
  }\href {\doibase 10.1103/PhysRevD.96.076001} {\bibfield  {journal} {\bibinfo
  {journal} {Phys. Rev. D}\ }\textbf {\bibinfo {volume} {96}},\ \bibinfo
  {pages} {076001} (\bibinfo {year} {2017})},\ \Eprint
  {http://arxiv.org/abs/1701.08598} {arXiv:1701.08598 [hep-ph]} \BibitemShut
  {NoStop}%
\bibitem [{\citenamefont {Harrison}\ \emph {et~al.}(2020)\citenamefont
  {Harrison}, \citenamefont {Davies},\ and\ \citenamefont
  {Lytle}}]{Harrison:2020nrv}%
  \BibitemOpen
  \bibfield  {author} {\bibinfo {author} {\bibfnamefont {J.}~\bibnamefont
  {Harrison}}, \bibinfo {author} {\bibfnamefont {C.~T.}\ \bibnamefont
  {Davies}}, \ and\ \bibinfo {author} {\bibfnamefont {A.}~\bibnamefont {Lytle}}
  (\bibinfo {collaboration} {LATTICE-HPQCD}),\ }\href {\doibase
  10.1103/PhysRevLett.125.222003} {\bibfield  {journal} {\bibinfo  {journal}
  {Phys. Rev. Lett.}\ }\textbf {\bibinfo {volume} {125}},\ \bibinfo {pages}
  {222003} (\bibinfo {year} {2020})},\ \Eprint
  {http://arxiv.org/abs/2007.06956} {arXiv:2007.06956 [hep-lat]} \BibitemShut
  {NoStop}%
\bibitem [{\citenamefont {Dutta}(2016)}]{Dutta:2015ueb}%
  \BibitemOpen
  \bibfield  {author} {\bibinfo {author} {\bibfnamefont {R.}~\bibnamefont
  {Dutta}},\ }\href {\doibase 10.1103/PhysRevD.93.054003} {\bibfield  {journal}
  {\bibinfo  {journal} {Phys. Rev. D}\ }\textbf {\bibinfo {volume} {93}},\
  \bibinfo {pages} {054003} (\bibinfo {year} {2016})},\ \Eprint
  {http://arxiv.org/abs/1512.04034} {arXiv:1512.04034 [hep-ph]} \BibitemShut
  {NoStop}%
\bibitem [{\citenamefont {Shivashankara}\ \emph {et~al.}(2015)\citenamefont
  {Shivashankara}, \citenamefont {Wu},\ and\ \citenamefont
  {Datta}}]{Shivashankara:2015cta}%
  \BibitemOpen
  \bibfield  {author} {\bibinfo {author} {\bibfnamefont {S.}~\bibnamefont
  {Shivashankara}}, \bibinfo {author} {\bibfnamefont {W.}~\bibnamefont {Wu}}, \
  and\ \bibinfo {author} {\bibfnamefont {A.}~\bibnamefont {Datta}},\ }\href
  {\doibase 10.1103/PhysRevD.91.115003} {\bibfield  {journal} {\bibinfo
  {journal} {Phys. Rev. D}\ }\textbf {\bibinfo {volume} {91}},\ \bibinfo
  {pages} {115003} (\bibinfo {year} {2015})},\ \Eprint
  {http://arxiv.org/abs/1502.07230} {arXiv:1502.07230 [hep-ph]} \BibitemShut
  {NoStop}%
\bibitem [{\citenamefont {Li}\ \emph {et~al.}(2017)\citenamefont {Li},
  \citenamefont {Yang},\ and\ \citenamefont {Zhang}}]{Li:2016pdv}%
  \BibitemOpen
  \bibfield  {author} {\bibinfo {author} {\bibfnamefont {X.-Q.}\ \bibnamefont
  {Li}}, \bibinfo {author} {\bibfnamefont {Y.-D.}\ \bibnamefont {Yang}}, \ and\
  \bibinfo {author} {\bibfnamefont {X.}~\bibnamefont {Zhang}},\ }\href
  {\doibase 10.1007/JHEP02(2017)068} {\bibfield  {journal} {\bibinfo  {journal}
  {JHEP}\ }\textbf {\bibinfo {volume} {02}},\ \bibinfo {pages} {068} (\bibinfo
  {year} {2017})},\ \Eprint {http://arxiv.org/abs/1611.01635} {arXiv:1611.01635
  [hep-ph]} \BibitemShut {NoStop}%
\bibitem [{\citenamefont {Datta}\ \emph {et~al.}(2017)\citenamefont {Datta},
  \citenamefont {Kamali}, \citenamefont {Meinel},\ and\ \citenamefont
  {Rashed}}]{Datta:2017aue}%
  \BibitemOpen
  \bibfield  {author} {\bibinfo {author} {\bibfnamefont {A.}~\bibnamefont
  {Datta}}, \bibinfo {author} {\bibfnamefont {S.}~\bibnamefont {Kamali}},
  \bibinfo {author} {\bibfnamefont {S.}~\bibnamefont {Meinel}}, \ and\ \bibinfo
  {author} {\bibfnamefont {A.}~\bibnamefont {Rashed}},\ }\href {\doibase
  10.1007/JHEP08(2017)131} {\bibfield  {journal} {\bibinfo  {journal} {JHEP}\
  }\textbf {\bibinfo {volume} {08}},\ \bibinfo {pages} {131} (\bibinfo {year}
  {2017})},\ \Eprint {http://arxiv.org/abs/1702.02243} {arXiv:1702.02243
  [hep-ph]} \BibitemShut {NoStop}%
\bibitem [{\citenamefont {Ray}\ \emph {et~al.}(2019)\citenamefont {Ray},
  \citenamefont {Sahoo},\ and\ \citenamefont {Mohanta}}]{Ray:2018hrx}%
  \BibitemOpen
  \bibfield  {author} {\bibinfo {author} {\bibfnamefont {A.}~\bibnamefont
  {Ray}}, \bibinfo {author} {\bibfnamefont {S.}~\bibnamefont {Sahoo}}, \ and\
  \bibinfo {author} {\bibfnamefont {R.}~\bibnamefont {Mohanta}},\ }\href
  {\doibase 10.1103/PhysRevD.99.015015} {\bibfield  {journal} {\bibinfo
  {journal} {Phys. Rev.}\ }\textbf {\bibinfo {volume} {D99}},\ \bibinfo {pages}
  {015015} (\bibinfo {year} {2019})},\ \Eprint
  {http://arxiv.org/abs/1812.08314} {arXiv:1812.08314 [hep-ph]} \BibitemShut
  {NoStop}%
\bibitem [{\citenamefont {Bernlochner}\ \emph {et~al.}(2019)\citenamefont
  {Bernlochner}, \citenamefont {Ligeti}, \citenamefont {Robinson},\ and\
  \citenamefont {Sutcliffe}}]{Bernlochner:2018bfn}%
  \BibitemOpen
  \bibfield  {author} {\bibinfo {author} {\bibfnamefont {F.~U.}\ \bibnamefont
  {Bernlochner}}, \bibinfo {author} {\bibfnamefont {Z.}~\bibnamefont {Ligeti}},
  \bibinfo {author} {\bibfnamefont {D.~J.}\ \bibnamefont {Robinson}}, \ and\
  \bibinfo {author} {\bibfnamefont {W.~L.}\ \bibnamefont {Sutcliffe}},\ }\href
  {\doibase 10.1103/PhysRevD.99.055008} {\bibfield  {journal} {\bibinfo
  {journal} {Phys. Rev.}\ }\textbf {\bibinfo {volume} {D99}},\ \bibinfo {pages}
  {055008} (\bibinfo {year} {2019})},\ \Eprint
  {http://arxiv.org/abs/1812.07593} {arXiv:1812.07593 [hep-ph]} \BibitemShut
  {NoStop}%
\bibitem [{\citenamefont {Di~Salvo}\ \emph {et~al.}(2018)\citenamefont
  {Di~Salvo}, \citenamefont {Fontanelli},\ and\ \citenamefont
  {Ajaltouni}}]{DiSalvo:2018ngq}%
  \BibitemOpen
  \bibfield  {author} {\bibinfo {author} {\bibfnamefont {E.}~\bibnamefont
  {Di~Salvo}}, \bibinfo {author} {\bibfnamefont {F.}~\bibnamefont
  {Fontanelli}}, \ and\ \bibinfo {author} {\bibfnamefont {Z.~J.}\ \bibnamefont
  {Ajaltouni}},\ }\href {\doibase 10.1142/S0217751X18501695} {\bibfield
  {journal} {\bibinfo  {journal} {Int. J. Mod. Phys.}\ }\textbf {\bibinfo
  {volume} {A33}},\ \bibinfo {pages} {1850169} (\bibinfo {year} {2018})},\
  \Eprint {http://arxiv.org/abs/1804.05592} {arXiv:1804.05592 [hep-ph]}
  \BibitemShut {NoStop}%
\bibitem [{\citenamefont {Blanke}\ \emph
  {et~al.}(2019{\natexlab{b}})\citenamefont {Blanke}, \citenamefont
  {Crivellin}, \citenamefont {Kitahara}, \citenamefont {Moscati}, \citenamefont
  {Nierste},\ and\ \citenamefont {Ni\v{s}and\v{z}i\'c}}]{Blanke:2019qrx}%
  \BibitemOpen
  \bibfield  {author} {\bibinfo {author} {\bibfnamefont {M.}~\bibnamefont
  {Blanke}}, \bibinfo {author} {\bibfnamefont {A.}~\bibnamefont {Crivellin}},
  \bibinfo {author} {\bibfnamefont {T.}~\bibnamefont {Kitahara}}, \bibinfo
  {author} {\bibfnamefont {M.}~\bibnamefont {Moscati}}, \bibinfo {author}
  {\bibfnamefont {U.}~\bibnamefont {Nierste}}, \ and\ \bibinfo {author}
  {\bibfnamefont {I.}~\bibnamefont {Ni\v{s}and\v{z}i\'c}},\ }\href {\doibase
  10.1103/PhysRevD.100.035035} {\bibfield  {journal} {\bibinfo  {journal}
  {Phys. Rev.}\ }\textbf {\bibinfo {volume} {D100}},\ \bibinfo {pages} {035035}
  (\bibinfo {year} {2019}{\natexlab{b}})},\ \Eprint
  {http://arxiv.org/abs/1905.08253} {arXiv:1905.08253 [hep-ph]} \BibitemShut
  {NoStop}%
\bibitem [{\citenamefont {B\"oer}\ \emph {et~al.}(2019)\citenamefont {B\"oer},
  \citenamefont {Kokulu}, \citenamefont {Toelstede},\ and\ \citenamefont {van
  Dyk}}]{Boer:2019zmp}%
  \BibitemOpen
  \bibfield  {author} {\bibinfo {author} {\bibfnamefont {P.}~\bibnamefont
  {B\"oer}}, \bibinfo {author} {\bibfnamefont {A.}~\bibnamefont {Kokulu}},
  \bibinfo {author} {\bibfnamefont {J.-N.}\ \bibnamefont {Toelstede}}, \ and\
  \bibinfo {author} {\bibfnamefont {D.}~\bibnamefont {van Dyk}},\ }\href
  {\doibase 10.1007/JHEP12(2019)082} {\bibfield  {journal} {\bibinfo  {journal}
  {JHEP}\ }\textbf {\bibinfo {volume} {12}},\ \bibinfo {pages} {082} (\bibinfo
  {year} {2019})},\ \Eprint {http://arxiv.org/abs/1907.12554} {arXiv:1907.12554
  [hep-ph]} \BibitemShut {NoStop}%
\bibitem [{\citenamefont {Mu}\ \emph {et~al.}(2019)\citenamefont {Mu},
  \citenamefont {Li}, \citenamefont {Zou},\ and\ \citenamefont
  {Zhu}}]{Mu:2019bin}%
  \BibitemOpen
  \bibfield  {author} {\bibinfo {author} {\bibfnamefont {X.-L.}\ \bibnamefont
  {Mu}}, \bibinfo {author} {\bibfnamefont {Y.}~\bibnamefont {Li}}, \bibinfo
  {author} {\bibfnamefont {Z.-T.}\ \bibnamefont {Zou}}, \ and\ \bibinfo
  {author} {\bibfnamefont {B.}~\bibnamefont {Zhu}},\ }\href {\doibase
  10.1103/PhysRevD.100.113004} {\bibfield  {journal} {\bibinfo  {journal}
  {Phys. Rev. D}\ }\textbf {\bibinfo {volume} {100}},\ \bibinfo {pages}
  {113004} (\bibinfo {year} {2019})},\ \Eprint
  {http://arxiv.org/abs/1909.10769} {arXiv:1909.10769 [hep-ph]} \BibitemShut
  {NoStop}%
\bibitem [{\citenamefont {Hu}\ \emph {et~al.}(2021)\citenamefont {Hu},
  \citenamefont {Li}, \citenamefont {Yang},\ and\ \citenamefont
  {Zheng}}]{Hu:2020axt}%
  \BibitemOpen
  \bibfield  {author} {\bibinfo {author} {\bibfnamefont {Q.-Y.}\ \bibnamefont
  {Hu}}, \bibinfo {author} {\bibfnamefont {X.-Q.}\ \bibnamefont {Li}}, \bibinfo
  {author} {\bibfnamefont {Y.-D.}\ \bibnamefont {Yang}}, \ and\ \bibinfo
  {author} {\bibfnamefont {D.-H.}\ \bibnamefont {Zheng}},\ }\href {\doibase
  10.1007/JHEP02(2021)183} {\bibfield  {journal} {\bibinfo  {journal} {JHEP}\
  }\textbf {\bibinfo {volume} {02}},\ \bibinfo {pages} {183} (\bibinfo {year}
  {2021})},\ \Eprint {http://arxiv.org/abs/2011.05912} {arXiv:2011.05912
  [hep-ph]} \BibitemShut {NoStop}%
\bibitem [{\citenamefont {Penalva}\ \emph {et~al.}(2019)\citenamefont
  {Penalva}, \citenamefont {Hern\'andez},\ and\ \citenamefont
  {Nieves}}]{Penalva:2019rgt}%
  \BibitemOpen
  \bibfield  {author} {\bibinfo {author} {\bibfnamefont {N.}~\bibnamefont
  {Penalva}}, \bibinfo {author} {\bibfnamefont {E.}~\bibnamefont
  {Hern\'andez}}, \ and\ \bibinfo {author} {\bibfnamefont {J.}~\bibnamefont
  {Nieves}},\ }\href {\doibase 10.1103/PhysRevD.100.113007} {\bibfield
  {journal} {\bibinfo  {journal} {Phys. Rev.}\ }\textbf {\bibinfo {volume}
  {D100}},\ \bibinfo {pages} {113007} (\bibinfo {year} {2019})},\ \Eprint
  {http://arxiv.org/abs/1908.02328} {arXiv:1908.02328 [hep-ph]} \BibitemShut
  {NoStop}%
\bibitem [{\citenamefont {Penalva}\ \emph
  {et~al.}(2020{\natexlab{b}})\citenamefont {Penalva}, \citenamefont
  {Hern\'andez},\ and\ \citenamefont {Nieves}}]{Penalva:2020xup}%
  \BibitemOpen
  \bibfield  {author} {\bibinfo {author} {\bibfnamefont {N.}~\bibnamefont
  {Penalva}}, \bibinfo {author} {\bibfnamefont {E.}~\bibnamefont
  {Hern\'andez}}, \ and\ \bibinfo {author} {\bibfnamefont {J.}~\bibnamefont
  {Nieves}},\ }\href {\doibase 10.1103/PhysRevD.101.113004} {\bibfield
  {journal} {\bibinfo  {journal} {Phys. Rev. D}\ }\textbf {\bibinfo {volume}
  {101}},\ \bibinfo {pages} {113004} (\bibinfo {year} {2020}{\natexlab{b}})},\
  \Eprint {http://arxiv.org/abs/2004.08253} {arXiv:2004.08253 [hep-ph]}
  \BibitemShut {NoStop}%
\bibitem [{\citenamefont {Aaij}\ \emph {et~al.}(2022)\citenamefont {Aaij} \emph
  {et~al.}}]{LHCb:2022piu}%
  \BibitemOpen
  \bibfield  {author} {\bibinfo {author} {\bibfnamefont {R.}~\bibnamefont
  {Aaij}} \emph {et~al.} (\bibinfo {collaboration} {LHCb}),\ }\href {\doibase
  10.1103/PhysRevLett.128.191803} {\bibfield  {journal} {\bibinfo  {journal}
  {Phys. Rev. Lett.}\ }\textbf {\bibinfo {volume} {128}},\ \bibinfo {pages}
  {191803} (\bibinfo {year} {2022})},\ \Eprint
  {http://arxiv.org/abs/2201.03497} {arXiv:2201.03497 [hep-ex]} \BibitemShut
  {NoStop}%
\bibitem [{\citenamefont {Detmold}\ \emph {et~al.}(2015)\citenamefont
  {Detmold}, \citenamefont {Lehner},\ and\ \citenamefont
  {Meinel}}]{Detmold:2015aaa}%
  \BibitemOpen
  \bibfield  {author} {\bibinfo {author} {\bibfnamefont {W.}~\bibnamefont
  {Detmold}}, \bibinfo {author} {\bibfnamefont {C.}~\bibnamefont {Lehner}}, \
  and\ \bibinfo {author} {\bibfnamefont {S.}~\bibnamefont {Meinel}},\ }\href
  {\doibase 10.1103/PhysRevD.92.034503} {\bibfield  {journal} {\bibinfo
  {journal} {Phys. Rev.}\ }\textbf {\bibinfo {volume} {D92}},\ \bibinfo {pages}
  {034503} (\bibinfo {year} {2015})},\ \Eprint
  {http://arxiv.org/abs/1503.01421} {arXiv:1503.01421 [hep-lat]} \BibitemShut
  {NoStop}%
\bibitem [{Mar()}]{Marco}%
  \BibitemOpen
  \href@noop {} {\bibinfo  {journal} {Marco Pappagallo (LHCB deputy physics
  coordinator) private communication}\ }\BibitemShut {NoStop}%
\bibitem [{\citenamefont {Penalva}\ \emph {et~al.}(2022)\citenamefont
  {Penalva}, \citenamefont {Hern\'andez},\ and\ \citenamefont
  {Nieves}}]{Penalva:2022vxy}%
  \BibitemOpen
\bibfield  {journal} {  }\bibfield  {author} {\bibinfo {author} {\bibfnamefont
  {N.}~\bibnamefont {Penalva}}, \bibinfo {author} {\bibfnamefont
  {E.}~\bibnamefont {Hern\'andez}}, \ and\ \bibinfo {author} {\bibfnamefont
  {J.}~\bibnamefont {Nieves}},\ }\href {\doibase 10.1007/JHEP04(2022)026}
  {\bibfield  {journal} {\bibinfo  {journal} {JHEP}\ }\textbf {\bibinfo
  {volume} {04}},\ \bibinfo {pages} {026} (\bibinfo {year} {2022})},\ \Eprint
  {http://arxiv.org/abs/2201.05537} {arXiv:2201.05537 [hep-ph]} \BibitemShut
  {NoStop}%
\bibitem [{\citenamefont {Bernlochner}\ \emph {et~al.}(2022)\citenamefont
  {Bernlochner}, \citenamefont {Ligeti}, \citenamefont {Papucci},\ and\
  \citenamefont {Robinson}}]{Bernlochner:2022hyz}%
  \BibitemOpen
  \bibfield  {author} {\bibinfo {author} {\bibfnamefont {F.~U.}\ \bibnamefont
  {Bernlochner}}, \bibinfo {author} {\bibfnamefont {Z.}~\bibnamefont {Ligeti}},
  \bibinfo {author} {\bibfnamefont {M.}~\bibnamefont {Papucci}}, \ and\
  \bibinfo {author} {\bibfnamefont {D.~J.}\ \bibnamefont {Robinson}},\
  }\href@noop {} {\  (\bibinfo {year} {2022})},\ \Eprint
  {http://arxiv.org/abs/2206.11282} {arXiv:2206.11282 [hep-ph]} \BibitemShut
  {NoStop}%
\bibitem [{\citenamefont {Penalva}\ \emph
  {et~al.}(2021{\natexlab{b}})\citenamefont {Penalva}, \citenamefont
  {Hern\'andez},\ and\ \citenamefont {Nieves}}]{Penalva:2021wye}%
  \BibitemOpen
  \bibfield  {author} {\bibinfo {author} {\bibfnamefont {N.}~\bibnamefont
  {Penalva}}, \bibinfo {author} {\bibfnamefont {E.}~\bibnamefont
  {Hern\'andez}}, \ and\ \bibinfo {author} {\bibfnamefont {J.}~\bibnamefont
  {Nieves}},\ }\href {\doibase 10.1007/JHEP10(2021)122} {\bibfield  {journal}
  {\bibinfo  {journal} {JHEP}\ }\textbf {\bibinfo {volume} {10}},\ \bibinfo
  {pages} {122} (\bibinfo {year} {2021}{\natexlab{b}})},\ \Eprint
  {http://arxiv.org/abs/2107.13406} {arXiv:2107.13406 [hep-ph]} \BibitemShut
  {NoStop}%
\bibitem [{\citenamefont {Asadi}\ \emph {et~al.}(2020)\citenamefont {Asadi},
  \citenamefont {Hallin}, \citenamefont {Martin~Camalich}, \citenamefont
  {Shih},\ and\ \citenamefont {Westhoff}}]{Asadi:2020fdo}%
  \BibitemOpen
  \bibfield  {author} {\bibinfo {author} {\bibfnamefont {P.}~\bibnamefont
  {Asadi}}, \bibinfo {author} {\bibfnamefont {A.}~\bibnamefont {Hallin}},
  \bibinfo {author} {\bibfnamefont {J.}~\bibnamefont {Martin~Camalich}},
  \bibinfo {author} {\bibfnamefont {D.}~\bibnamefont {Shih}}, \ and\ \bibinfo
  {author} {\bibfnamefont {S.}~\bibnamefont {Westhoff}},\ }\href {\doibase
  10.1103/PhysRevD.102.095028} {\bibfield  {journal} {\bibinfo  {journal}
  {Phys. Rev. D}\ }\textbf {\bibinfo {volume} {102}},\ \bibinfo {pages}
  {095028} (\bibinfo {year} {2020})},\ \Eprint
  {http://arxiv.org/abs/2006.16416} {arXiv:2006.16416 [hep-ph]} \BibitemShut
  {NoStop}%
\bibitem [{\citenamefont {Meinel}\ and\ \citenamefont
  {Rendon}(2021)}]{Meinel:2021rbm}%
  \BibitemOpen
  \bibfield  {author} {\bibinfo {author} {\bibfnamefont {S.}~\bibnamefont
  {Meinel}}\ and\ \bibinfo {author} {\bibfnamefont {G.}~\bibnamefont
  {Rendon}},\ }\href {\doibase 10.1103/PhysRevD.103.094516} {\bibfield
  {journal} {\bibinfo  {journal} {Phys. Rev. D}\ }\textbf {\bibinfo {volume}
  {103}},\ \bibinfo {pages} {094516} (\bibinfo {year} {2021})},\ \Eprint
  {http://arxiv.org/abs/2103.08775} {arXiv:2103.08775 [hep-lat]} \BibitemShut
  {NoStop}%
\bibitem [{\citenamefont {Meinel}\ and\ \citenamefont
  {Rendon}(2022)}]{Meinel:2021mdj}%
  \BibitemOpen
  \bibfield  {author} {\bibinfo {author} {\bibfnamefont {S.}~\bibnamefont
  {Meinel}}\ and\ \bibinfo {author} {\bibfnamefont {G.}~\bibnamefont
  {Rendon}},\ }\href {\doibase 10.1103/PhysRevD.105.054511} {\bibfield
  {journal} {\bibinfo  {journal} {Phys. Rev. D}\ }\textbf {\bibinfo {volume}
  {105}},\ \bibinfo {pages} {054511} (\bibinfo {year} {2022})},\ \Eprint
  {http://arxiv.org/abs/2107.13140} {arXiv:2107.13140 [hep-lat]} \BibitemShut
  {NoStop}%
\bibitem [{\citenamefont {Leibovich}\ and\ \citenamefont
  {Stewart}(1998)}]{Leibovich:1997az}%
  \BibitemOpen
  \bibfield  {author} {\bibinfo {author} {\bibfnamefont {A.~K.}\ \bibnamefont
  {Leibovich}}\ and\ \bibinfo {author} {\bibfnamefont {I.~W.}\ \bibnamefont
  {Stewart}},\ }\href {\doibase 10.1103/PhysRevD.57.5620} {\bibfield  {journal}
  {\bibinfo  {journal} {Phys. Rev. D}\ }\textbf {\bibinfo {volume} {57}},\
  \bibinfo {pages} {5620} (\bibinfo {year} {1998})},\ \Eprint
  {http://arxiv.org/abs/hep-ph/9711257} {arXiv:hep-ph/9711257} \BibitemShut
  {NoStop}%
\bibitem [{\citenamefont {Du}\ \emph {et~al.}(2022)\citenamefont {Du},
  \citenamefont {Hern\'andez},\ and\ \citenamefont {Nieves}}]{Du:2022fxg}%
  \BibitemOpen
  \bibfield  {author} {\bibinfo {author} {\bibfnamefont {M.-L.}\ \bibnamefont
  {Du}}, \bibinfo {author} {\bibfnamefont {E.}~\bibnamefont {Hern\'andez}}, \
  and\ \bibinfo {author} {\bibfnamefont {J.}~\bibnamefont {Nieves}},\
  }\href@noop {} {\  (\bibinfo {year} {2022})},\ \Eprint
  {http://arxiv.org/abs/2207.02109} {arXiv:2207.02109 [hep-ph]} \BibitemShut
  {NoStop}%
\bibitem [{\citenamefont {Nieves}\ and\ \citenamefont
  {Pavao}(2020)}]{Nieves:2019nol}%
  \BibitemOpen
  \bibfield  {author} {\bibinfo {author} {\bibfnamefont {J.}~\bibnamefont
  {Nieves}}\ and\ \bibinfo {author} {\bibfnamefont {R.}~\bibnamefont {Pavao}},\
  }\href {\doibase 10.1103/PhysRevD.101.014018} {\bibfield  {journal} {\bibinfo
   {journal} {Phys. Rev. D}\ }\textbf {\bibinfo {volume} {101}},\ \bibinfo
  {pages} {014018} (\bibinfo {year} {2020})},\ \Eprint
  {http://arxiv.org/abs/1907.05747} {arXiv:1907.05747 [hep-ph]} \BibitemShut
  {NoStop}%
\bibitem [{\citenamefont {Nieves}\ \emph {et~al.}(2019)\citenamefont {Nieves},
  \citenamefont {Pavao},\ and\ \citenamefont {Sakai}}]{Nieves:2019kdh}%
  \BibitemOpen
  \bibfield  {author} {\bibinfo {author} {\bibfnamefont {J.}~\bibnamefont
  {Nieves}}, \bibinfo {author} {\bibfnamefont {R.}~\bibnamefont {Pavao}}, \
  and\ \bibinfo {author} {\bibfnamefont {S.}~\bibnamefont {Sakai}},\ }\href
  {\doibase 10.1140/epjc/s10052-019-6929-7} {\bibfield  {journal} {\bibinfo
  {journal} {Eur. Phys. J. C}\ }\textbf {\bibinfo {volume} {79}},\ \bibinfo
  {pages} {417} (\bibinfo {year} {2019})},\ \Eprint
  {http://arxiv.org/abs/1903.11911} {arXiv:1903.11911 [hep-ph]} \BibitemShut
  {NoStop}%
\bibitem [{\citenamefont {Itzykson}\ and\ \citenamefont
  {Zuber}(1980)}]{Itzykson:1980rh}%
  \BibitemOpen
  \bibfield  {author} {\bibinfo {author} {\bibfnamefont {C.}~\bibnamefont
  {Itzykson}}\ and\ \bibinfo {author} {\bibfnamefont {J.~B.}\ \bibnamefont
  {Zuber}},\ }\href {http://dx.doi.org/10.1063/1.2916419} {\emph {\bibinfo
  {title} {{Quantum Field Theory}}}},\ International Series In Pure and Applied
  Physics\ (\bibinfo  {publisher} {McGraw-Hill},\ \bibinfo {address} {New
  York},\ \bibinfo {year} {1980})\BibitemShut {NoStop}%
\bibitem [{\citenamefont {Kiers}\ and\ \citenamefont
  {Soni}(1997)}]{Kiers:1997zt}%
  \BibitemOpen
  \bibfield  {author} {\bibinfo {author} {\bibfnamefont {K.}~\bibnamefont
  {Kiers}}\ and\ \bibinfo {author} {\bibfnamefont {A.}~\bibnamefont {Soni}},\
  }\href {\doibase 10.1103/PhysRevD.56.5786} {\bibfield  {journal} {\bibinfo
  {journal} {Phys. Rev. D}\ }\textbf {\bibinfo {volume} {56}},\ \bibinfo
  {pages} {5786} (\bibinfo {year} {1997})},\ \Eprint
  {http://arxiv.org/abs/hep-ph/9706337} {arXiv:hep-ph/9706337} \BibitemShut
  {NoStop}%
\bibitem [{\citenamefont {Alonso}\ \emph {et~al.}(2016)\citenamefont {Alonso},
  \citenamefont {Kobach},\ and\ \citenamefont
  {Martin~Camalich}}]{Alonso:2016gym}%
  \BibitemOpen
  \bibfield  {author} {\bibinfo {author} {\bibfnamefont {R.}~\bibnamefont
  {Alonso}}, \bibinfo {author} {\bibfnamefont {A.}~\bibnamefont {Kobach}}, \
  and\ \bibinfo {author} {\bibfnamefont {J.}~\bibnamefont {Martin~Camalich}},\
  }\href {\doibase 10.1103/PhysRevD.94.094021} {\bibfield  {journal} {\bibinfo
  {journal} {Phys. Rev. D}\ }\textbf {\bibinfo {volume} {94}},\ \bibinfo
  {pages} {094021} (\bibinfo {year} {2016})},\ \Eprint
  {http://arxiv.org/abs/1602.07671} {arXiv:1602.07671 [hep-ph]} \BibitemShut
  {NoStop}%
\bibitem [{\citenamefont {Alonso}\ \emph
  {et~al.}(2017{\natexlab{b}})\citenamefont {Alonso}, \citenamefont
  {Martin~Camalich},\ and\ \citenamefont {Westhoff}}]{Alonso:2017ktd}%
  \BibitemOpen
  \bibfield  {author} {\bibinfo {author} {\bibfnamefont {R.}~\bibnamefont
  {Alonso}}, \bibinfo {author} {\bibfnamefont {J.}~\bibnamefont
  {Martin~Camalich}}, \ and\ \bibinfo {author} {\bibfnamefont {S.}~\bibnamefont
  {Westhoff}},\ }\href {\doibase 10.1103/PhysRevD.95.093006} {\bibfield
  {journal} {\bibinfo  {journal} {Phys. Rev. D}\ }\textbf {\bibinfo {volume}
  {95}},\ \bibinfo {pages} {093006} (\bibinfo {year} {2017}{\natexlab{b}})},\
  \Eprint {http://arxiv.org/abs/1702.02773} {arXiv:1702.02773 [hep-ph]}
  \BibitemShut {NoStop}%
\bibitem [{\citenamefont {Tanaka}\ and\ \citenamefont
  {Watanabe}(2010)}]{Tanaka:2010se}%
  \BibitemOpen
  \bibfield  {author} {\bibinfo {author} {\bibfnamefont {M.}~\bibnamefont
  {Tanaka}}\ and\ \bibinfo {author} {\bibfnamefont {R.}~\bibnamefont
  {Watanabe}},\ }\href {\doibase 10.1103/PhysRevD.82.034027} {\bibfield
  {journal} {\bibinfo  {journal} {Phys. Rev. D}\ }\textbf {\bibinfo {volume}
  {82}},\ \bibinfo {pages} {034027} (\bibinfo {year} {2010})},\ \Eprint
  {http://arxiv.org/abs/1005.4306} {arXiv:1005.4306 [hep-ph]} \BibitemShut
  {NoStop}%
\bibitem [{\citenamefont {Papucci}\ and\ \citenamefont
  {Robinson}(2022)}]{Papucci:2021pmj}%
  \BibitemOpen
  \bibfield  {author} {\bibinfo {author} {\bibfnamefont {M.}~\bibnamefont
  {Papucci}}\ and\ \bibinfo {author} {\bibfnamefont {D.~J.}\ \bibnamefont
  {Robinson}},\ }\href {\doibase 10.1103/PhysRevD.105.016027} {\bibfield
  {journal} {\bibinfo  {journal} {Phys. Rev. D}\ }\textbf {\bibinfo {volume}
  {105}},\ \bibinfo {pages} {016027} (\bibinfo {year} {2022})},\ \Eprint
  {http://arxiv.org/abs/2105.09330} {arXiv:2105.09330 [hep-ph]} \BibitemShut
  {NoStop}%
\bibitem [{\citenamefont {Mertig}\ \emph {et~al.}(1991)\citenamefont {Mertig},
  \citenamefont {B$\ddot{\rm o}$hm},\ and\ \citenamefont
  {Denner}}]{MERTIG1991345}%
  \BibitemOpen
  \bibfield  {author} {\bibinfo {author} {\bibfnamefont {R.}~\bibnamefont
  {Mertig}}, \bibinfo {author} {\bibfnamefont {M.}~\bibnamefont {B$\ddot{\rm
  o}$hm}}, \ and\ \bibinfo {author} {\bibfnamefont {A.}~\bibnamefont
  {Denner}},\ }\href@noop {} {\bibfield  {journal} {\bibinfo  {journal}
  {Computer Physics Communications}\ }\textbf {\bibinfo {volume} {64}},\
  \bibinfo {pages} {345} (\bibinfo {year} {1991})}\BibitemShut {NoStop}%
\bibitem [{\citenamefont {Shtabovenko}\ \emph {et~al.}(2016)\citenamefont
  {Shtabovenko}, \citenamefont {Mertig},\ and\ \citenamefont
  {Orellana}}]{Shtabovenko:2016sxi}%
  \BibitemOpen
  \bibfield  {author} {\bibinfo {author} {\bibfnamefont {V.}~\bibnamefont
  {Shtabovenko}}, \bibinfo {author} {\bibfnamefont {R.}~\bibnamefont {Mertig}},
  \ and\ \bibinfo {author} {\bibfnamefont {F.}~\bibnamefont {Orellana}},\
  }\href {\doibase 10.1016/j.cpc.2016.06.008} {\bibfield  {journal} {\bibinfo
  {journal} {Comput. Phys. Commun.}\ }\textbf {\bibinfo {volume} {207}},\
  \bibinfo {pages} {432} (\bibinfo {year} {2016})},\ \Eprint
  {http://arxiv.org/abs/1601.01167} {arXiv:1601.01167 [hep-ph]} \BibitemShut
  {NoStop}%
\bibitem [{\citenamefont {Shtabovenko}\ \emph {et~al.}(2020)\citenamefont
  {Shtabovenko}, \citenamefont {Mertig},\ and\ \citenamefont
  {Orellana}}]{Shtabovenko:2020gxv}%
  \BibitemOpen
  \bibfield  {author} {\bibinfo {author} {\bibfnamefont {V.}~\bibnamefont
  {Shtabovenko}}, \bibinfo {author} {\bibfnamefont {R.}~\bibnamefont {Mertig}},
  \ and\ \bibinfo {author} {\bibfnamefont {F.}~\bibnamefont {Orellana}},\
  }\href {\doibase 10.1016/j.cpc.2020.107478} {\bibfield  {journal} {\bibinfo
  {journal} {Comput. Phys. Commun.}\ }\textbf {\bibinfo {volume} {256}},\
  \bibinfo {pages} {107478} (\bibinfo {year} {2020})},\ \Eprint
  {http://arxiv.org/abs/2001.04407} {arXiv:2001.04407 [hep-ph]} \BibitemShut
  {NoStop}%
\bibitem [{\citenamefont {{Wolfram Research, Inc.}}()}]{Mathematica}%
  \BibitemOpen
  \bibfield  {author} {\bibinfo {author} {\bibnamefont {{Wolfram Research,
  Inc.}}},\ }\href@noop {} {\enquote {\bibinfo {title} {{Mathematica, {V}ersion
  12.3}},}\ }\bibinfo {note} {{Champaign, IL, 2021}}\BibitemShut {NoStop}%
\end{thebibliography}%
\end{document}